\documentclass[%
 reprint,
superscriptaddress,
 amsmath,amssymb,
 aps,
linenumbers,
]{revtex4-2}

\usepackage{graphicx}% Include figure files
\usepackage{dcolumn}% Align table columns on decimal point
\usepackage{bm}% bold math
\usepackage[english]{babel}
\begin{document}

\preprint{APS/123-QED}

\title{Climate change and human mobility will shape dengue emergence risk in Europe}% Force line breaks with \\

\author{Charley Presigny}
 \affiliation{Department of Physics \& Astronomy ‘Galileo Galilei’, University of Padua, Padua, Italy} \affiliation{Istituto Nazionale di Fisica Nucleare, Sez. Padova, Padua, Italy}
\author{Paolo Baglioni}%
\affiliation{Istituto Nazionale di Fisica Nucleare, Sez. Milano Bicocca, Milan, Italy
}%
\author{Pietro Rotondo}
\affiliation{Department of Mathematical Sciences, University of Parma, Parma, Italy}%
\author{Michele Allegra}
\affiliation{Department of Physics \& Astronomy ‘Galileo Galilei’, University of Padua, Padua, Italy} \affiliation{Padova Neuroscience Center, University of Padua, Padua, Italy}
\author{Annalisa Barla} \affiliation{Department of Informatics, Bioengineering, Robotics, and Systems Engineering, University of Genoa, Genoa, Italy} \affiliation{Machine Learning Genoa Center, University of Genoa, Genoa, Italy}
\author{Manlio De Domenico} \affiliation{Department of Physics \& Astronomy ‘Galileo Galilei’, University of Padua, Padua, Italy} \affiliation{Istituto Nazionale di Fisica Nucleare, Sez. Padova, Padua, Italy} \affiliation{Padua Center for Network Medicine, University of Padua, Padua, Italy}

\date{\today}% It is always \today, today,
             %  but any date may be explicitly specified

\begin{abstract}
The risk of local arbovirus outbreaks in Europe is expected to increase due to climate change, as suggested by the multiplication of arbovirus outbreaks in the last decades. Europe has historically been a non-endemic region, making it vital to pinpoint which populations are potentially exposed—and under which conditions- so we can build truly robust epidemic preparedness capabilities. We introduce an integrated, multi-scale model that fuses a mechanistic transmission engine with a vector abundance framework, all embedded in a mobility-driven metapopulation system capturing human, vector, and air-traffic movement. To this end, we combine climate and population projections with mobility data to estimate and map dengue emergence risk in Europe throughout the 21st century. Additionally, we introduce a dedicated migration model that explores how climate-driven population redistribution could alter these risk estimates.Assuming the climate avoids major tipping points, model-derived risk indicators increase substantially under most emissions scenarios.
While the spatio-temporal risk will remain largely driven by importation, our results indicate a gradual transition toward an environment-driven regime, particularly under the worst-case emissions scenario. To better anticipate and manage recurrent arbovirus outbreaks, our findings highlight the need to integrate mobility pathways and climate-driven population redistribution into predictive models of vector-borne disease emergence in temperate regions.
\end{abstract}

%\keywords{Suggested keywords}%Use showkeys class option if keyword
                              %display desired
\nolinenumbers
\maketitle

%\tableofcontents

\section{Introduction}
%\linenumbers
A direct consequence of the ongoing climate change is the unprecedented increase in number and intensity of vector-borne disease outbreaks worldwide, now and in the future. Arboroviruses such as dengue, chikungunya, Zika or yellow fever already represent a potential burden for more than 3.9 billion human beings \cite{noauthor_vector-borne_nodate}. The expansion of the ecological range of invasive carriers, such as the mosquito \textit{Aedes albopictus}, and the multiplication of outbreaks in temperate areas (\cite{roche_spread_2015,manica_transmission_2017,sacco_autochthonous_2024,aranda_arbovirus_2018}) suggest that some current non-endemic areas are on the road to experience recurrent outbreaks within the current century (\cite{farooq_impact_2025,bouzid_climate_2014}). By assessing the current or future risk associated to an ensemble of local, regional or global areas, risk modeling is a crucial tool in guiding public health policy to target interventions or elaborate early warning systems (\cite{brady_why_2025,rees_risk_2019}).

The last decade has seen a tremendous scientific endeavor to assess and predict the vector-borne disease risk and understand its determinant co-variates from the regional to the global level \cite{messina_current_2019,lim_systematic_2023}. Common risk assessment uses statistical models based on disease and vector occurrence data, often combined with climate-data (e.g. temperature, precipitation, humidity,...) or land use data, to estimate spatial indices of vector suitability or vector abundance \cite{kraemer_global_2015,lim_overlapping_2025,rogers_using_2014,zhang_modeling_2020}. Then, the population at risk is estimated using geospatial projections aligned with the Shared Socioeconomic Pathways (SSPs), which encode assumptions on climate adaptation and mitigation (\cite{oneill_new_2014,jones_spatially_2016}). By translating hypotheses on migration and fertility into spatially explicit population maps, these scenarios are essential for assessing the future co-distribution of carriers and humans. More rarely, mechanistic models integrate climate-dependent epidemiological or biological parameters (e.g. mosquito life expectancy) into large-scale simulations to estimate the risk in terms of common epidemiological indicators such as the basic reproduction number or the number of infected cases \cite{mordecai_detecting_2017,nakase_population_2024,zardini_estimating_2024,de_souza_effects_2024,li_climate-driven_2019}.

While rising global temperatures are consistently linked to increased vector suitability (\cite{bouzid_climate_2014,colon-gonzalez_pnas_2018,childs_climate_2025}), local co-presence of arboviruses is required for this to translate into heightened transmission risk.
Actually, since the mobility range of a vector is limited to several hundreds of meters \cite{moore_estimating_2022}, human mobility was shown to be a primary driver of vector-borne disease transport within cities \cite{guzzetta_quantifying_2018,soriano-panos_vector-borne_2020}, countries  \cite{wesolowski_impact_2015,gibb_interactions_2023} or globally \cite{yang_mapping_2025,zhang_spread_2017}.

While integrating climate variables is standard and a growing number of studies integrate aspects of human mobility, a mechanistic model incorporating climate-dependent vector abundance, biological and epidemiological parameters and different routes of human mobility is still missing \cite{brady_why_2025}. In particular, while importation of cases has been suggested to be a relevant explanatory factor of the local emergence of vector-borne epidemics, it has never been integrated in data-driven mechanistic models \cite{menegale_risk_2025}. Moreover, while current risk estimates accounting for human mobility exist (\cite{wesolowski_impact_2015,zhang_modeling_2020,guzzetta_quantifying_2018}), future risk estimates accounting for mobility projections remain scarce.
In the same vein, the most popular population projections rely on historical migration patterns, accounting for migrations due to economic causes but largely overlooking climate, a future key driver of migration (\cite{abel_estimating_2013,abel_quantifying_2014}). In this regard, current projections could lead to overestimation of the population at risk of vector-borne diseases in areas that will become less suitable for human life or vice versa.

Here, we address these gaps by proposing an integrated approach to to scenario-based epidemic risk mapping. We first develop a joint model that combines information about vector abundance and mobility patterns. To this aim, we use a metapopulation framework to obtain a mechanistic transmission model accounting for different mobility sources (human mobility, vector mobility and air-traffic-based infection importation), vector-abundance, and population and climate projections. Next, we combine a generalization of the radiation model \cite{simini_universal_2012,raimondo_network_2022} with a human suitability index \cite{xu_human_niche_2020} derived from climate projections to obtain updated population projections accounting for climate-related migration. Finally, we  combine the two models. 

Our unified approach is applied to 
map scenario-dependent dengue emergence risk in Europe during the 21st century. Moreover, we assess the influence of infection importation on scenario-dependent risk estimates.
Risk indicators are provided at the NUTS3 level -- i.e., the smallest geographical breakdown in the European Union's Nomenclature of Territorial units for Statistics system -- for dengue carried by $\textit{Aedes albopictus}$, an invasive mosquito species which is already highly established in Europe.

\begin{figure*}[ht!]
    \centering
    \includegraphics[width=1\textwidth]{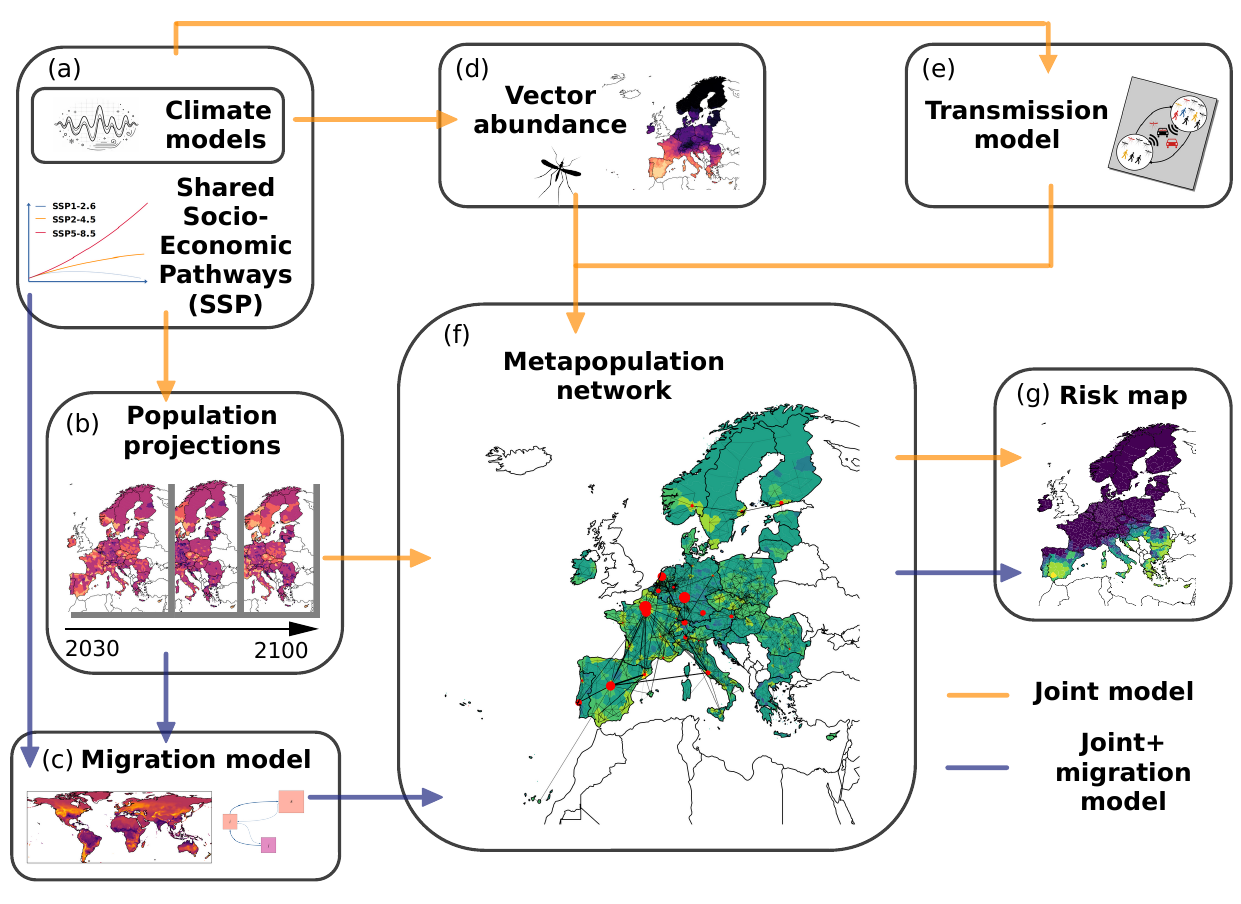}
    \caption{\textbf{Integrated framework for scenario-based dengue emergence risk mapping.}
    \textbf{a-b)} For each Shared Socio-Economic Pathways (SSPs) and each year, we use available population projections \cite{jones_spatially_2016} that we associate with each NUTS3 patch in the metapopulation network. \textbf{c)} From climate and population projections, the migration model produces alternative population distributions that can be associated with each NUTS3 patch.
    Within a given SSPs, climate models provide temperature and precipitation projections that are inputs to the \textbf{d)} vector abundance model and \textbf{e)} the transmission model whose outcome is the spatio-temporal distributions of vectors and the values of biological and epidemiological parameters, respectively. \textbf{f)} The European Union metapopulation network has 1200 patches, all of them being associated with a climate-dependent vector distribution and biological/epidemiological parameters. NUTS3 patches are linked with the mobility matrix $\mathbf{P}$ whose strongest directed weights ($\mathbf{P}>1000$ daily individuals) are represented in black lines. Red dots represent airports and their size is correlated with their incoming air traffic from dengue-endemic territories. The colormap illustrates the present-day population within each patch. \textbf{g)} From the previously described elements, risk maps of dengue-emergence risk are produced. Orange arrows indicate the data processing flow within the joint model. Blue arrows indicate additional processing associated with the migration model. More technical details are to be found in the Methods section and Supplementary Information.
    }
    \label{fig:schematic}
\end{figure*}

\section{Results}
The metapopulation model is divided into 1200 NUTS3 patches covering the European Union (EU), Switzerland, Liechtenstein and Norway. Current population for each patch is obtained through official census statistics. Short-range human mobility between patches is inferred from a gravity model \cite{balcan_2009}, while its long-range component is obtained from OAG air traffic data. Vector mobility flows are inferred from the human ones. Extra-EU air traffic from countries endemic with dengue are inferred from OAG air traffic data while the importation rate of infected cases is inferred from historical data from the European Center for Disease Control (ECDC), accounting for underreporting of cases \cite{hitchings_2025}. A schematic representation of the joint model is shown in Fig. \ref{fig:schematic}. For each year, each estimation is constructed by updating the population of each patch according to the population projections \cite{jones_spatially_2016}. Furthermore, we use the climate projections to associate a time-dependent value of temperature and precipitation to each patch from which we infer the time-dependent vector abundance. Note that the short-range human mobility is a function of the human population, which makes it evolve in time according to the population projections. Furthermore, the long-range  human mobility and the importation of infected cases rate rise by 1.1 \% each year (from 2024) following the predictions on the European air traffic provided by EuroControl \cite{EUROCONTROL2024}. All mobility parameters for the transmission model and parameters of the vector abundance model are calibrated from European data.

An overview of the transmission model's structure and the proposed methodological approach is available in \textit{Material and Methods}, while technical details, description and sources of data are to be found in Supplementary Information.
\begin{figure*}[ht!]
    \centering
    \includegraphics[width=\textwidth]{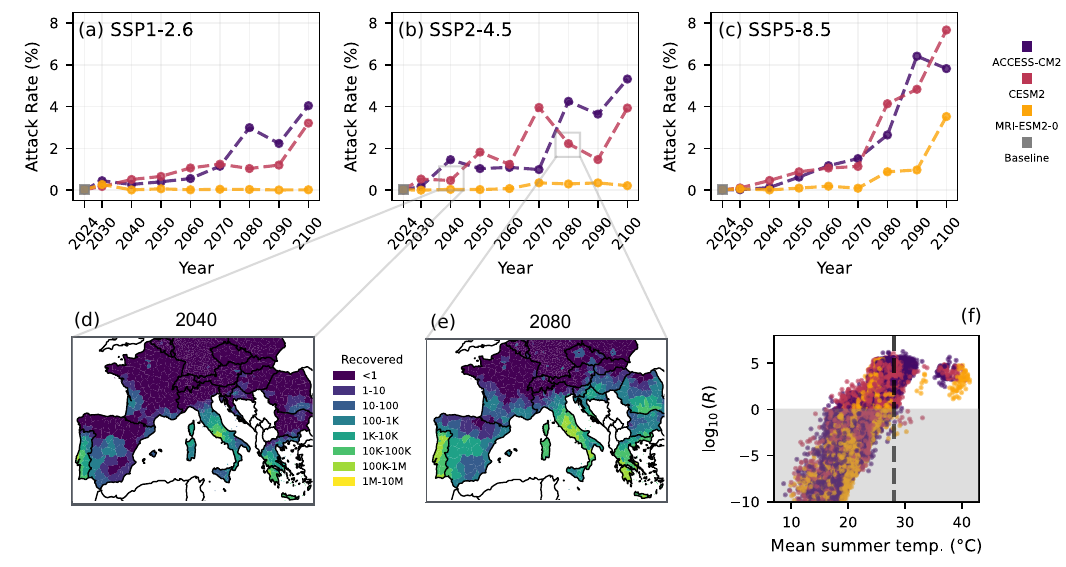}
    \caption{\textbf{Evolution of the estimated proportion of the population-at-risk indices of dengue infection in Europe from 2024 to 2100 based on the ACCESS-CM2, CESM2 and MRI-ESM2-0 climate projections.}  Model-derived annual attack-rate proxy from the joint model in the \textbf{a)} SSP1-2.6, \textbf{b)} SSP2-4.5 and \textbf{c)} SSP5-8.5 scenarios. Magnitude of the cumulative recovered individuals for the CESM2 climate projections (SSP2-4.5) in \textbf{d)} 2040 and \textbf{e)} 2080. \textbf{f)} Magnitude of the population-at-risk index as a function of the daily mean summer temperature for each NUTS3 patch of the joint model in each year.
    The dashed black line represents the isotherm \text{$28^\circ$}C which is approximately the upper bound of the optimal transmission temperature for \textit{Aedes albopictus} in the transmission model \cite{mordecai_detecting_2017}. We also highlighted with a shaded grey area the nonphysical region where the magnitude of the population-at-risk index takes negative values due to the mean-field approximation. The mean summer temperature is the daily average temperature from 1st June to 1st September (31st May-31st August in leap years). The magnitude of the population-at-risk index is computed as the logarithm in base 10 of the cumulative recovered individuals in a given year. The epidemics starts by importation of infected cases (no infected individual seeded at time $t=0$).  Population is estimated for every year according to the population projections \cite{jones_spatially_2016}. Attack rate is computed as the total number of infected individual in a year over the total projected population in this very year. This quantity is a comparative, scenario-dependent risk indicator rather than as a point estimate of future incidence or population at risk. Parameter values used in the simulations are specified in Methods and Tab. S2.
    }
    \label{fig:main_result_DENV}
\end{figure*}

\begin{figure*}[ht!]
    \centering
    \includegraphics[width=1\textwidth]{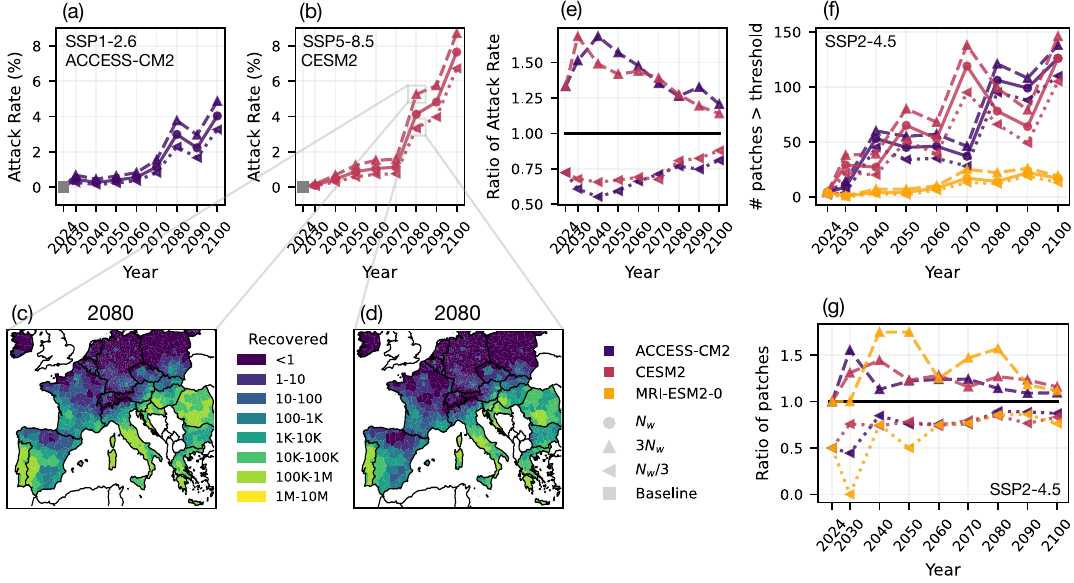}
    \caption{\textbf{Sensitivity of the estimated proportion of the population-at-risk indices of dengue infection in Europe to different importation scenarios.} \textbf{a)} Attack rate in the base (solid line), low (dotted line) and high importation (dashed line) scenario in the SSP1-2.6 using the ACCESS-CM2 climate projection. \textbf{b)} Attack rate in the base (solid line), low (dotted line) and high importation (dashed line) scenario in the SSP5-8.5 using the CESM2 climate projection. \textbf{c)} Magnitude of the cumulative recovered individuals for the CESM2 climate
    projection (SSP5-8.5) in 2080 for the  high importation scenario \textbf{d)} as c), for the low importation scenario. \textbf{e)} Ratio of the attack rate of the high importation scenario (dashed lines) and the low importation scenario (dotted lines) using ACCESS-CM2/SSP1-2.6 (purple) and CESM2/SSP5-8.5 (red) climate projections and scenarios. \textbf{f)} Number of NUTS3 patches whose attack rate exceeds $20\%$ i.e. high-risk patches in the SSP2-4.5 using the ACCESS-CM2, CESM2 and MRI-ESM2-0 climate projections. \textbf{g)} Ratio of the number of high-risk patches in the high importation scenario (dashed lines) and the low importation scenario (dotted lines) over the base importation scenario using the ACCESS-CM2, CESM2 and MRI-ESM2-0 climate projections (SSP2-4.5). In panels e) and g) the base importation scenario is represented with a black line. In all panels, the base, low and high importation scenarios are defined by a daily importation rate of $N_w$ (circles), $N_w/{3}$ (left-pointing triangle) and $3N_W$ (upward triangle), respectively. Colors indicate different climate projections: ACCESS-CM2 (purple), CESM2 (red) and MRI-ESM2-0 (yellow).
    The epidemics start by importation of infected cases (no infected individual seeded at time $t=0$). Base attack rate is estimated for the year 2024. Population is estimated from every year according to population projections \cite{jones_spatially_2016}.
    Parameter values used in the simulations are specified in Methods and Tab. S2.
    }
    \label{fig:sensitivity_result_DENV}
\end{figure*}

\subsection{Scenario-based dengue risk indicators increase across almost all scenarios.}
We estimate dengue emergence
risk from the mean-field prediction of the cumulative number of recovered individuals in each patch over one simulated year. From this, we compute a model-derived annual attack-rate proxy, i.e. the fraction of the local population reached by the deterministic transmission model.
Note that throughout the analysis, this quantity is interpreted as a comparative, scenario-dependent risk indicator rather than as a point forecast of future incidence or population at risk.
Present-day risk is evaluated using the climate record for dengue in 2024, a year marked by major outbreaks in Europe \cite{sacco_autochthonous_2024,menegale_risk_2025} (we refer to these estimations in all the figures as ``baseline'' and represented by grey squares). Climate projections are derived from three independent models—ACCESS-CM2, CESM2, and MRI-ESM2-0—selected for their distinct model lineages and their ability to reproduce historical summer conditions across Mediterranean and central Europe \cite{merrifield_climate_2023,palmer_performance-based_2023} (see TextS1.6 for details). Future risk is assessed under three SSPs—SSP1-2.6, SSP2-4.5, and SSP5-8.5—which couple socioeconomic trajectories with greenhouse gas scenarios, corresponding to approximately 2°C, 3°C, and 5°C global warming above pre-industrial levels by 2100, respectively \cite{van2011representative}.

The joint model indicates a clear increase in the scenario-dependent dengue risk indicator during the 21st century across all considered SSPs based on the ACCESS-CM2 and CESM2 climate projections (Fig. \ref{fig:main_result_DENV}). 
Based on the latter,   
the model-derived annual attack-rate proxy increases, as compared to present day, by at least a factor 150 in SSP1-2.6 and SSP2-4.5 (reaching 3.2 \% of the population) and at least a factor 300 in SSP5-8.5 (reaching 6 \% of the population). Considering the differential evolution of the population in the SSPs, this corresponds to a population-at-risk index exceeding 10 million individuals
in the optimistic (SSP1-2.6) and intermediate (SSP2-4.5) scenarios, and more than 40 millions in the worst-case scenario, SSP5-8.5 (see Fig. S7). On the contrary, results based on the MRI-ESM2-0 model predict a stable proportion of the population at risk in SSP1-2.6  during the century, and a slight increase in SSP2-4.5 (by a factor 10 from 2070). In SSP5-8.5, the proportion slightly increases until 2070, before undergoing a fast increase reaching almost 4\% of the European population in 2100. These results are confirmed by the evolution of the duration of the epidemic season: for MRI-ESM2-0, it decreases (in the SSP1-2.6 scenario) or remains stable (in the SSP2-4.5 scenario), while for the other climate projections it continuously increases across all scenarios (see Fig.S8-10).

Differences between predictions obtained with different climate projections may be explained by 
temperature, which is known to be an important predictor of vector abundance and dengue transmission \cite{farooq_impact_2025,li_big_2022}. Actually, we observe that the mean summer temperature at the European patch-level is consistently lower in the MRI-ESM2-0 climate projections as compared with the others, which could explain the difference in epidemic outcome observed in the joint model (see Figs. \ref{fig:main_result_DENV}f,S11). 

\subsection{Spatio-temporal evolution of dengue emergence risk.}
By 2030 and across all SSPs, the highest values of the dengue emergence risk indicator are
concentrated around international airports, i.e. sources of imported cases located in southern Europe, including Italy, Spain and Portugal (see Figs. \ref{fig:schematic}, \ref{fig:main_result_DENV}d-e). A notable exception is Greece, where elevated risk indicators emerge despite the absence of a major source of imported cases from endemic countries, suggesting a key role of internal mobility within Europe for areas that are suitable for vector development. The progressive evolution of the spatial expansion of the risk indicator during the century is SSP-dependent. In SSP1-2.6 and SSP2-4.5, the increase in the risk indicator is mostly driven by the intensification in areas that already present an enhanced risk rather than by a significant spatial expansion (see Figs. S12-14). In contrast, SSP5-8.5 also shows substantial spatial expansion of elevated emergence risk,
extending to  Cyprus, Bulgaria, Romania, Hungary, Croatia, Slovenia, the south and west of France and the north of Spain (see Figs.~S12-14). We note that while the increase of the population-at-risk index is mostly located in the western part of Europe in ACCESS-CM2, it is mostly located in eastern Europe for CESM2 and MRI-ESM2-0 (where it predicts an increase) suggesting precise spatial patterns are also climate-projection-dependent.

\subsection{From importation-driven to environment-driven risk.} Historical records of case importation in Europe vary greatly from one year to another depending on  the epidemic situation in the endemic areas.  We perform a sensitivity analysis on the impact of the daily importation rate of infected cases on the scenario-dependent dengue risk indicator.

Together with our base scenario of case importation, we consider two scenarios: one in which the daily importation rate is $N_w^{low}= \frac{N_w}{3}$ and the other one with $N_w^{high} = 3N_w$. $N_w^{low}$ corresponds to importation rates that were common in Europe before 2024 (see ECDC data in Tab.S1).

We show that the importation rate of infected cases is a major driver of the dengue emergence risk indicator across all climate projections and SSPs (see Fig. \ref{fig:sensitivity_result_DENV}). Indeed, the scenarios $N_w^{low}$ and $N_w^{high}$ respectively decrease and increase the model-derived risk indicator 
by at least $25\%$, almost consistently over the century. This leads to an uncertainty on the population-at-risk index amounting to millions of individuals in the second part of the century for predictions based on ACCESS-CM2 and CESM2 (see Fig.S15). These results are confirmed by the number of high-risk patches i.e. patches that present an attack rate superior to 20 \% of their population (see Figs. \ref{fig:sensitivity_result_DENV}f-g, S17). Actually, the uncertainty of the number of such patches is also of the order of $25\%$ compared to the base importation scenario. Note that this uncertainty proceeds from areas that are already highly affected in the low importation scenario rather than a significant spatial expansion of the population at risk to other areas (see Fig. \ref{fig:sensitivity_result_DENV}c-d).

We observe that the sensitivity of the estimated population-at-risk index and the estimated number of high-risk patches to the importation rate dampens over the course of the century (see Figs. \ref{fig:sensitivity_result_DENV}e-g and S16, S18). Furthermore, the worse the SSPs, the stronger the sensitivity dampening. At the same time, we observe that the number of high-risk patches increases from almost zero in 2024 to at most 150 patches in SSP1-2.6 and SSP2-4.5, and up to more than 250 in SSP5-8.5. This suggests a gradual shift from an importation-driven regime to a more environment-driven regime, where increasingly permissive local conditions reduce the relative dependence of emergence risk on the number of imported cases.

\subsection{Sensitivity to climate-related migration.} Current population projections depend on hypotheses about future migration based on historical migration patterns, therefore covering essentially economic aspects of migration, not climatic ones \cite{abel_estimating_2013,niva_worlds_2023,KC2017181}. Here, we build alternative population distributions that explore the potential effect of climate-related migration within the large-scale migration model. Essentially, the model quantifies the migration flow between areas by using climatic projections to compute an index of human suitability encompassing the entire world. We consider an extreme scenario in which every individuals migrate according to a heterogeneous diffusion pattern (see Methods and Text S.1.8-1.9 for more details). It is worth remarking that this component should be interpreted as a stress test of how climate-driven population redistribution could affect dengue emergence risk, rather than as a demographic forecast.
As a result, simulations from the joint model combined with the large-scale migration component should be interpreted as an upper bound on the impact of realistic climate-driven migration on the estimated risk.

\begin{figure*}[t]
    \centering
    \includegraphics[width=1\textwidth]{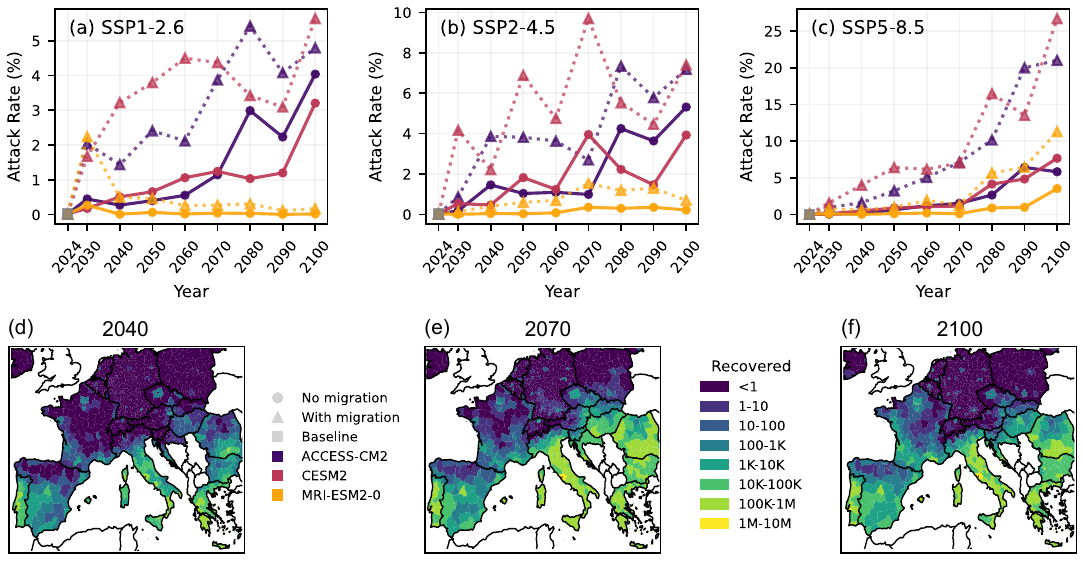}
    \caption{\textbf{Evolution of the estimated proportion of the population-at-risk indices of dengue infection in Europe from 2024 to 2100 based on alternative population projections.}  Attack rate in the \textbf{a)} SSP1-2.6, \textbf{b)} SSP2-4.5 and \textbf{c)} SSP5-8.5 scenarios on the ACCESS-CM2, CESM2 and MRI-ESM2-0 climate projections using the alternative population projections (dotted lines) from the migration model and the classic population projections \cite{jones_spatially_2016} (solid lines). Magnitude of the cumulative recovered individuals for the CESM2 climate projections (SSP2-4.5) in \textbf{d)} 2040, \textbf{e)} 2070 and \textbf{f)} 2100 using the alternative population projections. The epidemics starts by importation of infected cases (no infected individual seeded at time $t=0$). Attack rate is computed as the total number of infected individual in a year over the total projected population in this very year. Alternative population projections are defined and computed in Text S1.8.
    Parameter values used in the simulations are specified in Methods and Tab. S2.    }
    \label{fig:migration_result_DENV}
\end{figure*}

Under our assumptions, alternative population projections in Europe present a similar trend across SSPs, with population values highly different from the population projections used so far (see Fig. S5). Using these alternative population distributions, the model-derived risk indicator systematically increases across all considered SSPs, compared to the population projections used until now (see Fig. \ref{fig:migration_result_DENV}). 
We globally observe that the more pessimistic the scenario, the higher the discrepancy between the alternative population projections and the population projections used in the previous sections. While this discrepancy is of the order of few percentage points of the population-at-risk index for SSP1-2,6 and SSP2-4.5, it could reach tens of percentage points in the SSP5-8.5, where the effects of climate change are expected to be the strongest. Overall, these results suggest that climate-related population redistribution could amplify dengue emergence risk in Europe by the end of the century, particularly under high-emission scenarios.
While spatial pattern of the population-at-risk index stays globally unchanged, it increases markedly in the south and west of France, the north of Spain and in Hungary in the SSP5-8.5 compared to previous analyses (see Fig. S19-21).

\section{Discussion}

We presented a mechanistic model that encompasses several routes of human and vector mobility, climate-dependent vector abundance and transmission models and population projections into a consistent data-driven metapopulation framework. Furthermore, we proposed a migration model that leverages an index of human suitability to estimate alternative population projections including climate-related migration, based on current population projections. We estimated how climate change, mobility and potential climate-related migration jointly shape scenario-dependent dengue emergence risk carried by \textit{Aedes albopictus} in Europe during the 21st century. Mobility and vector-abundance parameters were calibrated using experimental evidence and surveillance records from Europe \cite{eritja_direct_2017,zardini_estimating_2024}. Accordingly, our results are best interpreted as comparative risk maps across climate, mobility and population scenarios rather than as point forecasts of future incidence.

Our scenario-based analysis supports the view that mobility, and in particular importation of cases through air traffic,
remains a major driver of vector-borne disease emergence risk in Europe in the 21st century \cite{zhang_modeling_2020,PISANESCHI_2026}, alongside rising temperature \cite{lim_systematic_2023,li_big_2022,adams_man_2009}. However, our simulations suggest that dengue emergence risk
in Europe may shift from an importation-driven regime, associated with sporadic outbreaks, to an increasingly environment-driven regime by the end of the century, in particular in the SSP5-8.5, increasing the potential for recurrent local transmission.
In both regimes, robust risk assessment requires time-dependent models of human mobility that account for the coupling between importation sources, internally connected regions and environmentally suitable areas.

Moreover, our stress-test analysis suggests that climate-related population redistribution could further amplify dengue emergence risk in the European Union, particularly under SSP5-8.5, for instance by altering local vector-to-host ratios in areas where human suitability becomes degraded
\cite{menegale_risk_2025}. 
Because the assumptions of the proposed migration model are deliberately simplified and may overestimate migration responses, these results should not be interpreted as demographic forecasts. Rather, they emphasize the need for a new generation of global population projection models that account for climate-related migration given their potential impact on the worldwide risk assessment for vector-borne diseases \cite{schewel_how_2024,mathew_pnas_2024}.

In our scenario ensemble, the model-derived dengue emergence risk indicator can increase by at least 150-fold by the end of the century compared to present-day estimates, with high values extending across the Iberian Peninsula, the Mediterranean basin and the Black Sea basin. At the same time, the lowest-emission scenarios are systematically associated with a smaller increase in the risk indicator and a more limited spatial extent.

From this perspective, mitigation pathways consistent with the Paris Agreement remain directly relevant for climate-sensitive infectious disease preparedness
~\cite{ryan_global_2019,messina_current_2019}.

However, this estimate is contingent on the climate projection used: in the MRI-ESM2-0 scenario, climate change is not necessarily associated with an increase of the population at risk. Actually, this climate projection is known to be associated with a strong weakening of the Atlantic Meridional Overturning Circulation (AMOC) that would significantly cool down Europe and therefore limit the transmission of vector-borne diseases \cite{gomez_amoc_2025}. This confirms that climate tipping points constitute a major source of uncertainty in estimating the vector-borne risk in  Europe during the century \cite{vamwesten_2024}.

Comparison with previous literature giving predictions of the population at risk and its spatial extent is rendered difficult by the multiplicity of approaches and more or less conservative definitions of the risk estimator. 
Models based on the vector suitability or capacity of dengue transmission provide spatial extensions in Europe that are globally consistent with our results \cite{kraemer_past_2019,liu-helmersson_climate_2016,cunze_aedes_2016,colon-gonzalez_projecting_2021}. In contradiction with our findings, these studies predict a non-negligible risk by the end of the century in northern Europe while predicting no risk in central and southern Spain. We suppose that the discrepancy in Spain might arise from the threshold they impose on the minimal annual precipitation to consider a non-zero vector suitability whereas our vector abundance model does not impose any threshold.
On the contrary, in the intermediate SSP2 scenario, combining statistical mapping techniques and population projections,  Ref. \cite{messina_current_2019} predicts a spatially limited extension of the dengue risk to the south of Spain by the end of the century compared to current times. However, these conclusions seem incompatible with the dengue outbreaks that happened in 2024 in Italy, Spain and France, already \cite{sacco_autochthonous_2024}. In the literature, definitions of the presence and intensity of risk in an area are often threshold-based. By assessing the risk in terms of the cumulative number of recovered individuals, a variable that naturally belongs to the transmission model, we obtained a continuous, graded, and more direct risk assessment, integrating the spatio-temporal dynamics of vector and transmission parameters.

\subsection{Limitations and future work.} Several limitations should be considered in interpreting the results. First, the metapopulation framework supposes a homogeneous mixing of vector and human population within each patch (here at the NUTS3-level). Yet, current evidence shows that both populations are highly heterogeneously distributed within the same NUTS-3 level with maximal risk concentrated in coastal areas \cite{menegale_risk_2025} or across neighborhoods of the same municipality \cite{guzzetta_quantifying_2018,soriano-panos_vector-borne_2020,knoblauch_modeling_2025,romeo-aznar_fine-scale_2022}. 
While we correct for low human density areas within each NUT3-level, the homogeneous mixing assumptions could therefore lead to under- or over-estimation of the risk depending on whether the estimated node-level vector-to-host ratio is locally under- or over-estimated compared to the ones of geographical subcomponents.
Under this assumption, the cumulated recovered individuals should be interpreted as an index rather than an exact prediction of prevalence. \\
Second, the transmission model relies on a mean-field approximation which treats the number of infected individuals as a continuous number, so that epidemic dynamics may be triggered  even when this number is below one - i.e., in presence of fictitious infected ``nano-individuals'' \cite{pastor-satorras_epidemic_2015}. In fact, importation of infected individuals is a rare event at the NUTS-3 level, with most airports hosting less than one infected individual on average within a simulated day. Therefore, when a highly suitable environment for disease transmission and small importation rate jointly occur in a NUTS-3 patch, the imported infected nano-individual could trigger an epidemic. On the contrary, the discrete and rare nature of cases importation would forbid any outbreak in the real case. This could explain why our framework returns relatively high and early risk for countries like Greece or Portugal that experienced no major outbreaks in recent time, while being currently associated with very few detected importation of cases. This limitation primarily affects absolute estimates of local incidence, epidemic size and the apparent timing of local outbreak establishment, while the main aim of our framework is to compare how climate, mobility, importation and population scenarios reshape relative dengue emergence risk. \\
Third, while the joint model accounts for precipitation in the habitat suitability of \textit{Aedes albopictus}, scarcity or absence of annual precipitations is not prohibitive for the vector development in the model. Yet, precipitation has been predicted to be the main limiting factor for the habitat suitability of \textit{Aedes albopictus} in the 21st century in Europe \cite{cunze_aedes_2016-1}. Nonetheless, the impact of precipitation scarcity is still under discussion, as the development of the vector could be sustained by artificial sources of water in dry areas \cite{MERKENSCHLAGER_2025}.\\
Fourth, while we model the evolution of mobility patterns and the expected growth of airline traffic, we do not account for the evolution of the comparative size of the airline routes from endemic areas and the potential appearance of new routes. 
Fifth, the heterogeneous size of the NUTS3 patches within Europe could bias the estimation of the commuting mobility from the gravity model \cite{balcan_multiscale_2009}.
Sixth, our model does not consider immunity or interventions such as vaccines that could gradually alleviate the effective population at risk compared to our predictions.

Despite these limitations, our results show that dengue emergence risk in Europe cannot be assessed from climatic suitability alone. Rather than providing point forecasts of future dengue incidence or population at risk, the main advantage of our framework is that it allows one to identify where and under which combinations of climate change, vector abundance, human mobility, case importation and population redistribution Europe may become increasingly permissive to dengue emergence. Future work should integrate stochastic outbreak establishment, finer-scale heterogeneity and short-term surveillance data to support early-warning systems and advance a new generation of vector-borne disease risk mapping~\cite{brady_why_2025}, potentially through hybrid approaches combining mechanistic epidemiological models with machine learning, deep learning, and graph-based methods to capture complex spatial and climatic transmission dynamics~\cite{kraemer2025artificial}. In this sense, scenario-based risk mapping can help inform epidemic preparedness by identifying regions where vector monitoring, importation-aware surveillance and adaptive public-health responses may become increasingly relevant under different climate futures.

\section{Methods}

\subsection{Transmission model}
To model the vector-borne transmission, we assume that human and vector populations interact in $M$ geographically distributed patches. At the patch-level, the epidemiological dynamics is a standard compartmental model whose compartments are Susceptible ($S_h$), Exposed ($E_h$), Infected ($I_h$), Recovered ($R_h$) for humans and Susceptible ($S_v$), Exposed ($E_v$) and Infected ($I_v$) for vectors \cite{newton_model_1992}.

The contribution of the human and vector mobility in the spatial spread of epidemics is encoded in a force of infection formulation \cite{keeling_modeling_2008,nguyen_modelling_2023,sumner_quantifying_2017,guzzetta_quantifying_2018,zhang_spread_2017,bosetti_heterogeneity_2020}. The mobility matrix $\mathbf{P}$ characterizes the transition rate between patches  where $P_{ij}$ represents the daily fraction of total population $N_i$ in patch $i$ moving from patch $i$ 
to patch $j$ ($\sum_j P_{ij}=1$).
The force of infection integrates four routes of mobility: (i) host mobility within the system, (ii) host-mediated vector transport (e.g. vectors in cars/trains),
(iii) short range vector dispersal, and (iv) long-range airline travel i.e. cases importation. While the host-mediated vector-transport is characterized by $p_0$ (the average number of vector per host trip), the characteristic length $d_0$ and the average daily infected airline passengers arriving in a node $i$, $f_iN_w$ characterize the vector dispersal and the long-range air traffic mobility from endemic countries, respectively. Here $f_i$ represents the air mobility flow vector in node $i$ such as $\sum_i f_i=1$ and $N_w$ represents the daily importation rate into the all system. Altogether, the force of infection $\lambda_v^i$ that infected hosts(-vectors, $\lambda_h^i$) in node $j$, $I_h^j$(-$I_v^j$) exerts on susceptible vectors(-hosts) in node $i$, $S_v^i$(-$S_h^i$) read:
{\small
\begin{align}
  \label{eq:1}
  \lambda_v^i(T_r,P_r) &= \frac{S_v^i(T_r,P_r) c_{hv}^i(T_r)}{{N_h^i}} \sum_j \left(I_h^j+ \delta_{ij}  f_iN_w \right)  P_{j i} \, ,
  \\
  \lambda_h^i(T_r,P_r) &= \frac{S_h^i(T_r,P_r)c_{vh}^i(T_r)}{N_h^i} \sum_j I_v^j K_h^{ji} \, ,
  \\
  K_h^{ji} &= \left(\frac{e^{-\frac{d_{j i}}{d_0}}+ p_0P_{j i}}{1+\sum_{i\neq j }e^{-\frac{d_{j i}}{d_0}}+p_0}\right),
\end{align}
}with $\delta_{ij}$, $N_h^i$, $c_{vh}^i$, $c_{hv}^i$ the Kronecker delta, the number of hosts in node $i$, the effective vector-to-host (host-to-vector) contact rate in node $i$, respectively. $T_r$ and $P_r$ represent variables of temperature and precipitations at the considered point, respectively. The air flow mobility vector $f_i$ and $N_w$ are calibrated from air traffic data and European epidemic surveillance data, respectively. Mobility parameters $d_0$ and $p_0$ are set from the literature \cite{eritja_direct_2017,marini_effectiveness_2019}. The complete set of equations and temperature-dependence of parameters and data sources are to be found in Text S1.
We use the recent work of \cite{zardini_estimating_2024} to provide a calibrated and tested model for the abundance of vector at a given patch $i$, which determines its carrying capacity $\kappa_i(T_r,P_r)$ with $\kappa_i(T_r,P_r)= S_v^i(T_r,P_r)$ in absence of epidemics. The model assumes that 
the relative abundance for a given patch is related to its annual mean temperature, maximum temperature of the warmest month, annual precipitation and precipitation of the warmest quarter, which define a vector suitability index. Furthermore, the model assumes that the absolute vector abundance at time $t$ depends on sustained favorable conditions of temperature i.e. the average daily temperature between $t$ and $t-w$ ($w$ being a particular number of days). Consistent with the system we investigate, we consider the average parameters calibrated for Europe.
In order to avoid an overestimation of the vector-to-host ratio at the NUTS3 level, we follow the original methodology in \cite{zardini_estimating_2024} by reducing our analysis to areas where the population density is higher than 100 inhabitants per square kilometers as assessed by current population data (see Text S1.3). Further details on the vector-abundance models are to be found in \cite{zardini_estimating_2024}.

\subsection{Large-scale migration model}

To model the climate-related migrations, we leverage the concept of human niche developed by \cite{xu_human_niche_2020} which converts the relation between current local human density and local climatic variables (mean annual precipitation, mean annual temperature) into a local index of human suitability. The more the human density at a given location for a given local climatic variables, the more suitable the location is. For any location $n$, we define $\Delta H_n$, the difference in the local human suitability between projected climate and current climate. In other words, it quantifies the difference between current human density at location $n$ and the expected human density for the given projected climate variables, extracted from the current relation human density-climate variables.

We make the hypothesis that the difference in the local human suitability between projected climate and current climate $\Delta H$ quantifies a potential of migration.
Therefore, population  tends to migrate where the human suitability is comparatively less degraded i.e. population at a given place $i$ tends to migrate in a place $j$ if $\Delta H_j>\Delta H_i$. 

We quantify the migration flow $M_{ij}$ between each place $i$ and $j$ using the Feature-Enriched Radiation Model (FERM) where $\Delta H$ is the fitness measure with $\sum_j M_{ij}=1$ (see Text S1.9 for a description of the FERM) \cite{raimondo_network_2022}. The general idea being that migration occurs not only in a distance-dependent fashion but also based on opportunities i.e., towards highly populated place and least degraded areas.

Finally, the large-scale migration model consists in updating the population $N_h^i$ by accounting for the amount of migrated population in a diffusive manner: $N_h^i \to  N^i_{mig}-N_h^i$,  $N_{mig}^i = \sum_j M_{ji}N_h^j$.

For each year and SSPs, we consider the initial population provided by the population projections to take into account fertility rate and economic migrations. The proposed alternative projections are computed from approximately 70,000 ground cells encompassing the entire world and are integrated at the patch level for the Europe model. They are obtained for every year, climate projections and scenarios considered in the study. We refer to the Text S1.8 for more technical details.

%\bibliography{PRIN_bibliography}% Produces the bibliography via BibTeX.

\begin{thebibliography}{71}%
\makeatletter
\providecommand \@ifxundefined [1]{%
 \@ifx{#1\undefined}
}%
\providecommand \@ifnum [1]{%
 \ifnum #1\expandafter \@firstoftwo
 \else \expandafter \@secondoftwo
 \fi
}%
\providecommand \@ifx [1]{%
 \ifx #1\expandafter \@firstoftwo
 \else \expandafter \@secondoftwo
 \fi
}%
\providecommand \natexlab [1]{#1}%
\providecommand \enquote  [1]{``#1''}%
\providecommand \bibnamefont  [1]{#1}%
\providecommand \bibfnamefont [1]{#1}%
\providecommand \citenamefont [1]{#1}%
\providecommand \href@noop [0]{\@secondoftwo}%
\providecommand \href [0]{\begingroup \@sanitize@url \@href}%
\providecommand \@href[1]{\@@startlink{#1}\@@href}%
\providecommand \@@href[1]{\endgroup#1\@@endlink}%
\providecommand \@sanitize@url [0]{\catcode `\\12\catcode `\$12\catcode `\&12\catcode `\#12\catcode `\^12\catcode `\_12\catcode `\%12\relax}%
\providecommand \@@startlink[1]{}%
\providecommand \@@endlink[0]{}%
\providecommand \url  [0]{\begingroup\@sanitize@url \@url }%
\providecommand \@url [1]{\endgroup\@href {#1}{\urlprefix }}%
\providecommand \urlprefix  [0]{URL }%
\providecommand \Eprint [0]{\href }%
\providecommand \doibase [0]{https://doi.org/}%
\providecommand \selectlanguage [0]{\@gobble}%
\providecommand \bibinfo  [0]{\@secondoftwo}%
\providecommand \bibfield  [0]{\@secondoftwo}%
\providecommand \translation [1]{[#1]}%
\providecommand \BibitemOpen [0]{}%
\providecommand \bibitemStop [0]{}%
\providecommand \bibitemNoStop [0]{.\EOS\space}%
\providecommand \EOS [0]{\spacefactor3000\relax}%
\providecommand \BibitemShut  [1]{\csname bibitem#1\endcsname}%
\let\auto@bib@innerbib\@empty
%</preamble>
\bibitem [{noa(2024)}]{noauthor_vector-borne_nodate}%
  \BibitemOpen
  \bibfield  {title} {\bibinfo {title} {Who fact sheet on vector-borne diseases, including key facts, overview, and who response},\ }\href@noop {} {\bibfield  {journal} {\bibinfo  {journal} {WHO Fact Sheets}\ } (\bibinfo {year} {2024})},\ \Eprint {https://arxiv.org/abs/https://www.who.int/news-room/fact-sheets/detail/vector-borne-diseases} {https://www.who.int/news-room/fact-sheets/detail/vector-borne-diseases} \BibitemShut {NoStop}%
\bibitem [{\citenamefont {Roche}\ \emph {et~al.}(2015)\citenamefont {Roche}, \citenamefont {Léger}, \citenamefont {L’Ambert}, \citenamefont {Lacour}, \citenamefont {Foussadier}, \citenamefont {Besnard}, \citenamefont {Barré-Cardi}, \citenamefont {Simard},\ and\ \citenamefont {Fontenille}}]{roche_spread_2015}%
  \BibitemOpen
  \bibfield  {author} {\bibinfo {author} {\bibfnamefont {B.}~\bibnamefont {Roche}}, \bibinfo {author} {\bibfnamefont {L.}~\bibnamefont {Léger}}, \bibinfo {author} {\bibfnamefont {G.}~\bibnamefont {L’Ambert}}, \bibinfo {author} {\bibfnamefont {G.}~\bibnamefont {Lacour}}, \bibinfo {author} {\bibfnamefont {R.}~\bibnamefont {Foussadier}}, \bibinfo {author} {\bibfnamefont {G.}~\bibnamefont {Besnard}}, \bibinfo {author} {\bibfnamefont {H.}~\bibnamefont {Barré-Cardi}}, \bibinfo {author} {\bibfnamefont {F.}~\bibnamefont {Simard}},\ and\ \bibinfo {author} {\bibfnamefont {D.}~\bibnamefont {Fontenille}},\ }\bibfield  {title} {\bibinfo {title} {The {Spread} of {Aedes} albopictus in {Metropolitan} {France}: {Contribution} of {Environmental} {Drivers} and {Human} {Activities} and {Predictions} for a {Near} {Future}},\ }\href {https://doi.org/10.1371/journal.pone.0125600} {\bibfield  {journal} {\bibinfo  {journal} {PLOS ONE}\ }\textbf {\bibinfo {volume} {10}},\ \bibinfo {pages} {e0125600} (\bibinfo {year} {2015})},\
  \bibinfo {note} {publisher: Public Library of Science}\BibitemShut {NoStop}%
\bibitem [{\citenamefont {Manica}\ \emph {et~al.}(2017)\citenamefont {Manica}, \citenamefont {Guzzetta}, \citenamefont {Poletti}, \citenamefont {Filipponi}, \citenamefont {Solimini}, \citenamefont {Caputo}, \citenamefont {Torre}, \citenamefont {Rosà},\ and\ \citenamefont {Merler}}]{manica_transmission_2017}%
  \BibitemOpen
  \bibfield  {author} {\bibinfo {author} {\bibfnamefont {M.}~\bibnamefont {Manica}}, \bibinfo {author} {\bibfnamefont {G.}~\bibnamefont {Guzzetta}}, \bibinfo {author} {\bibfnamefont {P.}~\bibnamefont {Poletti}}, \bibinfo {author} {\bibfnamefont {F.}~\bibnamefont {Filipponi}}, \bibinfo {author} {\bibfnamefont {A.}~\bibnamefont {Solimini}}, \bibinfo {author} {\bibfnamefont {B.}~\bibnamefont {Caputo}}, \bibinfo {author} {\bibfnamefont {A.~d.}\ \bibnamefont {Torre}}, \bibinfo {author} {\bibfnamefont {R.}~\bibnamefont {Rosà}},\ and\ \bibinfo {author} {\bibfnamefont {S.}~\bibnamefont {Merler}},\ }\bibfield  {title} {\bibinfo {title} {Transmission dynamics of the ongoing chikungunya outbreak in {Central} {Italy}: from coastal areas to the metropolitan city of {Rome}, summer 2017},\ }\href {https://doi.org/10.2807/1560-7917.ES.2017.22.44.17-00685} {\bibfield  {journal} {\bibinfo  {journal} {Eurosurveillance}\ }\textbf {\bibinfo {volume} {22}},\ \bibinfo {pages} {17} (\bibinfo {year} {2017})},\ \bibinfo {note}
  {publisher: European Centre for Disease Prevention and Control}\BibitemShut {NoStop}%
\bibitem [{\citenamefont {Sacco}\ \emph {et~al.}(2024)\citenamefont {Sacco}, \citenamefont {Liverani}, \citenamefont {Venturi}, \citenamefont {Gavaudan}, \citenamefont {Riccardo}, \citenamefont {Salvoni}, \citenamefont {Fortuna}, \citenamefont {Marinelli}, \citenamefont {Marsili}, \citenamefont {Pesaresi}, \citenamefont {Grané}, \citenamefont {Mercuri}, \citenamefont {Manica}, \citenamefont {Caucci}, \citenamefont {Morelli}, \citenamefont {Sebastianelli}, \citenamefont {Marcacci}, \citenamefont {Ferraro}, \citenamefont {Luca}, \citenamefont {Pascucci}, \citenamefont {Merakou}, \citenamefont {Duranti}, \citenamefont {Pati}, \citenamefont {Lombardini}, \citenamefont {Fiacchini}, \citenamefont {Filipponi}, \citenamefont {Maraglino}, \citenamefont {Palamara}, \citenamefont {Poletti}, \citenamefont {Pezzotti}, \citenamefont {Filippetti}, \citenamefont {Merler}, \citenamefont {Manso}, \citenamefont {Menzo},\ and\ \citenamefont {Group}}]{sacco_autochthonous_2024}%
  \BibitemOpen
  \bibfield  {author} {\bibinfo {author} {\bibfnamefont {C.}~\bibnamefont {Sacco}}, \bibinfo {author} {\bibfnamefont {A.}~\bibnamefont {Liverani}}, \bibinfo {author} {\bibfnamefont {G.}~\bibnamefont {Venturi}}, \bibinfo {author} {\bibfnamefont {S.}~\bibnamefont {Gavaudan}}, \bibinfo {author} {\bibfnamefont {F.}~\bibnamefont {Riccardo}}, \bibinfo {author} {\bibfnamefont {G.}~\bibnamefont {Salvoni}}, \bibinfo {author} {\bibfnamefont {C.}~\bibnamefont {Fortuna}}, \bibinfo {author} {\bibfnamefont {K.}~\bibnamefont {Marinelli}}, \bibinfo {author} {\bibfnamefont {G.}~\bibnamefont {Marsili}}, \bibinfo {author} {\bibfnamefont {A.}~\bibnamefont {Pesaresi}}, \bibinfo {author} {\bibfnamefont {C.~M.}\ \bibnamefont {Grané}}, \bibinfo {author} {\bibfnamefont {I.}~\bibnamefont {Mercuri}}, \bibinfo {author} {\bibfnamefont {M.}~\bibnamefont {Manica}}, \bibinfo {author} {\bibfnamefont {S.}~\bibnamefont {Caucci}}, \bibinfo {author} {\bibfnamefont {D.}~\bibnamefont {Morelli}}, \bibinfo {author} {\bibfnamefont {L.}~\bibnamefont
  {Sebastianelli}}, \bibinfo {author} {\bibfnamefont {M.}~\bibnamefont {Marcacci}}, \bibinfo {author} {\bibfnamefont {F.}~\bibnamefont {Ferraro}}, \bibinfo {author} {\bibfnamefont {M.~D.}\ \bibnamefont {Luca}}, \bibinfo {author} {\bibfnamefont {I.}~\bibnamefont {Pascucci}}, \bibinfo {author} {\bibfnamefont {C.}~\bibnamefont {Merakou}}, \bibinfo {author} {\bibfnamefont {A.}~\bibnamefont {Duranti}}, \bibinfo {author} {\bibfnamefont {I.}~\bibnamefont {Pati}}, \bibinfo {author} {\bibfnamefont {L.}~\bibnamefont {Lombardini}}, \bibinfo {author} {\bibfnamefont {D.}~\bibnamefont {Fiacchini}}, \bibinfo {author} {\bibfnamefont {G.}~\bibnamefont {Filipponi}}, \bibinfo {author} {\bibfnamefont {F.}~\bibnamefont {Maraglino}}, \bibinfo {author} {\bibfnamefont {A.~T.}\ \bibnamefont {Palamara}}, \bibinfo {author} {\bibfnamefont {P.}~\bibnamefont {Poletti}}, \bibinfo {author} {\bibfnamefont {P.}~\bibnamefont {Pezzotti}}, \bibinfo {author} {\bibfnamefont {F.}~\bibnamefont {Filippetti}}, \bibinfo {author} {\bibfnamefont
  {S.}~\bibnamefont {Merler}}, \bibinfo {author} {\bibfnamefont {M.~D.}\ \bibnamefont {Manso}}, \bibinfo {author} {\bibfnamefont {S.}~\bibnamefont {Menzo}},\ and\ \bibinfo {author} {\bibfnamefont {M.~d.~o.}\ \bibnamefont {Group}},\ }\bibfield  {title} {\bibinfo {title} {Autochthonous dengue outbreak in {Marche} {Region}, {Central} {Italy}, {August} to {October} 2024},\ }\href {https://doi.org/10.2807/1560-7917.ES.2024.29.47.2400713} {\bibfield  {journal} {\bibinfo  {journal} {Eurosurveillance}\ }\textbf {\bibinfo {volume} {29}},\ \bibinfo {pages} {2400713} (\bibinfo {year} {2024})},\ \bibinfo {note} {publisher: European Centre for Disease Prevention and Control}\BibitemShut {NoStop}%
\bibitem [{\citenamefont {Aranda}\ \emph {et~al.}(2018)\citenamefont {Aranda}, \citenamefont {Martínez}, \citenamefont {Montalvo}, \citenamefont {Eritja}, \citenamefont {Navero-Castillejos}, \citenamefont {Herreros}, \citenamefont {Marqués}, \citenamefont {Escosa}, \citenamefont {Corbella}, \citenamefont {Bigas}, \citenamefont {Picart}, \citenamefont {Jané}, \citenamefont {Barrabeig}, \citenamefont {Torner}, \citenamefont {Talavera}, \citenamefont {Vázquez}, \citenamefont {Sánchez-Seco},\ and\ \citenamefont {Busquets}}]{aranda_arbovirus_2018}%
  \BibitemOpen
  \bibfield  {author} {\bibinfo {author} {\bibfnamefont {C.}~\bibnamefont {Aranda}}, \bibinfo {author} {\bibfnamefont {M.~J.}\ \bibnamefont {Martínez}}, \bibinfo {author} {\bibfnamefont {T.}~\bibnamefont {Montalvo}}, \bibinfo {author} {\bibfnamefont {R.}~\bibnamefont {Eritja}}, \bibinfo {author} {\bibfnamefont {J.}~\bibnamefont {Navero-Castillejos}}, \bibinfo {author} {\bibfnamefont {E.}~\bibnamefont {Herreros}}, \bibinfo {author} {\bibfnamefont {E.}~\bibnamefont {Marqués}}, \bibinfo {author} {\bibfnamefont {R.}~\bibnamefont {Escosa}}, \bibinfo {author} {\bibfnamefont {I.}~\bibnamefont {Corbella}}, \bibinfo {author} {\bibfnamefont {E.}~\bibnamefont {Bigas}}, \bibinfo {author} {\bibfnamefont {L.}~\bibnamefont {Picart}}, \bibinfo {author} {\bibfnamefont {M.}~\bibnamefont {Jané}}, \bibinfo {author} {\bibfnamefont {I.}~\bibnamefont {Barrabeig}}, \bibinfo {author} {\bibfnamefont {N.}~\bibnamefont {Torner}}, \bibinfo {author} {\bibfnamefont {S.}~\bibnamefont {Talavera}}, \bibinfo {author} {\bibfnamefont
  {A.}~\bibnamefont {Vázquez}}, \bibinfo {author} {\bibfnamefont {M.~P.}\ \bibnamefont {Sánchez-Seco}},\ and\ \bibinfo {author} {\bibfnamefont {N.}~\bibnamefont {Busquets}},\ }\bibfield  {title} {\bibinfo {title} {Arbovirus surveillance: first dengue virus detection in local {Aedes} albopictus mosquitoes in {Europe}, {Catalonia}, {Spain}, 2015},\ }\href {https://doi.org/10.2807/1560-7917.ES.2018.23.47.1700837} {\bibfield  {journal} {\bibinfo  {journal} {Eurosurveillance}\ }\textbf {\bibinfo {volume} {23}},\ \bibinfo {pages} {1700837} (\bibinfo {year} {2018})}\BibitemShut {NoStop}%
\bibitem [{\citenamefont {Farooq}\ \emph {et~al.}(2025)\citenamefont {Farooq}, \citenamefont {Segelmark}, \citenamefont {Rocklöv}, \citenamefont {Lillepold}, \citenamefont {Sewe}, \citenamefont {Briet},\ and\ \citenamefont {Semenza}}]{farooq_impact_2025}%
  \BibitemOpen
  \bibfield  {author} {\bibinfo {author} {\bibfnamefont {Z.}~\bibnamefont {Farooq}}, \bibinfo {author} {\bibfnamefont {L.}~\bibnamefont {Segelmark}}, \bibinfo {author} {\bibfnamefont {J.}~\bibnamefont {Rocklöv}}, \bibinfo {author} {\bibfnamefont {K.}~\bibnamefont {Lillepold}}, \bibinfo {author} {\bibfnamefont {M.~O.}\ \bibnamefont {Sewe}}, \bibinfo {author} {\bibfnamefont {O.~J.~T.}\ \bibnamefont {Briet}},\ and\ \bibinfo {author} {\bibfnamefont {J.~C.}\ \bibnamefont {Semenza}},\ }\bibfield  {title} {{\selectlanguage {English}\bibinfo {title} {Impact of climate and {Aedes} albopictus establishment on dengue and chikungunya outbreaks in {Europe}: a time-to-event analysis}},\ }\href {https://doi.org/10.1016/S2542-5196(25)00059-2} {\bibfield  {journal} {\bibinfo  {journal} {The Lancet Planetary Health}\ }\textbf {\bibinfo {volume} {9}},\ \bibinfo {pages} {e374} (\bibinfo {year} {2025})},\ \bibinfo {note} {publisher: Elsevier}\BibitemShut {NoStop}%
\bibitem [{\citenamefont {Bouzid}\ \emph {et~al.}(2014)\citenamefont {Bouzid}, \citenamefont {Colón-González}, \citenamefont {Lung}, \citenamefont {Lake},\ and\ \citenamefont {Hunter}}]{bouzid_climate_2014}%
  \BibitemOpen
  \bibfield  {author} {\bibinfo {author} {\bibfnamefont {M.}~\bibnamefont {Bouzid}}, \bibinfo {author} {\bibfnamefont {F.~J.}\ \bibnamefont {Colón-González}}, \bibinfo {author} {\bibfnamefont {T.}~\bibnamefont {Lung}}, \bibinfo {author} {\bibfnamefont {I.~R.}\ \bibnamefont {Lake}},\ and\ \bibinfo {author} {\bibfnamefont {P.~R.}\ \bibnamefont {Hunter}},\ }\bibfield  {title} {\bibinfo {title} {Climate change and the emergence of vector-borne diseases in {Europe}: case study of dengue fever},\ }\href {https://doi.org/10.1186/1471-2458-14-781} {\bibfield  {journal} {\bibinfo  {journal} {BMC Public Health}\ }\textbf {\bibinfo {volume} {14}},\ \bibinfo {pages} {781} (\bibinfo {year} {2014})}\BibitemShut {NoStop}%
\bibitem [{\citenamefont {Brady}\ \emph {et~al.}(2025)\citenamefont {Brady}, \citenamefont {Bastos}, \citenamefont {Caldwell}, \citenamefont {Cauchemez}, \citenamefont {Clapham}, \citenamefont {Dorigatti}, \citenamefont {Gaythorpe}, \citenamefont {Hu}, \citenamefont {Hussain-Alkhateeb}, \citenamefont {Johansson}, \citenamefont {Lim}, \citenamefont {Lopez}, \citenamefont {Maude}, \citenamefont {Messina}, \citenamefont {Mordecai}, \citenamefont {Peterson}, \citenamefont {Rodriquez-Barraquer}, \citenamefont {Rabe}, \citenamefont {Rojas}, \citenamefont {Ryan}, \citenamefont {Salje}, \citenamefont {Semenza},\ and\ \citenamefont {Tran}}]{brady_why_2025}%
  \BibitemOpen
  \bibfield  {author} {\bibinfo {author} {\bibfnamefont {O.~J.}\ \bibnamefont {Brady}}, \bibinfo {author} {\bibfnamefont {L.~S.}\ \bibnamefont {Bastos}}, \bibinfo {author} {\bibfnamefont {J.~M.}\ \bibnamefont {Caldwell}}, \bibinfo {author} {\bibfnamefont {S.}~\bibnamefont {Cauchemez}}, \bibinfo {author} {\bibfnamefont {H.~E.}\ \bibnamefont {Clapham}}, \bibinfo {author} {\bibfnamefont {I.}~\bibnamefont {Dorigatti}}, \bibinfo {author} {\bibfnamefont {K.~A.~M.}\ \bibnamefont {Gaythorpe}}, \bibinfo {author} {\bibfnamefont {W.}~\bibnamefont {Hu}}, \bibinfo {author} {\bibfnamefont {L.}~\bibnamefont {Hussain-Alkhateeb}}, \bibinfo {author} {\bibfnamefont {M.~A.}\ \bibnamefont {Johansson}}, \bibinfo {author} {\bibfnamefont {A.}~\bibnamefont {Lim}}, \bibinfo {author} {\bibfnamefont {V.~K.}\ \bibnamefont {Lopez}}, \bibinfo {author} {\bibfnamefont {R.~J.}\ \bibnamefont {Maude}}, \bibinfo {author} {\bibfnamefont {J.~P.}\ \bibnamefont {Messina}}, \bibinfo {author} {\bibfnamefont {E.~A.}\ \bibnamefont {Mordecai}}, \bibinfo
  {author} {\bibfnamefont {A.~T.}\ \bibnamefont {Peterson}}, \bibinfo {author} {\bibfnamefont {I.}~\bibnamefont {Rodriquez-Barraquer}}, \bibinfo {author} {\bibfnamefont {I.~B.}\ \bibnamefont {Rabe}}, \bibinfo {author} {\bibfnamefont {D.~P.}\ \bibnamefont {Rojas}}, \bibinfo {author} {\bibfnamefont {S.~J.}\ \bibnamefont {Ryan}}, \bibinfo {author} {\bibfnamefont {H.}~\bibnamefont {Salje}}, \bibinfo {author} {\bibfnamefont {J.~C.}\ \bibnamefont {Semenza}},\ and\ \bibinfo {author} {\bibfnamefont {Q.~M.}\ \bibnamefont {Tran}},\ }\bibfield  {title} {\bibinfo {title} {Why the growth of arboviral diseases necessitates a new generation of global risk maps and future projections},\ }\href {https://doi.org/10.1371/journal.pcbi.1012771} {\bibfield  {journal} {\bibinfo  {journal} {PLOS Computational Biology}\ }\textbf {\bibinfo {volume} {21}},\ \bibinfo {pages} {e1012771} (\bibinfo {year} {2025})},\ \bibinfo {note} {publisher: Public Library of Science}\BibitemShut {NoStop}%
\bibitem [{\citenamefont {Rees}\ \emph {et~al.}(2019)\citenamefont {Rees}, \citenamefont {Ng}, \citenamefont {Gachon}, \citenamefont {Mawudeku}, \citenamefont {McKenney}, \citenamefont {Pedlar}, \citenamefont {Yemshanov}, \citenamefont {Parmely},\ and\ \citenamefont {Knox}}]{rees_risk_2019}%
  \BibitemOpen
  \bibfield  {author} {\bibinfo {author} {\bibfnamefont {E.}~\bibnamefont {Rees}}, \bibinfo {author} {\bibfnamefont {V.}~\bibnamefont {Ng}}, \bibinfo {author} {\bibfnamefont {P.}~\bibnamefont {Gachon}}, \bibinfo {author} {\bibfnamefont {A.}~\bibnamefont {Mawudeku}}, \bibinfo {author} {\bibfnamefont {D.}~\bibnamefont {McKenney}}, \bibinfo {author} {\bibfnamefont {J.}~\bibnamefont {Pedlar}}, \bibinfo {author} {\bibfnamefont {D.}~\bibnamefont {Yemshanov}}, \bibinfo {author} {\bibfnamefont {J.}~\bibnamefont {Parmely}},\ and\ \bibinfo {author} {\bibfnamefont {J.}~\bibnamefont {Knox}},\ }\bibfield  {title} {\bibinfo {title} {Risk assessment strategies for early detection and prediction of infectious disease outbreaks associated with climate change},\ }\href {https://doi.org/10.14745/ccdr.v45i05a02} {\bibfield  {journal} {\bibinfo  {journal} {Canada Communicable Disease Report}\ }\textbf {\bibinfo {volume} {45}},\ \bibinfo {pages} {119} (\bibinfo {year} {2019})}\BibitemShut {NoStop}%
\bibitem [{\citenamefont {Messina}\ \emph {et~al.}(2019)\citenamefont {Messina}, \citenamefont {Brady}, \citenamefont {Golding}, \citenamefont {Kraemer}, \citenamefont {Wint}, \citenamefont {Ray}, \citenamefont {Pigott}, \citenamefont {Shearer}, \citenamefont {Johnson}, \citenamefont {Earl}, \citenamefont {Marczak}, \citenamefont {Shirude}, \citenamefont {Davis~Weaver}, \citenamefont {Gilbert}, \citenamefont {Velayudhan}, \citenamefont {Jones}, \citenamefont {Jaenisch}, \citenamefont {Scott}, \citenamefont {Reiner},\ and\ \citenamefont {Hay}}]{messina_current_2019}%
  \BibitemOpen
  \bibfield  {author} {\bibinfo {author} {\bibfnamefont {J.~P.}\ \bibnamefont {Messina}}, \bibinfo {author} {\bibfnamefont {O.~J.}\ \bibnamefont {Brady}}, \bibinfo {author} {\bibfnamefont {N.}~\bibnamefont {Golding}}, \bibinfo {author} {\bibfnamefont {M.~U.~G.}\ \bibnamefont {Kraemer}}, \bibinfo {author} {\bibfnamefont {G.~R.~W.}\ \bibnamefont {Wint}}, \bibinfo {author} {\bibfnamefont {S.~E.}\ \bibnamefont {Ray}}, \bibinfo {author} {\bibfnamefont {D.~M.}\ \bibnamefont {Pigott}}, \bibinfo {author} {\bibfnamefont {F.~M.}\ \bibnamefont {Shearer}}, \bibinfo {author} {\bibfnamefont {K.}~\bibnamefont {Johnson}}, \bibinfo {author} {\bibfnamefont {L.}~\bibnamefont {Earl}}, \bibinfo {author} {\bibfnamefont {L.~B.}\ \bibnamefont {Marczak}}, \bibinfo {author} {\bibfnamefont {S.}~\bibnamefont {Shirude}}, \bibinfo {author} {\bibfnamefont {N.}~\bibnamefont {Davis~Weaver}}, \bibinfo {author} {\bibfnamefont {M.}~\bibnamefont {Gilbert}}, \bibinfo {author} {\bibfnamefont {R.}~\bibnamefont {Velayudhan}}, \bibinfo {author}
  {\bibfnamefont {P.}~\bibnamefont {Jones}}, \bibinfo {author} {\bibfnamefont {T.}~\bibnamefont {Jaenisch}}, \bibinfo {author} {\bibfnamefont {T.~W.}\ \bibnamefont {Scott}}, \bibinfo {author} {\bibfnamefont {R.~C.}\ \bibnamefont {Reiner}},\ and\ \bibinfo {author} {\bibfnamefont {S.~I.}\ \bibnamefont {Hay}},\ }\bibfield  {title} {\bibinfo {title} {The current and future global distribution and population at risk of dengue},\ }\href {https://doi.org/10.1038/s41564-019-0476-8} {\bibfield  {journal} {\bibinfo  {journal} {Nature Microbiology}\ }\textbf {\bibinfo {volume} {4}},\ \bibinfo {pages} {1508} (\bibinfo {year} {2019})},\ \bibinfo {note} {publisher: Nature Publishing Group}\BibitemShut {NoStop}%
\bibitem [{\citenamefont {Lim}\ \emph {et~al.}(2023)\citenamefont {Lim}, \citenamefont {Jafari}, \citenamefont {Caldwell}, \citenamefont {Clapham}, \citenamefont {Gaythorpe}, \citenamefont {Hussain-Alkhateeb}, \citenamefont {Johansson}, \citenamefont {Kraemer}, \citenamefont {Maude}, \citenamefont {McCormack}, \citenamefont {Messina}, \citenamefont {Mordecai}, \citenamefont {Rabe}, \citenamefont {Reiner}, \citenamefont {Ryan}, \citenamefont {Salje}, \citenamefont {Semenza}, \citenamefont {Rojas},\ and\ \citenamefont {Brady}}]{lim_systematic_2023}%
  \BibitemOpen
  \bibfield  {author} {\bibinfo {author} {\bibfnamefont {A.-Y.}\ \bibnamefont {Lim}}, \bibinfo {author} {\bibfnamefont {Y.}~\bibnamefont {Jafari}}, \bibinfo {author} {\bibfnamefont {J.~M.}\ \bibnamefont {Caldwell}}, \bibinfo {author} {\bibfnamefont {H.~E.}\ \bibnamefont {Clapham}}, \bibinfo {author} {\bibfnamefont {K.~A.~M.}\ \bibnamefont {Gaythorpe}}, \bibinfo {author} {\bibfnamefont {L.}~\bibnamefont {Hussain-Alkhateeb}}, \bibinfo {author} {\bibfnamefont {M.~A.}\ \bibnamefont {Johansson}}, \bibinfo {author} {\bibfnamefont {M.~U.~G.}\ \bibnamefont {Kraemer}}, \bibinfo {author} {\bibfnamefont {R.~J.}\ \bibnamefont {Maude}}, \bibinfo {author} {\bibfnamefont {C.~P.}\ \bibnamefont {McCormack}}, \bibinfo {author} {\bibfnamefont {J.~P.}\ \bibnamefont {Messina}}, \bibinfo {author} {\bibfnamefont {E.~A.}\ \bibnamefont {Mordecai}}, \bibinfo {author} {\bibfnamefont {I.~B.}\ \bibnamefont {Rabe}}, \bibinfo {author} {\bibfnamefont {R.~C.}\ \bibnamefont {Reiner}}, \bibinfo {author} {\bibfnamefont {S.~J.}\ \bibnamefont
  {Ryan}}, \bibinfo {author} {\bibfnamefont {H.}~\bibnamefont {Salje}}, \bibinfo {author} {\bibfnamefont {J.~C.}\ \bibnamefont {Semenza}}, \bibinfo {author} {\bibfnamefont {D.~P.}\ \bibnamefont {Rojas}},\ and\ \bibinfo {author} {\bibfnamefont {O.~J.}\ \bibnamefont {Brady}},\ }\bibfield  {title} {\bibinfo {title} {A systematic review of the data, methods and environmental covariates used to map {Aedes}-borne arbovirus transmission risk},\ }\href {https://doi.org/10.1186/s12879-023-08717-8} {\bibfield  {journal} {\bibinfo  {journal} {BMC Infectious Diseases}\ }\textbf {\bibinfo {volume} {23}},\ \bibinfo {pages} {708} (\bibinfo {year} {2023})}\BibitemShut {NoStop}%
\bibitem [{\citenamefont {Kraemer}\ \emph {et~al.}(2015)\citenamefont {Kraemer}, \citenamefont {Sinka}, \citenamefont {Duda}, \citenamefont {Mylne}, \citenamefont {Shearer}, \citenamefont {Barker}, \citenamefont {Moore}, \citenamefont {Carvalho}, \citenamefont {Coelho}, \citenamefont {Van~Bortel}, \citenamefont {Hendrickx}, \citenamefont {Schaffner}, \citenamefont {Elyazar}, \citenamefont {Teng}, \citenamefont {Brady}, \citenamefont {Messina}, \citenamefont {Pigott}, \citenamefont {Scott}, \citenamefont {Smith}, \citenamefont {Wint}, \citenamefont {Golding},\ and\ \citenamefont {Hay}}]{kraemer_global_2015}%
  \BibitemOpen
  \bibfield  {author} {\bibinfo {author} {\bibfnamefont {M.~U.}\ \bibnamefont {Kraemer}}, \bibinfo {author} {\bibfnamefont {M.~E.}\ \bibnamefont {Sinka}}, \bibinfo {author} {\bibfnamefont {K.~A.}\ \bibnamefont {Duda}}, \bibinfo {author} {\bibfnamefont {A.~Q.}\ \bibnamefont {Mylne}}, \bibinfo {author} {\bibfnamefont {F.~M.}\ \bibnamefont {Shearer}}, \bibinfo {author} {\bibfnamefont {C.~M.}\ \bibnamefont {Barker}}, \bibinfo {author} {\bibfnamefont {C.~G.}\ \bibnamefont {Moore}}, \bibinfo {author} {\bibfnamefont {R.~G.}\ \bibnamefont {Carvalho}}, \bibinfo {author} {\bibfnamefont {G.~E.}\ \bibnamefont {Coelho}}, \bibinfo {author} {\bibfnamefont {W.}~\bibnamefont {Van~Bortel}}, \bibinfo {author} {\bibfnamefont {G.}~\bibnamefont {Hendrickx}}, \bibinfo {author} {\bibfnamefont {F.}~\bibnamefont {Schaffner}}, \bibinfo {author} {\bibfnamefont {I.~R.}\ \bibnamefont {Elyazar}}, \bibinfo {author} {\bibfnamefont {H.-J.}\ \bibnamefont {Teng}}, \bibinfo {author} {\bibfnamefont {O.~J.}\ \bibnamefont {Brady}}, \bibinfo
  {author} {\bibfnamefont {J.~P.}\ \bibnamefont {Messina}}, \bibinfo {author} {\bibfnamefont {D.~M.}\ \bibnamefont {Pigott}}, \bibinfo {author} {\bibfnamefont {T.~W.}\ \bibnamefont {Scott}}, \bibinfo {author} {\bibfnamefont {D.~L.}\ \bibnamefont {Smith}}, \bibinfo {author} {\bibfnamefont {G.~W.}\ \bibnamefont {Wint}}, \bibinfo {author} {\bibfnamefont {N.}~\bibnamefont {Golding}},\ and\ \bibinfo {author} {\bibfnamefont {S.~I.}\ \bibnamefont {Hay}},\ }\bibfield  {title} {\bibinfo {title} {The global distribution of the arbovirus vectors {Aedes} aegypti and {Ae}. albopictus},\ }\href {https://doi.org/10.7554/eLife.08347} {\bibfield  {journal} {\bibinfo  {journal} {eLife}\ }\textbf {\bibinfo {volume} {4}},\ \bibinfo {pages} {e08347} (\bibinfo {year} {2015})},\ \bibinfo {note} {publisher: eLife Sciences Publications, Ltd}\BibitemShut {NoStop}%
\bibitem [{\citenamefont {Lim}\ \emph {et~al.}(2025)\citenamefont {Lim}, \citenamefont {Shearer}, \citenamefont {Sewalk}, \citenamefont {Pigott}, \citenamefont {Clarke}, \citenamefont {Ghouse}, \citenamefont {Judge}, \citenamefont {Kang}, \citenamefont {Messina}, \citenamefont {Kraemer}, \citenamefont {Gaythorpe}, \citenamefont {de~Souza}, \citenamefont {Nsoesie}, \citenamefont {Celone}, \citenamefont {Faria}, \citenamefont {Ryan}, \citenamefont {Rabe}, \citenamefont {Rojas}, \citenamefont {Hay}, \citenamefont {Brownstein}, \citenamefont {Golding},\ and\ \citenamefont {Brady}}]{lim_overlapping_2025}%
  \BibitemOpen
  \bibfield  {author} {\bibinfo {author} {\bibfnamefont {A.}~\bibnamefont {Lim}}, \bibinfo {author} {\bibfnamefont {F.~M.}\ \bibnamefont {Shearer}}, \bibinfo {author} {\bibfnamefont {K.}~\bibnamefont {Sewalk}}, \bibinfo {author} {\bibfnamefont {D.~M.}\ \bibnamefont {Pigott}}, \bibinfo {author} {\bibfnamefont {J.}~\bibnamefont {Clarke}}, \bibinfo {author} {\bibfnamefont {A.}~\bibnamefont {Ghouse}}, \bibinfo {author} {\bibfnamefont {C.}~\bibnamefont {Judge}}, \bibinfo {author} {\bibfnamefont {H.}~\bibnamefont {Kang}}, \bibinfo {author} {\bibfnamefont {J.~P.}\ \bibnamefont {Messina}}, \bibinfo {author} {\bibfnamefont {M.~U.~G.}\ \bibnamefont {Kraemer}}, \bibinfo {author} {\bibfnamefont {K.~A.~M.}\ \bibnamefont {Gaythorpe}}, \bibinfo {author} {\bibfnamefont {W.~M.}\ \bibnamefont {de~Souza}}, \bibinfo {author} {\bibfnamefont {E.~O.}\ \bibnamefont {Nsoesie}}, \bibinfo {author} {\bibfnamefont {M.}~\bibnamefont {Celone}}, \bibinfo {author} {\bibfnamefont {N.}~\bibnamefont {Faria}}, \bibinfo {author} {\bibfnamefont
  {S.~J.}\ \bibnamefont {Ryan}}, \bibinfo {author} {\bibfnamefont {I.~B.}\ \bibnamefont {Rabe}}, \bibinfo {author} {\bibfnamefont {D.~P.}\ \bibnamefont {Rojas}}, \bibinfo {author} {\bibfnamefont {S.~I.}\ \bibnamefont {Hay}}, \bibinfo {author} {\bibfnamefont {J.~S.}\ \bibnamefont {Brownstein}}, \bibinfo {author} {\bibfnamefont {N.}~\bibnamefont {Golding}},\ and\ \bibinfo {author} {\bibfnamefont {O.~J.}\ \bibnamefont {Brady}},\ }\bibfield  {title} {\bibinfo {title} {The overlapping global distribution of dengue, chikungunya, {Zika} and yellow fever},\ }\href {https://doi.org/10.1038/s41467-025-58609-5} {\bibfield  {journal} {\bibinfo  {journal} {Nature Communications}\ }\textbf {\bibinfo {volume} {16}},\ \bibinfo {pages} {3418} (\bibinfo {year} {2025})},\ \bibinfo {note} {publisher: Nature Publishing Group}\BibitemShut {NoStop}%
\bibitem [{\citenamefont {Rogers}\ \emph {et~al.}(2014)\citenamefont {Rogers}, \citenamefont {Suk},\ and\ \citenamefont {Semenza}}]{rogers_using_2014}%
  \BibitemOpen
  \bibfield  {author} {\bibinfo {author} {\bibfnamefont {D.~J.}\ \bibnamefont {Rogers}}, \bibinfo {author} {\bibfnamefont {J.~E.}\ \bibnamefont {Suk}},\ and\ \bibinfo {author} {\bibfnamefont {J.~C.}\ \bibnamefont {Semenza}},\ }\bibfield  {title} {\bibinfo {title} {Using global maps to predict the risk of dengue in {Europe}},\ }\href {https://doi.org/10.1016/j.actatropica.2013.08.008} {\bibfield  {journal} {\bibinfo  {journal} {Acta Tropica}\ }\bibinfo {series} {Human {Infectious} {Diseases} and {Environmental} {Changes}},\ \textbf {\bibinfo {volume} {129}},\ \bibinfo {pages} {1} (\bibinfo {year} {2014})}\BibitemShut {NoStop}%
\bibitem [{\citenamefont {Zhang}\ \emph {et~al.}(2020)\citenamefont {Zhang}, \citenamefont {Riera}, \citenamefont {Ostrow}, \citenamefont {Siddiqui}, \citenamefont {de~Silva}, \citenamefont {Sarkar}, \citenamefont {Fernando},\ and\ \citenamefont {Gardner}}]{zhang_modeling_2020}%
  \BibitemOpen
  \bibfield  {author} {\bibinfo {author} {\bibfnamefont {Y.}~\bibnamefont {Zhang}}, \bibinfo {author} {\bibfnamefont {J.}~\bibnamefont {Riera}}, \bibinfo {author} {\bibfnamefont {K.}~\bibnamefont {Ostrow}}, \bibinfo {author} {\bibfnamefont {S.}~\bibnamefont {Siddiqui}}, \bibinfo {author} {\bibfnamefont {H.}~\bibnamefont {de~Silva}}, \bibinfo {author} {\bibfnamefont {S.}~\bibnamefont {Sarkar}}, \bibinfo {author} {\bibfnamefont {L.}~\bibnamefont {Fernando}},\ and\ \bibinfo {author} {\bibfnamefont {L.}~\bibnamefont {Gardner}},\ }\bibfield  {title} {\bibinfo {title} {Modeling the relative role of human mobility, land-use and climate factors on dengue outbreak emergence in {Sri} {Lanka}},\ }\href {https://doi.org/10.1186/s12879-020-05369-w} {\bibfield  {journal} {\bibinfo  {journal} {BMC Infectious Diseases}\ }\textbf {\bibinfo {volume} {20}},\ \bibinfo {pages} {649} (\bibinfo {year} {2020})}\BibitemShut {NoStop}%
\bibitem [{\citenamefont {O’Neill}\ \emph {et~al.}(2014)\citenamefont {O’Neill}, \citenamefont {Kriegler}, \citenamefont {Riahi}, \citenamefont {Ebi}, \citenamefont {Hallegatte}, \citenamefont {Carter}, \citenamefont {Mathur},\ and\ \citenamefont {van Vuuren}}]{oneill_new_2014}%
  \BibitemOpen
  \bibfield  {author} {\bibinfo {author} {\bibfnamefont {B.~C.}\ \bibnamefont {O’Neill}}, \bibinfo {author} {\bibfnamefont {E.}~\bibnamefont {Kriegler}}, \bibinfo {author} {\bibfnamefont {K.}~\bibnamefont {Riahi}}, \bibinfo {author} {\bibfnamefont {K.~L.}\ \bibnamefont {Ebi}}, \bibinfo {author} {\bibfnamefont {S.}~\bibnamefont {Hallegatte}}, \bibinfo {author} {\bibfnamefont {T.~R.}\ \bibnamefont {Carter}}, \bibinfo {author} {\bibfnamefont {R.}~\bibnamefont {Mathur}},\ and\ \bibinfo {author} {\bibfnamefont {D.~P.}\ \bibnamefont {van Vuuren}},\ }\bibfield  {title} {\bibinfo {title} {A new scenario framework for climate change research: the concept of shared socioeconomic pathways},\ }\href {https://doi.org/10.1007/s10584-013-0905-2} {\bibfield  {journal} {\bibinfo  {journal} {Climatic Change}\ }\textbf {\bibinfo {volume} {122}},\ \bibinfo {pages} {387} (\bibinfo {year} {2014})}\BibitemShut {NoStop}%
\bibitem [{\citenamefont {Jones}\ and\ \citenamefont {O’Neill}(2016)}]{jones_spatially_2016}%
  \BibitemOpen
  \bibfield  {author} {\bibinfo {author} {\bibfnamefont {B.}~\bibnamefont {Jones}}\ and\ \bibinfo {author} {\bibfnamefont {B.~C.}\ \bibnamefont {O’Neill}},\ }\bibfield  {title} {\bibinfo {title} {Spatially explicit global population scenarios consistent with the {Shared} {Socioeconomic} {Pathways}},\ }\href {https://doi.org/10.1088/1748-9326/11/8/084003} {\bibfield  {journal} {\bibinfo  {journal} {Environmental Research Letters}\ }\textbf {\bibinfo {volume} {11}},\ \bibinfo {pages} {084003} (\bibinfo {year} {2016})},\ \bibinfo {note} {publisher: IOP Publishing}\BibitemShut {NoStop}%
\bibitem [{\citenamefont {Mordecai}\ \emph {et~al.}(2017)\citenamefont {Mordecai}, \citenamefont {Cohen}, \citenamefont {Evans}, \citenamefont {Gudapati}, \citenamefont {Johnson}, \citenamefont {Lippi}, \citenamefont {Miazgowicz}, \citenamefont {Murdock}, \citenamefont {Rohr}, \citenamefont {Ryan}, \citenamefont {Savage}, \citenamefont {Shocket}, \citenamefont {Ibarra}, \citenamefont {Thomas},\ and\ \citenamefont {Weikel}}]{mordecai_detecting_2017}%
  \BibitemOpen
  \bibfield  {author} {\bibinfo {author} {\bibfnamefont {E.~A.}\ \bibnamefont {Mordecai}}, \bibinfo {author} {\bibfnamefont {J.~M.}\ \bibnamefont {Cohen}}, \bibinfo {author} {\bibfnamefont {M.~V.}\ \bibnamefont {Evans}}, \bibinfo {author} {\bibfnamefont {P.}~\bibnamefont {Gudapati}}, \bibinfo {author} {\bibfnamefont {L.~R.}\ \bibnamefont {Johnson}}, \bibinfo {author} {\bibfnamefont {C.~A.}\ \bibnamefont {Lippi}}, \bibinfo {author} {\bibfnamefont {K.}~\bibnamefont {Miazgowicz}}, \bibinfo {author} {\bibfnamefont {C.~C.}\ \bibnamefont {Murdock}}, \bibinfo {author} {\bibfnamefont {J.~R.}\ \bibnamefont {Rohr}}, \bibinfo {author} {\bibfnamefont {S.~J.}\ \bibnamefont {Ryan}}, \bibinfo {author} {\bibfnamefont {V.}~\bibnamefont {Savage}}, \bibinfo {author} {\bibfnamefont {M.~S.}\ \bibnamefont {Shocket}}, \bibinfo {author} {\bibfnamefont {A.~S.}\ \bibnamefont {Ibarra}}, \bibinfo {author} {\bibfnamefont {M.~B.}\ \bibnamefont {Thomas}},\ and\ \bibinfo {author} {\bibfnamefont {D.~P.}\ \bibnamefont {Weikel}},\ }\bibfield
  {title} {\bibinfo {title} {Detecting the impact of temperature on transmission of {Zika}, dengue, and chikungunya using mechanistic models},\ }\href {https://doi.org/10.1371/journal.pntd.0005568} {\bibfield  {journal} {\bibinfo  {journal} {PLOS Neglected Tropical Diseases}\ }\textbf {\bibinfo {volume} {11}},\ \bibinfo {pages} {e0005568} (\bibinfo {year} {2017})},\ \bibinfo {note} {publisher: Public Library of Science}\BibitemShut {NoStop}%
\bibitem [{\citenamefont {Nakase}\ \emph {et~al.}(2024)\citenamefont {Nakase}, \citenamefont {Giovanetti}, \citenamefont {Obolski},\ and\ \citenamefont {Lourenço}}]{nakase_population_2024}%
  \BibitemOpen
  \bibfield  {author} {\bibinfo {author} {\bibfnamefont {T.}~\bibnamefont {Nakase}}, \bibinfo {author} {\bibfnamefont {M.}~\bibnamefont {Giovanetti}}, \bibinfo {author} {\bibfnamefont {U.}~\bibnamefont {Obolski}},\ and\ \bibinfo {author} {\bibfnamefont {J.}~\bibnamefont {Lourenço}},\ }\bibfield  {title} {\bibinfo {title} {Population at risk of dengue virus transmission has increased due to coupled climate factors and population growth},\ }\href {https://doi.org/10.1038/s43247-024-01639-6} {\bibfield  {journal} {\bibinfo  {journal} {Communications Earth \& Environment}\ }\textbf {\bibinfo {volume} {5}},\ \bibinfo {pages} {475} (\bibinfo {year} {2024})},\ \bibinfo {note} {publisher: Nature Publishing Group}\BibitemShut {NoStop}%
\bibitem [{\citenamefont {Zardini}\ \emph {et~al.}(2024)\citenamefont {Zardini}, \citenamefont {Menegale}, \citenamefont {Gobbi}, \citenamefont {Manica}, \citenamefont {Guzzetta}, \citenamefont {d'Andrea}, \citenamefont {Marziano}, \citenamefont {Trentini}, \citenamefont {Montarsi}, \citenamefont {Caputo}, \citenamefont {Solimini}, \citenamefont {Marques-Toledo}, \citenamefont {Wilke}, \citenamefont {Rosà}, \citenamefont {Marini}, \citenamefont {Arnoldi}, \citenamefont {Piontti}, \citenamefont {Pugliese}, \citenamefont {Capelli}, \citenamefont {Torre}, \citenamefont {Teixeira}, \citenamefont {Beier}, \citenamefont {Rizzoli}, \citenamefont {Vespignani}, \citenamefont {Ajelli}, \citenamefont {Merler},\ and\ \citenamefont {Poletti}}]{zardini_estimating_2024}%
  \BibitemOpen
  \bibfield  {author} {\bibinfo {author} {\bibfnamefont {A.}~\bibnamefont {Zardini}}, \bibinfo {author} {\bibfnamefont {F.}~\bibnamefont {Menegale}}, \bibinfo {author} {\bibfnamefont {A.}~\bibnamefont {Gobbi}}, \bibinfo {author} {\bibfnamefont {M.}~\bibnamefont {Manica}}, \bibinfo {author} {\bibfnamefont {G.}~\bibnamefont {Guzzetta}}, \bibinfo {author} {\bibfnamefont {V.}~\bibnamefont {d'Andrea}}, \bibinfo {author} {\bibfnamefont {V.}~\bibnamefont {Marziano}}, \bibinfo {author} {\bibfnamefont {F.}~\bibnamefont {Trentini}}, \bibinfo {author} {\bibfnamefont {F.}~\bibnamefont {Montarsi}}, \bibinfo {author} {\bibfnamefont {B.}~\bibnamefont {Caputo}}, \bibinfo {author} {\bibfnamefont {A.}~\bibnamefont {Solimini}}, \bibinfo {author} {\bibfnamefont {C.}~\bibnamefont {Marques-Toledo}}, \bibinfo {author} {\bibfnamefont {A.~B.~B.}\ \bibnamefont {Wilke}}, \bibinfo {author} {\bibfnamefont {R.}~\bibnamefont {Rosà}}, \bibinfo {author} {\bibfnamefont {G.}~\bibnamefont {Marini}}, \bibinfo {author} {\bibfnamefont
  {D.}~\bibnamefont {Arnoldi}}, \bibinfo {author} {\bibfnamefont {A.~P.~y.}\ \bibnamefont {Piontti}}, \bibinfo {author} {\bibfnamefont {A.}~\bibnamefont {Pugliese}}, \bibinfo {author} {\bibfnamefont {G.}~\bibnamefont {Capelli}}, \bibinfo {author} {\bibfnamefont {A.~d.}\ \bibnamefont {Torre}}, \bibinfo {author} {\bibfnamefont {M.~M.}\ \bibnamefont {Teixeira}}, \bibinfo {author} {\bibfnamefont {J.~C.}\ \bibnamefont {Beier}}, \bibinfo {author} {\bibfnamefont {A.}~\bibnamefont {Rizzoli}}, \bibinfo {author} {\bibfnamefont {A.}~\bibnamefont {Vespignani}}, \bibinfo {author} {\bibfnamefont {M.}~\bibnamefont {Ajelli}}, \bibinfo {author} {\bibfnamefont {S.}~\bibnamefont {Merler}},\ and\ \bibinfo {author} {\bibfnamefont {P.}~\bibnamefont {Poletti}},\ }\bibfield  {title} {\bibinfo {title} {Estimating the potential risk of transmission of arboviruses in the {Americas} and {Europe}: a modelling study},\ }\href {https://doi.org/10.1016/S2542-5196(23)00252-8} {\bibfield  {journal} {\bibinfo  {journal} {The Lancet Planetary
  Health}\ }\textbf {\bibinfo {volume} {8}},\ \bibinfo {pages} {e30} (\bibinfo {year} {2024})},\ \bibinfo {note} {publisher: Elsevier}\BibitemShut {NoStop}%
\bibitem [{\citenamefont {de~Souza}\ and\ \citenamefont {Weaver}(2024)}]{de_souza_effects_2024}%
  \BibitemOpen
  \bibfield  {author} {\bibinfo {author} {\bibfnamefont {W.~M.}\ \bibnamefont {de~Souza}}\ and\ \bibinfo {author} {\bibfnamefont {S.~C.}\ \bibnamefont {Weaver}},\ }\bibfield  {title} {\bibinfo {title} {Effects of climate change and human activities on vector-borne diseases},\ }\href {https://doi.org/10.1038/s41579-024-01026-0} {\bibfield  {journal} {\bibinfo  {journal} {Nature Reviews Microbiology}\ }\textbf {\bibinfo {volume} {22}},\ \bibinfo {pages} {476} (\bibinfo {year} {2024})},\ \bibinfo {note} {publisher: Nature Publishing Group}\BibitemShut {NoStop}%
\bibitem [{\citenamefont {Li}\ \emph {et~al.}(2019)\citenamefont {Li}, \citenamefont {Xu}, \citenamefont {Bjørnstad}, \citenamefont {Liu}, \citenamefont {Song}, \citenamefont {Chen}, \citenamefont {Xu}, \citenamefont {Liu},\ and\ \citenamefont {Stenseth}}]{li_climate-driven_2019}%
  \BibitemOpen
  \bibfield  {author} {\bibinfo {author} {\bibfnamefont {R.}~\bibnamefont {Li}}, \bibinfo {author} {\bibfnamefont {L.}~\bibnamefont {Xu}}, \bibinfo {author} {\bibfnamefont {O.~N.}\ \bibnamefont {Bjørnstad}}, \bibinfo {author} {\bibfnamefont {K.}~\bibnamefont {Liu}}, \bibinfo {author} {\bibfnamefont {T.}~\bibnamefont {Song}}, \bibinfo {author} {\bibfnamefont {A.}~\bibnamefont {Chen}}, \bibinfo {author} {\bibfnamefont {B.}~\bibnamefont {Xu}}, \bibinfo {author} {\bibfnamefont {Q.}~\bibnamefont {Liu}},\ and\ \bibinfo {author} {\bibfnamefont {N.~C.}\ \bibnamefont {Stenseth}},\ }\bibfield  {title} {\bibinfo {title} {Climate-driven variation in mosquito density predicts the spatiotemporal dynamics of dengue},\ }\href {https://doi.org/10.1073/pnas.1806094116} {\bibfield  {journal} {\bibinfo  {journal} {Proceedings of the National Academy of Sciences}\ }\textbf {\bibinfo {volume} {116}},\ \bibinfo {pages} {3624} (\bibinfo {year} {2019})},\ \bibinfo {note} {company: National Academy of Sciences Distributor: National
  Academy of Sciences Institution: National Academy of Sciences Label: National Academy of Sciences Publisher: Proceedings of the National Academy of Sciences}\BibitemShut {NoStop}%
\bibitem [{\citenamefont {Colón-González}\ \emph {et~al.}(2018)\citenamefont {Colón-González}, \citenamefont {Harris}, \citenamefont {Osborn}, \citenamefont {Bernardo}, \citenamefont {Peres}, \citenamefont {Hunter}, \citenamefont {Warren}, \citenamefont {van Vuurene},\ and\ \citenamefont {Lake}}]{colon-gonzalez_pnas_2018}%
  \BibitemOpen
  \bibfield  {author} {\bibinfo {author} {\bibfnamefont {F.~J.}\ \bibnamefont {Colón-González}}, \bibinfo {author} {\bibfnamefont {I.}~\bibnamefont {Harris}}, \bibinfo {author} {\bibfnamefont {T.~J.}\ \bibnamefont {Osborn}}, \bibinfo {author} {\bibfnamefont {C.~S.~S.}\ \bibnamefont {Bernardo}}, \bibinfo {author} {\bibfnamefont {C.~A.}\ \bibnamefont {Peres}}, \bibinfo {author} {\bibfnamefont {P.~R.}\ \bibnamefont {Hunter}}, \bibinfo {author} {\bibfnamefont {R.}~\bibnamefont {Warren}}, \bibinfo {author} {\bibfnamefont {D.}~\bibnamefont {van Vuurene}},\ and\ \bibinfo {author} {\bibfnamefont {I.~R.}\ \bibnamefont {Lake}},\ }\bibfield  {title} {\bibinfo {title} {Limiting global-mean temperature increase to 1.5–2 °c could reduce the incidence and spatial spread of dengue fever in latin america},\ }\href {https://doi.org/10.1073/pnas.1718945115} {\bibfield  {journal} {\bibinfo  {journal} {Proceedings of the National Academy of Sciences}\ }\textbf {\bibinfo {volume} {115}},\ \bibinfo {pages} {6243} (\bibinfo
  {year} {2018})},\ \Eprint {https://arxiv.org/abs/https://www.pnas.org/doi/pdf/10.1073/pnas.1718945115} {https://www.pnas.org/doi/pdf/10.1073/pnas.1718945115} \BibitemShut {NoStop}%
\bibitem [{\citenamefont {Childs}\ \emph {et~al.}(2025)\citenamefont {Childs}, \citenamefont {Lyberger}, \citenamefont {Harris}, \citenamefont {Burke},\ and\ \citenamefont {Mordecai}}]{childs_climate_2025}%
  \BibitemOpen
  \bibfield  {author} {\bibinfo {author} {\bibfnamefont {M.~L.}\ \bibnamefont {Childs}}, \bibinfo {author} {\bibfnamefont {K.}~\bibnamefont {Lyberger}}, \bibinfo {author} {\bibfnamefont {M.~J.}\ \bibnamefont {Harris}}, \bibinfo {author} {\bibfnamefont {M.}~\bibnamefont {Burke}},\ and\ \bibinfo {author} {\bibfnamefont {E.~A.}\ \bibnamefont {Mordecai}},\ }\bibfield  {title} {\bibinfo {title} {Climate warming is expanding dengue burden in the {Americas} and {Asia}},\ }\href {https://doi.org/10.1073/pnas.2512350122} {\bibfield  {journal} {\bibinfo  {journal} {Proceedings of the National Academy of Sciences}\ }\textbf {\bibinfo {volume} {122}},\ \bibinfo {pages} {e2512350122} (\bibinfo {year} {2025})},\ \bibinfo {note} {publisher: Proceedings of the National Academy of Sciences}\BibitemShut {NoStop}%
\bibitem [{\citenamefont {Moore}\ and\ \citenamefont {Brown}(2022)}]{moore_estimating_2022}%
  \BibitemOpen
  \bibfield  {author} {\bibinfo {author} {\bibfnamefont {T.~C.}\ \bibnamefont {Moore}}\ and\ \bibinfo {author} {\bibfnamefont {H.~E.}\ \bibnamefont {Brown}},\ }\bibfield  {title} {\bibinfo {title} {Estimating {Aedes} aegypti ({Diptera}: {Culicidae}) {Flight} {Distance}: {Meta}-{Data} {Analysis}},\ }\href {https://doi.org/10.1093/jme/tjac070} {\bibfield  {journal} {\bibinfo  {journal} {Journal of Medical Entomology}\ }\textbf {\bibinfo {volume} {59}},\ \bibinfo {pages} {1164} (\bibinfo {year} {2022})}\BibitemShut {NoStop}%
\bibitem [{\citenamefont {Guzzetta}\ \emph {et~al.}(2018)\citenamefont {Guzzetta}, \citenamefont {Marques-Toledo}, \citenamefont {Rosà}, \citenamefont {Teixeira},\ and\ \citenamefont {Merler}}]{guzzetta_quantifying_2018}%
  \BibitemOpen
  \bibfield  {author} {\bibinfo {author} {\bibfnamefont {G.}~\bibnamefont {Guzzetta}}, \bibinfo {author} {\bibfnamefont {C.~A.}\ \bibnamefont {Marques-Toledo}}, \bibinfo {author} {\bibfnamefont {R.}~\bibnamefont {Rosà}}, \bibinfo {author} {\bibfnamefont {M.}~\bibnamefont {Teixeira}},\ and\ \bibinfo {author} {\bibfnamefont {S.}~\bibnamefont {Merler}},\ }\bibfield  {title} {\bibinfo {title} {Quantifying the spatial spread of dengue in a non-endemic {Brazilian} metropolis via transmission chain reconstruction},\ }\href {https://doi.org/10.1038/s41467-018-05230-4} {\bibfield  {journal} {\bibinfo  {journal} {Nature Communications}\ }\textbf {\bibinfo {volume} {9}},\ \bibinfo {pages} {2837} (\bibinfo {year} {2018})},\ \bibinfo {note} {number: 1 Publisher: Nature Publishing Group}\BibitemShut {NoStop}%
\bibitem [{\citenamefont {Soriano-Paños}\ \emph {et~al.}(2020)\citenamefont {Soriano-Paños}, \citenamefont {Arias-Castro}, \citenamefont {Reyna-Lara}, \citenamefont {Martínez}, \citenamefont {Meloni},\ and\ \citenamefont {Gómez-Gardeñes}}]{soriano-panos_vector-borne_2020}%
  \BibitemOpen
  \bibfield  {author} {\bibinfo {author} {\bibfnamefont {D.}~\bibnamefont {Soriano-Paños}}, \bibinfo {author} {\bibfnamefont {J.~H.}\ \bibnamefont {Arias-Castro}}, \bibinfo {author} {\bibfnamefont {A.}~\bibnamefont {Reyna-Lara}}, \bibinfo {author} {\bibfnamefont {H.~J.}\ \bibnamefont {Martínez}}, \bibinfo {author} {\bibfnamefont {S.}~\bibnamefont {Meloni}},\ and\ \bibinfo {author} {\bibfnamefont {J.}~\bibnamefont {Gómez-Gardeñes}},\ }\bibfield  {title} {\bibinfo {title} {Vector-borne epidemics driven by human mobility},\ }\href {https://doi.org/10.1103/PhysRevResearch.2.013312} {\bibfield  {journal} {\bibinfo  {journal} {Physical Review Research}\ }\textbf {\bibinfo {volume} {2}},\ \bibinfo {pages} {013312} (\bibinfo {year} {2020})},\ \bibinfo {note} {publisher: American Physical Society}\BibitemShut {NoStop}%
\bibitem [{\citenamefont {Wesolowski}\ \emph {et~al.}(2015)\citenamefont {Wesolowski}, \citenamefont {Qureshi}, \citenamefont {Boni}, \citenamefont {Sundsøy}, \citenamefont {Johansson}, \citenamefont {Rasheed}, \citenamefont {Engø-Monsen},\ and\ \citenamefont {Buckee}}]{wesolowski_impact_2015}%
  \BibitemOpen
  \bibfield  {author} {\bibinfo {author} {\bibfnamefont {A.}~\bibnamefont {Wesolowski}}, \bibinfo {author} {\bibfnamefont {T.}~\bibnamefont {Qureshi}}, \bibinfo {author} {\bibfnamefont {M.~F.}\ \bibnamefont {Boni}}, \bibinfo {author} {\bibfnamefont {P.~R.}\ \bibnamefont {Sundsøy}}, \bibinfo {author} {\bibfnamefont {M.~A.}\ \bibnamefont {Johansson}}, \bibinfo {author} {\bibfnamefont {S.~B.}\ \bibnamefont {Rasheed}}, \bibinfo {author} {\bibfnamefont {K.}~\bibnamefont {Engø-Monsen}},\ and\ \bibinfo {author} {\bibfnamefont {C.~O.}\ \bibnamefont {Buckee}},\ }\bibfield  {title} {\bibinfo {title} {Impact of human mobility on the emergence of dengue epidemics in {Pakistan}},\ }\href {https://doi.org/10.1073/pnas.1504964112} {\bibfield  {journal} {\bibinfo  {journal} {Proceedings of the National Academy of Sciences}\ }\textbf {\bibinfo {volume} {112}},\ \bibinfo {pages} {11887} (\bibinfo {year} {2015})},\ \bibinfo {note} {publisher: Proceedings of the National Academy of Sciences}\BibitemShut {NoStop}%
\bibitem [{\citenamefont {Gibb}\ \emph {et~al.}(2023)\citenamefont {Gibb}, \citenamefont {Colón-González}, \citenamefont {Lan}, \citenamefont {Huong}, \citenamefont {Nam}, \citenamefont {Duoc}, \citenamefont {Hung}, \citenamefont {Dong}, \citenamefont {Chien}, \citenamefont {Trang}, \citenamefont {Kien~Quoc}, \citenamefont {Hoa}, \citenamefont {Tai}, \citenamefont {Hang}, \citenamefont {Tsarouchi}, \citenamefont {Ainscoe}, \citenamefont {Harpham}, \citenamefont {Hofmann}, \citenamefont {Lumbroso}, \citenamefont {Brady},\ and\ \citenamefont {Lowe}}]{gibb_interactions_2023}%
  \BibitemOpen
  \bibfield  {author} {\bibinfo {author} {\bibfnamefont {R.}~\bibnamefont {Gibb}}, \bibinfo {author} {\bibfnamefont {F.~J.}\ \bibnamefont {Colón-González}}, \bibinfo {author} {\bibfnamefont {P.~T.}\ \bibnamefont {Lan}}, \bibinfo {author} {\bibfnamefont {P.~T.}\ \bibnamefont {Huong}}, \bibinfo {author} {\bibfnamefont {V.~S.}\ \bibnamefont {Nam}}, \bibinfo {author} {\bibfnamefont {V.~T.}\ \bibnamefont {Duoc}}, \bibinfo {author} {\bibfnamefont {D.~T.}\ \bibnamefont {Hung}}, \bibinfo {author} {\bibfnamefont {N.~T.}\ \bibnamefont {Dong}}, \bibinfo {author} {\bibfnamefont {V.~C.}\ \bibnamefont {Chien}}, \bibinfo {author} {\bibfnamefont {L.~T.~T.}\ \bibnamefont {Trang}}, \bibinfo {author} {\bibfnamefont {D.}~\bibnamefont {Kien~Quoc}}, \bibinfo {author} {\bibfnamefont {T.~M.}\ \bibnamefont {Hoa}}, \bibinfo {author} {\bibfnamefont {N.~H.}\ \bibnamefont {Tai}}, \bibinfo {author} {\bibfnamefont {T.~T.}\ \bibnamefont {Hang}}, \bibinfo {author} {\bibfnamefont {G.}~\bibnamefont {Tsarouchi}}, \bibinfo {author}
  {\bibfnamefont {E.}~\bibnamefont {Ainscoe}}, \bibinfo {author} {\bibfnamefont {Q.}~\bibnamefont {Harpham}}, \bibinfo {author} {\bibfnamefont {B.}~\bibnamefont {Hofmann}}, \bibinfo {author} {\bibfnamefont {D.}~\bibnamefont {Lumbroso}}, \bibinfo {author} {\bibfnamefont {O.~J.}\ \bibnamefont {Brady}},\ and\ \bibinfo {author} {\bibfnamefont {R.}~\bibnamefont {Lowe}},\ }\bibfield  {title} {\bibinfo {title} {Interactions between climate change, urban infrastructure and mobility are driving dengue emergence in {Vietnam}},\ }\href {https://doi.org/10.1038/s41467-023-43954-0} {\bibfield  {journal} {\bibinfo  {journal} {Nature Communications}\ }\textbf {\bibinfo {volume} {14}},\ \bibinfo {pages} {8179} (\bibinfo {year} {2023})},\ \bibinfo {note} {publisher: Nature Publishing Group}\BibitemShut {NoStop}%
\bibitem [{\citenamefont {Yang}\ \emph {et~al.}(2025)\citenamefont {Yang}, \citenamefont {Qiu}, \citenamefont {Xu}, \citenamefont {Gao}, \citenamefont {Tang}, \citenamefont {Tian}, \citenamefont {Wang}, \citenamefont {Lin}, \citenamefont {Shi}, \citenamefont {Chen}, \citenamefont {Zhang}, \citenamefont {Ma}, \citenamefont {Lv}, \citenamefont {Wang}, \citenamefont {Pan}, \citenamefont {Liu},\ and\ \citenamefont {Fang}}]{yang_mapping_2025}%
  \BibitemOpen
  \bibfield  {author} {\bibinfo {author} {\bibfnamefont {Y.-F.}\ \bibnamefont {Yang}}, \bibinfo {author} {\bibfnamefont {Y.-B.}\ \bibnamefont {Qiu}}, \bibinfo {author} {\bibfnamefont {Q.}~\bibnamefont {Xu}}, \bibinfo {author} {\bibfnamefont {R.-C.}\ \bibnamefont {Gao}}, \bibinfo {author} {\bibfnamefont {T.}~\bibnamefont {Tang}}, \bibinfo {author} {\bibfnamefont {Y.}~\bibnamefont {Tian}}, \bibinfo {author} {\bibfnamefont {Y.-H.}\ \bibnamefont {Wang}}, \bibinfo {author} {\bibfnamefont {S.-H.}\ \bibnamefont {Lin}}, \bibinfo {author} {\bibfnamefont {Y.-D.}\ \bibnamefont {Shi}}, \bibinfo {author} {\bibfnamefont {L.-T.}\ \bibnamefont {Chen}}, \bibinfo {author} {\bibfnamefont {Y.}~\bibnamefont {Zhang}}, \bibinfo {author} {\bibfnamefont {J.}~\bibnamefont {Ma}}, \bibinfo {author} {\bibfnamefont {C.-L.}\ \bibnamefont {Lv}}, \bibinfo {author} {\bibfnamefont {G.-L.}\ \bibnamefont {Wang}}, \bibinfo {author} {\bibfnamefont {H.-F.}\ \bibnamefont {Pan}}, \bibinfo {author} {\bibfnamefont {W.}~\bibnamefont {Liu}},\ and\
  \bibinfo {author} {\bibfnamefont {L.-Q.}\ \bibnamefont {Fang}},\ }\bibfield  {title} {\bibinfo {title} {Mapping the global risk of chikungunya virus endemicity and autochthonous transmission following importation},\ }\href {https://doi.org/10.1016/j.tmaid.2025.102892} {\bibfield  {journal} {\bibinfo  {journal} {Travel Medicine and Infectious Disease}\ }\textbf {\bibinfo {volume} {67}},\ \bibinfo {pages} {102892} (\bibinfo {year} {2025})}\BibitemShut {NoStop}%
\bibitem [{\citenamefont {Zhang}\ \emph {et~al.}(2017)\citenamefont {Zhang}, \citenamefont {Sun}, \citenamefont {Chinazzi}, \citenamefont {Pastore~y Piontti}, \citenamefont {Dean}, \citenamefont {Rojas}, \citenamefont {Merler}, \citenamefont {Mistry}, \citenamefont {Poletti}, \citenamefont {Rossi}, \citenamefont {Bray}, \citenamefont {Halloran}, \citenamefont {Longini},\ and\ \citenamefont {Vespignani}}]{zhang_spread_2017}%
  \BibitemOpen
  \bibfield  {author} {\bibinfo {author} {\bibfnamefont {Q.}~\bibnamefont {Zhang}}, \bibinfo {author} {\bibfnamefont {K.}~\bibnamefont {Sun}}, \bibinfo {author} {\bibfnamefont {M.}~\bibnamefont {Chinazzi}}, \bibinfo {author} {\bibfnamefont {A.}~\bibnamefont {Pastore~y Piontti}}, \bibinfo {author} {\bibfnamefont {N.~E.}\ \bibnamefont {Dean}}, \bibinfo {author} {\bibfnamefont {D.~P.}\ \bibnamefont {Rojas}}, \bibinfo {author} {\bibfnamefont {S.}~\bibnamefont {Merler}}, \bibinfo {author} {\bibfnamefont {D.}~\bibnamefont {Mistry}}, \bibinfo {author} {\bibfnamefont {P.}~\bibnamefont {Poletti}}, \bibinfo {author} {\bibfnamefont {L.}~\bibnamefont {Rossi}}, \bibinfo {author} {\bibfnamefont {M.}~\bibnamefont {Bray}}, \bibinfo {author} {\bibfnamefont {M.~E.}\ \bibnamefont {Halloran}}, \bibinfo {author} {\bibfnamefont {I.~M.}\ \bibnamefont {Longini}},\ and\ \bibinfo {author} {\bibfnamefont {A.}~\bibnamefont {Vespignani}},\ }\bibfield  {title} {\bibinfo {title} {Spread of {Zika} virus in the {Americas}},\ }\href
  {https://doi.org/10.1073/pnas.1620161114} {\bibfield  {journal} {\bibinfo  {journal} {Proceedings of the National Academy of Sciences}\ }\textbf {\bibinfo {volume} {114}},\ \bibinfo {pages} {E4334} (\bibinfo {year} {2017})},\ \bibinfo {note} {publisher: Proceedings of the National Academy of Sciences}\BibitemShut {NoStop}%
\bibitem [{\citenamefont {Menegale}\ \emph {et~al.}(2025)\citenamefont {Menegale}, \citenamefont {Manica}, \citenamefont {Del~Manso}, \citenamefont {Bella}, \citenamefont {Zardini}, \citenamefont {Gobbi}, \citenamefont {Mignuoli}, \citenamefont {Mattei}, \citenamefont {Vairo}, \citenamefont {Vezzosi}, \citenamefont {Russo}, \citenamefont {Ferraro}, \citenamefont {Maraglino}, \citenamefont {Palamara}, \citenamefont {Poletti}, \citenamefont {Pezzotti}, \citenamefont {Merler},\ and\ \citenamefont {Riccardo}}]{menegale_risk_2025}%
  \BibitemOpen
  \bibfield  {author} {\bibinfo {author} {\bibfnamefont {F.}~\bibnamefont {Menegale}}, \bibinfo {author} {\bibfnamefont {M.}~\bibnamefont {Manica}}, \bibinfo {author} {\bibfnamefont {M.}~\bibnamefont {Del~Manso}}, \bibinfo {author} {\bibfnamefont {A.}~\bibnamefont {Bella}}, \bibinfo {author} {\bibfnamefont {A.}~\bibnamefont {Zardini}}, \bibinfo {author} {\bibfnamefont {A.}~\bibnamefont {Gobbi}}, \bibinfo {author} {\bibfnamefont {A.~D.}\ \bibnamefont {Mignuoli}}, \bibinfo {author} {\bibfnamefont {G.}~\bibnamefont {Mattei}}, \bibinfo {author} {\bibfnamefont {F.}~\bibnamefont {Vairo}}, \bibinfo {author} {\bibfnamefont {L.}~\bibnamefont {Vezzosi}}, \bibinfo {author} {\bibfnamefont {F.}~\bibnamefont {Russo}}, \bibinfo {author} {\bibfnamefont {F.}~\bibnamefont {Ferraro}}, \bibinfo {author} {\bibfnamefont {F.}~\bibnamefont {Maraglino}}, \bibinfo {author} {\bibfnamefont {A.~T.}\ \bibnamefont {Palamara}}, \bibinfo {author} {\bibfnamefont {P.}~\bibnamefont {Poletti}}, \bibinfo {author} {\bibfnamefont {P.}~\bibnamefont
  {Pezzotti}}, \bibinfo {author} {\bibfnamefont {S.}~\bibnamefont {Merler}},\ and\ \bibinfo {author} {\bibfnamefont {F.}~\bibnamefont {Riccardo}},\ }\bibfield  {title} {\bibinfo {title} {Risk assessment and perspectives of local transmission of chikungunya and dengue in {Italy}, a {European} forerunner},\ }\href {https://doi.org/10.1038/s41467-025-61109-1} {\bibfield  {journal} {\bibinfo  {journal} {Nature Communications}\ }\textbf {\bibinfo {volume} {16}},\ \bibinfo {pages} {6237} (\bibinfo {year} {2025})},\ \bibinfo {note} {publisher: Nature Publishing Group}\BibitemShut {NoStop}%
\bibitem [{\citenamefont {Abel}(2013)}]{abel_estimating_2013}%
  \BibitemOpen
  \bibfield  {author} {\bibinfo {author} {\bibfnamefont {G.}~\bibnamefont {Abel}},\ }\bibfield  {title} {\bibinfo {title} {Estimating global migration flow tables using place of birth data},\ }\href {https://doi.org/10.4054/DemRes.2013.28.18} {\bibfield  {journal} {\bibinfo  {journal} {Demographic Research}\ }\textbf {\bibinfo {volume} {28}},\ \bibinfo {pages} {505} (\bibinfo {year} {2013})}\BibitemShut {NoStop}%
\bibitem [{\citenamefont {Abel}\ and\ \citenamefont {Sander}(2014)}]{abel_quantifying_2014}%
  \BibitemOpen
  \bibfield  {author} {\bibinfo {author} {\bibfnamefont {G.~J.}\ \bibnamefont {Abel}}\ and\ \bibinfo {author} {\bibfnamefont {N.}~\bibnamefont {Sander}},\ }\bibfield  {title} {\bibinfo {title} {Quantifying {Global} {International} {Migration} {Flows}},\ }\href {https://doi.org/10.1126/science.1248676} {\bibfield  {journal} {\bibinfo  {journal} {Science}\ }\textbf {\bibinfo {volume} {343}},\ \bibinfo {pages} {1520} (\bibinfo {year} {2014})},\ \bibinfo {note} {publisher: American Association for the Advancement of Science}\BibitemShut {NoStop}%
\bibitem [{\citenamefont {Simini}\ \emph {et~al.}(2012)\citenamefont {Simini}, \citenamefont {González}, \citenamefont {Maritan},\ and\ \citenamefont {Barabási}}]{simini_universal_2012}%
  \BibitemOpen
  \bibfield  {author} {\bibinfo {author} {\bibfnamefont {F.}~\bibnamefont {Simini}}, \bibinfo {author} {\bibfnamefont {M.~C.}\ \bibnamefont {González}}, \bibinfo {author} {\bibfnamefont {A.}~\bibnamefont {Maritan}},\ and\ \bibinfo {author} {\bibfnamefont {A.-L.}\ \bibnamefont {Barabási}},\ }\bibfield  {title} {\bibinfo {title} {A universal model for mobility and migration patterns},\ }\href {https://doi.org/10.1038/nature10856} {\bibfield  {journal} {\bibinfo  {journal} {Nature}\ }\textbf {\bibinfo {volume} {484}},\ \bibinfo {pages} {96} (\bibinfo {year} {2012})},\ \bibinfo {note} {publisher: Nature Publishing Group}\BibitemShut {NoStop}%
\bibitem [{\citenamefont {Raimondo}(2022)}]{raimondo_network_2022}%
  \BibitemOpen
  \bibfield  {author} {\bibinfo {author} {\bibfnamefont {S.}~\bibnamefont {Raimondo}},\ }\emph {\bibinfo {title} {Network {Models} for {Large}-{Scale} {Human} {Mobility}}},\ \href {https://doi.org/10.15168/11572_346543} {Ph.D. thesis},\ \bibinfo  {school} {UniTrento} (\bibinfo {year} {2022})\BibitemShut {NoStop}%
\bibitem [{\citenamefont {Xu}\ \emph {et~al.}(2020)\citenamefont {Xu}, \citenamefont {Kohler}, \citenamefont {Lenton}, \citenamefont {Svenning},\ and\ \citenamefont {Scheffer}}]{xu_human_niche_2020}%
  \BibitemOpen
  \bibfield  {author} {\bibinfo {author} {\bibfnamefont {C.}~\bibnamefont {Xu}}, \bibinfo {author} {\bibfnamefont {T.~A.}\ \bibnamefont {Kohler}}, \bibinfo {author} {\bibfnamefont {T.~M.}\ \bibnamefont {Lenton}}, \bibinfo {author} {\bibfnamefont {J.-C.}\ \bibnamefont {Svenning}},\ and\ \bibinfo {author} {\bibfnamefont {M.}~\bibnamefont {Scheffer}},\ }\bibfield  {title} {\bibinfo {title} {Future of the human climate niche},\ }\href {https://doi.org/10.1073/pnas.1910114117} {\bibfield  {journal} {\bibinfo  {journal} {Proceedings of the National Academy of Sciences}\ }\textbf {\bibinfo {volume} {117}},\ \bibinfo {pages} {11350} (\bibinfo {year} {2020})},\ \Eprint {https://arxiv.org/abs/https://www.pnas.org/doi/pdf/10.1073/pnas.1910114117} {https://www.pnas.org/doi/pdf/10.1073/pnas.1910114117} \BibitemShut {NoStop}%
\bibitem [{\citenamefont {Balcan}\ \emph {et~al.}(2009{\natexlab{a}})\citenamefont {Balcan}, \citenamefont {Colizza}, \citenamefont {Gonçalves}, \citenamefont {Hu}, \citenamefont {Ramasco},\ and\ \citenamefont {Vespignani}}]{balcan_2009}%
  \BibitemOpen
  \bibfield  {author} {\bibinfo {author} {\bibfnamefont {D.}~\bibnamefont {Balcan}}, \bibinfo {author} {\bibfnamefont {V.}~\bibnamefont {Colizza}}, \bibinfo {author} {\bibfnamefont {B.}~\bibnamefont {Gonçalves}}, \bibinfo {author} {\bibfnamefont {H.}~\bibnamefont {Hu}}, \bibinfo {author} {\bibfnamefont {J.~J.}\ \bibnamefont {Ramasco}},\ and\ \bibinfo {author} {\bibfnamefont {A.}~\bibnamefont {Vespignani}},\ }\bibfield  {title} {\bibinfo {title} {Multiscale mobility networks and the spatial spreading of infectious diseases},\ }\href {https://doi.org/10.1073/pnas.0906910106} {\bibfield  {journal} {\bibinfo  {journal} {Proceedings of the National Academy of Sciences}\ }\textbf {\bibinfo {volume} {106}},\ \bibinfo {pages} {21484} (\bibinfo {year} {2009}{\natexlab{a}})},\ \Eprint {https://arxiv.org/abs/https://www.pnas.org/doi/pdf/10.1073/pnas.0906910106} {https://www.pnas.org/doi/pdf/10.1073/pnas.0906910106} \BibitemShut {NoStop}%
\bibitem [{\citenamefont {Hitchings}\ \emph {et~al.}(2025)\citenamefont {Hitchings}, \citenamefont {Xu}, \citenamefont {García-Carreras}, \citenamefont {Gallagher}, \citenamefont {O’Hagan},\ and\ \citenamefont {Cummings}}]{hitchings_2025}%
  \BibitemOpen
  \bibfield  {author} {\bibinfo {author} {\bibfnamefont {M.~D.~T.}\ \bibnamefont {Hitchings}}, \bibinfo {author} {\bibfnamefont {Y.}~\bibnamefont {Xu}}, \bibinfo {author} {\bibfnamefont {B.}~\bibnamefont {García-Carreras}}, \bibinfo {author} {\bibfnamefont {A.}~\bibnamefont {Gallagher}}, \bibinfo {author} {\bibfnamefont {J.~J.}\ \bibnamefont {O’Hagan}},\ and\ \bibinfo {author} {\bibfnamefont {D.~A.~T.}\ \bibnamefont {Cummings}},\ }\bibfield  {title} {\bibinfo {title} {Estimating the incidence of dengue in international air travelers from non-endemic countries between 2010–2019},\ }\href {https://doi.org/10.1371/journal.pntd.0013291} {\bibfield  {journal} {\bibinfo  {journal} {PLOS Neglected Tropical Diseases}\ }\textbf {\bibinfo {volume} {19}},\ \bibinfo {pages} {1} (\bibinfo {year} {2025})}\BibitemShut {NoStop}%
\bibitem [{\citenamefont {{EUROCONTROL}}(2024)}]{EUROCONTROL2024}%
  \BibitemOpen
  \bibfield  {author} {\bibinfo {author} {\bibnamefont {{EUROCONTROL}}},\ }\href {https://www.eurocontrol.int/publication/eurocontrol-forecast-2024-2050} {\bibinfo {title} {Eurocontrol aviation long-term outlook: Flights and co$_2$ emissions forecast 2024--2050}} (\bibinfo {year} {2024}),\ \bibinfo {note} {published 19 December 2024}\BibitemShut {NoStop}%
\bibitem [{\citenamefont {Merrifield}\ \emph {et~al.}(2023)\citenamefont {Merrifield}, \citenamefont {Brunner}, \citenamefont {Lorenz}, \citenamefont {Humphrey},\ and\ \citenamefont {Knutti}}]{merrifield_climate_2023}%
  \BibitemOpen
  \bibfield  {author} {\bibinfo {author} {\bibfnamefont {A.~L.}\ \bibnamefont {Merrifield}}, \bibinfo {author} {\bibfnamefont {L.}~\bibnamefont {Brunner}}, \bibinfo {author} {\bibfnamefont {R.}~\bibnamefont {Lorenz}}, \bibinfo {author} {\bibfnamefont {V.}~\bibnamefont {Humphrey}},\ and\ \bibinfo {author} {\bibfnamefont {R.}~\bibnamefont {Knutti}},\ }\bibfield  {title} {\bibinfo {title} {Climate model {Selection} by {Independence}, {Performance}, and {Spread} ({ClimSIPS} v1.0.1) for regional applications},\ }\href {https://doi.org/10.5194/gmd-16-4715-2023} {\bibfield  {journal} {\bibinfo  {journal} {Geoscientific Model Development}\ }\textbf {\bibinfo {volume} {16}},\ \bibinfo {pages} {4715} (\bibinfo {year} {2023})},\ \bibinfo {note} {publisher: Copernicus GmbH}\BibitemShut {NoStop}%
\bibitem [{\citenamefont {Palmer}\ \emph {et~al.}(2023)\citenamefont {Palmer}, \citenamefont {McSweeney}, \citenamefont {Booth}, \citenamefont {Priestley}, \citenamefont {Davini}, \citenamefont {Brunner}, \citenamefont {Borchert},\ and\ \citenamefont {Menary}}]{palmer_performance-based_2023}%
  \BibitemOpen
  \bibfield  {author} {\bibinfo {author} {\bibfnamefont {T.~E.}\ \bibnamefont {Palmer}}, \bibinfo {author} {\bibfnamefont {C.~F.}\ \bibnamefont {McSweeney}}, \bibinfo {author} {\bibfnamefont {B.~B.~B.}\ \bibnamefont {Booth}}, \bibinfo {author} {\bibfnamefont {M.~D.~K.}\ \bibnamefont {Priestley}}, \bibinfo {author} {\bibfnamefont {P.}~\bibnamefont {Davini}}, \bibinfo {author} {\bibfnamefont {L.}~\bibnamefont {Brunner}}, \bibinfo {author} {\bibfnamefont {L.}~\bibnamefont {Borchert}},\ and\ \bibinfo {author} {\bibfnamefont {M.~B.}\ \bibnamefont {Menary}},\ }\bibfield  {title} {\bibinfo {title} {Performance-based sub-selection of {CMIP6} models for impact assessments in {Europe}},\ }\href {https://doi.org/10.5194/esd-14-457-2023} {\bibfield  {journal} {\bibinfo  {journal} {Earth System Dynamics}\ }\textbf {\bibinfo {volume} {14}},\ \bibinfo {pages} {457} (\bibinfo {year} {2023})},\ \bibinfo {note} {publisher: Copernicus GmbH}\BibitemShut {NoStop}%
\bibitem [{\citenamefont {Van~Vuuren}\ \emph {et~al.}(2011)\citenamefont {Van~Vuuren}, \citenamefont {Edmonds}, \citenamefont {Kainuma}, \citenamefont {Riahi}, \citenamefont {Thomson}, \citenamefont {Hibbard}, \citenamefont {Hurtt}, \citenamefont {Kram}, \citenamefont {Krey}, \citenamefont {Lamarque} \emph {et~al.}}]{van2011representative}%
  \BibitemOpen
  \bibfield  {author} {\bibinfo {author} {\bibfnamefont {D.~P.}\ \bibnamefont {Van~Vuuren}}, \bibinfo {author} {\bibfnamefont {J.}~\bibnamefont {Edmonds}}, \bibinfo {author} {\bibfnamefont {M.}~\bibnamefont {Kainuma}}, \bibinfo {author} {\bibfnamefont {K.}~\bibnamefont {Riahi}}, \bibinfo {author} {\bibfnamefont {A.}~\bibnamefont {Thomson}}, \bibinfo {author} {\bibfnamefont {K.}~\bibnamefont {Hibbard}}, \bibinfo {author} {\bibfnamefont {G.~C.}\ \bibnamefont {Hurtt}}, \bibinfo {author} {\bibfnamefont {T.}~\bibnamefont {Kram}}, \bibinfo {author} {\bibfnamefont {V.}~\bibnamefont {Krey}}, \bibinfo {author} {\bibfnamefont {J.-F.}\ \bibnamefont {Lamarque}}, \emph {et~al.},\ }\bibfield  {title} {\bibinfo {title} {The representative concentration pathways: an overview},\ }\href@noop {} {\bibfield  {journal} {\bibinfo  {journal} {Climatic change}\ }\textbf {\bibinfo {volume} {109}},\ \bibinfo {pages} {5} (\bibinfo {year} {2011})}\BibitemShut {NoStop}%
\bibitem [{\citenamefont {Li}\ \emph {et~al.}(2022)\citenamefont {Li}, \citenamefont {Dong}, \citenamefont {Li},\ and\ \citenamefont {Dong}}]{li_big_2022}%
  \BibitemOpen
  \bibfield  {author} {\bibinfo {author} {\bibfnamefont {Z.}~\bibnamefont {Li}}, \bibinfo {author} {\bibfnamefont {J.}~\bibnamefont {Dong}}, \bibinfo {author} {\bibfnamefont {Z.}~\bibnamefont {Li}},\ and\ \bibinfo {author} {\bibfnamefont {J.}~\bibnamefont {Dong}},\ }\bibfield  {title} {\bibinfo {title} {Big {Geospatial} {Data} and {Data}-{Driven} {Methods} for {Urban} {Dengue} {Risk} {Forecasting}: {A} {Review}},\ }\bibfield  {journal} {\bibinfo  {journal} {Remote Sensing}\ }\textbf {\bibinfo {volume} {14}},\ \href {https://doi.org/10.3390/rs14195052} {10.3390/rs14195052} (\bibinfo {year} {2022}),\ \bibinfo {note} {company: Multidisciplinary Digital Publishing Institute Distributor: Multidisciplinary Digital Publishing Institute Institution: Multidisciplinary Digital Publishing Institute Label: Multidisciplinary Digital Publishing Institute Publisher: publisher}\BibitemShut {NoStop}%
\bibitem [{\citenamefont {Niva}\ \emph {et~al.}(2023)\citenamefont {Niva}, \citenamefont {Horton}, \citenamefont {Virkki}, \citenamefont {Heino}, \citenamefont {Kosonen}, \citenamefont {Kallio}, \citenamefont {Kinnunen}, \citenamefont {Abel}, \citenamefont {Muttarak}, \citenamefont {Taka}, \citenamefont {Varis},\ and\ \citenamefont {Kummu}}]{niva_worlds_2023}%
  \BibitemOpen
  \bibfield  {author} {\bibinfo {author} {\bibfnamefont {V.}~\bibnamefont {Niva}}, \bibinfo {author} {\bibfnamefont {A.}~\bibnamefont {Horton}}, \bibinfo {author} {\bibfnamefont {V.}~\bibnamefont {Virkki}}, \bibinfo {author} {\bibfnamefont {M.}~\bibnamefont {Heino}}, \bibinfo {author} {\bibfnamefont {M.}~\bibnamefont {Kosonen}}, \bibinfo {author} {\bibfnamefont {M.}~\bibnamefont {Kallio}}, \bibinfo {author} {\bibfnamefont {P.}~\bibnamefont {Kinnunen}}, \bibinfo {author} {\bibfnamefont {G.~J.}\ \bibnamefont {Abel}}, \bibinfo {author} {\bibfnamefont {R.}~\bibnamefont {Muttarak}}, \bibinfo {author} {\bibfnamefont {M.}~\bibnamefont {Taka}}, \bibinfo {author} {\bibfnamefont {O.}~\bibnamefont {Varis}},\ and\ \bibinfo {author} {\bibfnamefont {M.}~\bibnamefont {Kummu}},\ }\bibfield  {title} {\bibinfo {title} {World’s human migration patterns in 2000–2019 unveiled by high-resolution data},\ }\href {https://doi.org/10.1038/s41562-023-01689-4} {\bibfield  {journal} {\bibinfo  {journal} {Nature Human Behaviour}\
  }\textbf {\bibinfo {volume} {7}},\ \bibinfo {pages} {2023} (\bibinfo {year} {2023})}\BibitemShut {NoStop}%
\bibitem [{\citenamefont {KC}\ and\ \citenamefont {Lutz}(2017)}]{KC2017181}%
  \BibitemOpen
  \bibfield  {author} {\bibinfo {author} {\bibfnamefont {S.}~\bibnamefont {KC}}\ and\ \bibinfo {author} {\bibfnamefont {W.}~\bibnamefont {Lutz}},\ }\bibfield  {title} {\bibinfo {title} {The human core of the shared socioeconomic pathways: Population scenarios by age, sex and level of education for all countries to 2100},\ }\href {https://doi.org/https://doi.org/10.1016/j.gloenvcha.2014.06.004} {\bibfield  {journal} {\bibinfo  {journal} {Global Environmental Change}\ }\textbf {\bibinfo {volume} {42}},\ \bibinfo {pages} {181} (\bibinfo {year} {2017})}\BibitemShut {NoStop}%
\bibitem [{\citenamefont {Eritja}\ \emph {et~al.}(2017)\citenamefont {Eritja}, \citenamefont {Palmer}, \citenamefont {Roiz}, \citenamefont {Sanpera-Calbet},\ and\ \citenamefont {Bartumeus}}]{eritja_direct_2017}%
  \BibitemOpen
  \bibfield  {author} {\bibinfo {author} {\bibfnamefont {R.}~\bibnamefont {Eritja}}, \bibinfo {author} {\bibfnamefont {J.~R.~B.}\ \bibnamefont {Palmer}}, \bibinfo {author} {\bibfnamefont {D.}~\bibnamefont {Roiz}}, \bibinfo {author} {\bibfnamefont {I.}~\bibnamefont {Sanpera-Calbet}},\ and\ \bibinfo {author} {\bibfnamefont {F.}~\bibnamefont {Bartumeus}},\ }\bibfield  {title} {\bibinfo {title} {Direct {Evidence} of {Adult} {Aedes} albopictus {Dispersal} by {Car}},\ }\href {https://doi.org/10.1038/s41598-017-12652-5} {\bibfield  {journal} {\bibinfo  {journal} {Scientific Reports}\ }\textbf {\bibinfo {volume} {7}},\ \bibinfo {pages} {14399} (\bibinfo {year} {2017})},\ \bibinfo {note} {publisher: Nature Publishing Group}\BibitemShut {NoStop}%
\bibitem [{\citenamefont {Pisaneschi}\ \emph {et~al.}(2026)\citenamefont {Pisaneschi}, \citenamefont {Manfredi}, \citenamefont {Landi}, \citenamefont {Stollenwerk},\ and\ \citenamefont {Aguiar}}]{PISANESCHI_2026}%
  \BibitemOpen
  \bibfield  {author} {\bibinfo {author} {\bibfnamefont {G.}~\bibnamefont {Pisaneschi}}, \bibinfo {author} {\bibfnamefont {P.}~\bibnamefont {Manfredi}}, \bibinfo {author} {\bibfnamefont {A.}~\bibnamefont {Landi}}, \bibinfo {author} {\bibfnamefont {N.}~\bibnamefont {Stollenwerk}},\ and\ \bibinfo {author} {\bibfnamefont {M.}~\bibnamefont {Aguiar}},\ }\bibfield  {title} {\bibinfo {title} {When few mosquitoes are enough: Dengue outbreaks in non-endemic areas},\ }\href {https://doi.org/https://doi.org/10.1016/j.onehlt.2025.101308} {\bibfield  {journal} {\bibinfo  {journal} {One Health}\ }\textbf {\bibinfo {volume} {22}},\ \bibinfo {pages} {101308} (\bibinfo {year} {2026})}\BibitemShut {NoStop}%
\bibitem [{\citenamefont {Adams}\ and\ \citenamefont {Kapan}(2009)}]{adams_man_2009}%
  \BibitemOpen
  \bibfield  {author} {\bibinfo {author} {\bibfnamefont {B.}~\bibnamefont {Adams}}\ and\ \bibinfo {author} {\bibfnamefont {D.~D.}\ \bibnamefont {Kapan}},\ }\bibfield  {title} {\bibinfo {title} {Man {Bites} {Mosquito}: {Understanding} the {Contribution} of {Human} {Movement} to {Vector}-{Borne} {Disease} {Dynamics}},\ }\href {https://doi.org/10.1371/journal.pone.0006763} {\bibfield  {journal} {\bibinfo  {journal} {PLOS ONE}\ }\textbf {\bibinfo {volume} {4}},\ \bibinfo {pages} {e6763} (\bibinfo {year} {2009})},\ \bibinfo {note} {publisher: Public Library of Science}\BibitemShut {NoStop}%
\bibitem [{\citenamefont {Schewel}\ \emph {et~al.}(2024)\citenamefont {Schewel}, \citenamefont {Dickerson}, \citenamefont {Madson},\ and\ \citenamefont {Nagle~Alverio}}]{schewel_how_2024}%
  \BibitemOpen
  \bibfield  {author} {\bibinfo {author} {\bibfnamefont {K.}~\bibnamefont {Schewel}}, \bibinfo {author} {\bibfnamefont {S.}~\bibnamefont {Dickerson}}, \bibinfo {author} {\bibfnamefont {B.}~\bibnamefont {Madson}},\ and\ \bibinfo {author} {\bibfnamefont {G.}~\bibnamefont {Nagle~Alverio}},\ }\bibfield  {title} {{\selectlanguage {English}\bibinfo {title} {How well can we predict climate migration? {A} review of forecasting models}},\ }\bibfield  {journal} {\bibinfo  {journal} {Frontiers in Climate}\ }\textbf {\bibinfo {volume} {5}},\ \href {https://doi.org/10.3389/fclim.2023.1189125} {10.3389/fclim.2023.1189125} (\bibinfo {year} {2024})\BibitemShut {NoStop}%
\bibitem [{\citenamefont {Hauer}\ \emph {et~al.}(2024)\citenamefont {Hauer}, \citenamefont {Jacobs},\ and\ \citenamefont {Kulp}}]{mathew_pnas_2024}%
  \BibitemOpen
  \bibfield  {author} {\bibinfo {author} {\bibfnamefont {M.~E.}\ \bibnamefont {Hauer}}, \bibinfo {author} {\bibfnamefont {S.~A.}\ \bibnamefont {Jacobs}},\ and\ \bibinfo {author} {\bibfnamefont {S.~A.}\ \bibnamefont {Kulp}},\ }\bibfield  {title} {\bibinfo {title} {Climate migration amplifies demographic change and population aging},\ }\href {https://doi.org/10.1073/pnas.2206192119} {\bibfield  {journal} {\bibinfo  {journal} {Proceedings of the National Academy of Sciences}\ }\textbf {\bibinfo {volume} {121}},\ \bibinfo {pages} {e2206192119} (\bibinfo {year} {2024})},\ \Eprint {https://arxiv.org/abs/https://www.pnas.org/doi/pdf/10.1073/pnas.2206192119} {https://www.pnas.org/doi/pdf/10.1073/pnas.2206192119} \BibitemShut {NoStop}%
\bibitem [{\citenamefont {Ryan}\ \emph {et~al.}(2019)\citenamefont {Ryan}, \citenamefont {Carlson}, \citenamefont {Mordecai},\ and\ \citenamefont {Johnson}}]{ryan_global_2019}%
  \BibitemOpen
  \bibfield  {author} {\bibinfo {author} {\bibfnamefont {S.~J.}\ \bibnamefont {Ryan}}, \bibinfo {author} {\bibfnamefont {C.~J.}\ \bibnamefont {Carlson}}, \bibinfo {author} {\bibfnamefont {E.~A.}\ \bibnamefont {Mordecai}},\ and\ \bibinfo {author} {\bibfnamefont {L.~R.}\ \bibnamefont {Johnson}},\ }\bibfield  {title} {\bibinfo {title} {Global expansion and redistribution of {Aedes}-borne virus transmission risk with climate change},\ }\href {https://doi.org/10.1371/journal.pntd.0007213} {\bibfield  {journal} {\bibinfo  {journal} {PLOS Neglected Tropical Diseases}\ }\textbf {\bibinfo {volume} {13}},\ \bibinfo {pages} {e0007213} (\bibinfo {year} {2019})},\ \bibinfo {note} {publisher: Public Library of Science}\BibitemShut {NoStop}%
\bibitem [{\citenamefont {Rold\'an-G\'omez}\ \emph {et~al.}(2025)\citenamefont {Rold\'an-G\'omez}, \citenamefont {Ortega},\ and\ \citenamefont {Donat}}]{gomez_amoc_2025}%
  \BibitemOpen
  \bibfield  {author} {\bibinfo {author} {\bibfnamefont {P.~J.}\ \bibnamefont {Rold\'an-G\'omez}}, \bibinfo {author} {\bibfnamefont {P.}~\bibnamefont {Ortega}},\ and\ \bibinfo {author} {\bibfnamefont {M.~G.}\ \bibnamefont {Donat}},\ }\bibfield  {title} {\bibinfo {title} {Contribution of meridional overturning circulation and sea ice changes to large-scale temperature asymmetries in cmip6 overshoot scenarios},\ }\href {https://doi.org/10.5194/os-21-2283-2025} {\bibfield  {journal} {\bibinfo  {journal} {Ocean Science}\ }\textbf {\bibinfo {volume} {21}},\ \bibinfo {pages} {2283} (\bibinfo {year} {2025})}\BibitemShut {NoStop}%
\bibitem [{\citenamefont {van Westen}\ \emph {et~al.}(2024)\citenamefont {van Westen}, \citenamefont {Kliphuis},\ and\ \citenamefont {Dijkstra}}]{vamwesten_2024}%
  \BibitemOpen
  \bibfield  {author} {\bibinfo {author} {\bibfnamefont {R.~M.}\ \bibnamefont {van Westen}}, \bibinfo {author} {\bibfnamefont {M.}~\bibnamefont {Kliphuis}},\ and\ \bibinfo {author} {\bibfnamefont {H.~A.}\ \bibnamefont {Dijkstra}},\ }\bibfield  {title} {\bibinfo {title} {Physics-based early warning signal shows that amoc is on tipping course},\ }\href {https://doi.org/10.1126/sciadv.adk1189} {\bibfield  {journal} {\bibinfo  {journal} {Science Advances}\ }\textbf {\bibinfo {volume} {10}},\ \bibinfo {pages} {eadk1189} (\bibinfo {year} {2024})},\ \Eprint {https://arxiv.org/abs/https://www.science.org/doi/pdf/10.1126/sciadv.adk1189} {https://www.science.org/doi/pdf/10.1126/sciadv.adk1189} \BibitemShut {NoStop}%
\bibitem [{\citenamefont {Kraemer}\ \emph {et~al.}(2019)\citenamefont {Kraemer}, \citenamefont {Reiner}, \citenamefont {Brady}, \citenamefont {Messina}, \citenamefont {Gilbert}, \citenamefont {Pigott}, \citenamefont {Yi}, \citenamefont {Johnson}, \citenamefont {Earl}, \citenamefont {Marczak}, \citenamefont {Shirude}, \citenamefont {Davis~Weaver}, \citenamefont {Bisanzio}, \citenamefont {Perkins}, \citenamefont {Lai}, \citenamefont {Lu}, \citenamefont {Jones}, \citenamefont {Coelho}, \citenamefont {Carvalho}, \citenamefont {Van~Bortel}, \citenamefont {Marsboom}, \citenamefont {Hendrickx}, \citenamefont {Schaffner}, \citenamefont {Moore}, \citenamefont {Nax}, \citenamefont {Bengtsson}, \citenamefont {Wetter}, \citenamefont {Tatem}, \citenamefont {Brownstein}, \citenamefont {Smith}, \citenamefont {Lambrechts}, \citenamefont {Cauchemez}, \citenamefont {Linard}, \citenamefont {Faria}, \citenamefont {Pybus}, \citenamefont {Scott}, \citenamefont {Liu}, \citenamefont {Yu}, \citenamefont {Wint}, \citenamefont {Hay},\
  and\ \citenamefont {Golding}}]{kraemer_past_2019}%
  \BibitemOpen
  \bibfield  {author} {\bibinfo {author} {\bibfnamefont {M.~U.~G.}\ \bibnamefont {Kraemer}}, \bibinfo {author} {\bibfnamefont {R.~C.}\ \bibnamefont {Reiner}}, \bibinfo {author} {\bibfnamefont {O.~J.}\ \bibnamefont {Brady}}, \bibinfo {author} {\bibfnamefont {J.~P.}\ \bibnamefont {Messina}}, \bibinfo {author} {\bibfnamefont {M.}~\bibnamefont {Gilbert}}, \bibinfo {author} {\bibfnamefont {D.~M.}\ \bibnamefont {Pigott}}, \bibinfo {author} {\bibfnamefont {D.}~\bibnamefont {Yi}}, \bibinfo {author} {\bibfnamefont {K.}~\bibnamefont {Johnson}}, \bibinfo {author} {\bibfnamefont {L.}~\bibnamefont {Earl}}, \bibinfo {author} {\bibfnamefont {L.~B.}\ \bibnamefont {Marczak}}, \bibinfo {author} {\bibfnamefont {S.}~\bibnamefont {Shirude}}, \bibinfo {author} {\bibfnamefont {N.}~\bibnamefont {Davis~Weaver}}, \bibinfo {author} {\bibfnamefont {D.}~\bibnamefont {Bisanzio}}, \bibinfo {author} {\bibfnamefont {T.~A.}\ \bibnamefont {Perkins}}, \bibinfo {author} {\bibfnamefont {S.}~\bibnamefont {Lai}}, \bibinfo {author} {\bibfnamefont
  {X.}~\bibnamefont {Lu}}, \bibinfo {author} {\bibfnamefont {P.}~\bibnamefont {Jones}}, \bibinfo {author} {\bibfnamefont {G.~E.}\ \bibnamefont {Coelho}}, \bibinfo {author} {\bibfnamefont {R.~G.}\ \bibnamefont {Carvalho}}, \bibinfo {author} {\bibfnamefont {W.}~\bibnamefont {Van~Bortel}}, \bibinfo {author} {\bibfnamefont {C.}~\bibnamefont {Marsboom}}, \bibinfo {author} {\bibfnamefont {G.}~\bibnamefont {Hendrickx}}, \bibinfo {author} {\bibfnamefont {F.}~\bibnamefont {Schaffner}}, \bibinfo {author} {\bibfnamefont {C.~G.}\ \bibnamefont {Moore}}, \bibinfo {author} {\bibfnamefont {H.~H.}\ \bibnamefont {Nax}}, \bibinfo {author} {\bibfnamefont {L.}~\bibnamefont {Bengtsson}}, \bibinfo {author} {\bibfnamefont {E.}~\bibnamefont {Wetter}}, \bibinfo {author} {\bibfnamefont {A.~J.}\ \bibnamefont {Tatem}}, \bibinfo {author} {\bibfnamefont {J.~S.}\ \bibnamefont {Brownstein}}, \bibinfo {author} {\bibfnamefont {D.~L.}\ \bibnamefont {Smith}}, \bibinfo {author} {\bibfnamefont {L.}~\bibnamefont {Lambrechts}}, \bibinfo {author}
  {\bibfnamefont {S.}~\bibnamefont {Cauchemez}}, \bibinfo {author} {\bibfnamefont {C.}~\bibnamefont {Linard}}, \bibinfo {author} {\bibfnamefont {N.~R.}\ \bibnamefont {Faria}}, \bibinfo {author} {\bibfnamefont {O.~G.}\ \bibnamefont {Pybus}}, \bibinfo {author} {\bibfnamefont {T.~W.}\ \bibnamefont {Scott}}, \bibinfo {author} {\bibfnamefont {Q.}~\bibnamefont {Liu}}, \bibinfo {author} {\bibfnamefont {H.}~\bibnamefont {Yu}}, \bibinfo {author} {\bibfnamefont {G.~R.~W.}\ \bibnamefont {Wint}}, \bibinfo {author} {\bibfnamefont {S.~I.}\ \bibnamefont {Hay}},\ and\ \bibinfo {author} {\bibfnamefont {N.}~\bibnamefont {Golding}},\ }\bibfield  {title} {\bibinfo {title} {Past and future spread of the arbovirus vectors {Aedes} aegypti and {Aedes} albopictus},\ }\href {https://doi.org/10.1038/s41564-019-0376-y} {\bibfield  {journal} {\bibinfo  {journal} {Nature Microbiology}\ }\textbf {\bibinfo {volume} {4}},\ \bibinfo {pages} {854} (\bibinfo {year} {2019})},\ \bibinfo {note} {publisher: Nature Publishing Group}\BibitemShut
  {NoStop}%
\bibitem [{\citenamefont {Liu-Helmersson}\ \emph {et~al.}(2016)\citenamefont {Liu-Helmersson}, \citenamefont {Quam}, \citenamefont {Wilder-Smith}, \citenamefont {Stenlund}, \citenamefont {Ebi}, \citenamefont {Massad},\ and\ \citenamefont {Rocklöv}}]{liu-helmersson_climate_2016}%
  \BibitemOpen
  \bibfield  {author} {\bibinfo {author} {\bibfnamefont {J.}~\bibnamefont {Liu-Helmersson}}, \bibinfo {author} {\bibfnamefont {M.}~\bibnamefont {Quam}}, \bibinfo {author} {\bibfnamefont {A.}~\bibnamefont {Wilder-Smith}}, \bibinfo {author} {\bibfnamefont {H.}~\bibnamefont {Stenlund}}, \bibinfo {author} {\bibfnamefont {K.}~\bibnamefont {Ebi}}, \bibinfo {author} {\bibfnamefont {E.}~\bibnamefont {Massad}},\ and\ \bibinfo {author} {\bibfnamefont {J.}~\bibnamefont {Rocklöv}},\ }\bibfield  {title} {{\selectlanguage {English}\bibinfo {title} {Climate {Change} and {Aedes} {Vectors}: 21st {Century} {Projections} for {Dengue} {Transmission} in {Europe}}},\ }\href {https://doi.org/10.1016/j.ebiom.2016.03.046} {\bibfield  {journal} {\bibinfo  {journal} {eBioMedicine}\ }\textbf {\bibinfo {volume} {7}},\ \bibinfo {pages} {267} (\bibinfo {year} {2016})},\ \bibinfo {note} {publisher: Elsevier}\BibitemShut {NoStop}%
\bibitem [{\citenamefont {Cunze}\ \emph {et~al.}(2016{\natexlab{a}})\citenamefont {Cunze}, \citenamefont {Koch}, \citenamefont {Kochmann},\ and\ \citenamefont {Klimpel}}]{cunze_aedes_2016}%
  \BibitemOpen
  \bibfield  {author} {\bibinfo {author} {\bibfnamefont {S.}~\bibnamefont {Cunze}}, \bibinfo {author} {\bibfnamefont {L.~K.}\ \bibnamefont {Koch}}, \bibinfo {author} {\bibfnamefont {J.}~\bibnamefont {Kochmann}},\ and\ \bibinfo {author} {\bibfnamefont {S.}~\bibnamefont {Klimpel}},\ }\bibfield  {title} {\bibinfo {title} {Aedes albopictus and {Aedes} japonicus - two invasive mosquito species with different temperature niches in {Europe}},\ }\href {https://doi.org/10.1186/s13071-016-1853-2} {\bibfield  {journal} {\bibinfo  {journal} {Parasites \& Vectors}\ }\textbf {\bibinfo {volume} {9}},\ \bibinfo {pages} {573} (\bibinfo {year} {2016}{\natexlab{a}})}\BibitemShut {NoStop}%
\bibitem [{\citenamefont {Colón-González}\ \emph {et~al.}(2021)\citenamefont {Colón-González}, \citenamefont {Sewe}, \citenamefont {Tompkins}, \citenamefont {Sjödin}, \citenamefont {Casallas}, \citenamefont {Rocklöv}, \citenamefont {Caminade},\ and\ \citenamefont {Lowe}}]{colon-gonzalez_projecting_2021}%
  \BibitemOpen
  \bibfield  {author} {\bibinfo {author} {\bibfnamefont {F.~J.}\ \bibnamefont {Colón-González}}, \bibinfo {author} {\bibfnamefont {M.~O.}\ \bibnamefont {Sewe}}, \bibinfo {author} {\bibfnamefont {A.~M.}\ \bibnamefont {Tompkins}}, \bibinfo {author} {\bibfnamefont {H.}~\bibnamefont {Sjödin}}, \bibinfo {author} {\bibfnamefont {A.}~\bibnamefont {Casallas}}, \bibinfo {author} {\bibfnamefont {J.}~\bibnamefont {Rocklöv}}, \bibinfo {author} {\bibfnamefont {C.}~\bibnamefont {Caminade}},\ and\ \bibinfo {author} {\bibfnamefont {R.}~\bibnamefont {Lowe}},\ }\bibfield  {title} {{\selectlanguage {English}\bibinfo {title} {Projecting the risk of mosquito-borne diseases in a warmer and more populated world: a multi-model, multi-scenario intercomparison modelling study}},\ }\href {https://doi.org/10.1016/S2542-5196(21)00132-7} {\bibfield  {journal} {\bibinfo  {journal} {The Lancet Planetary Health}\ }\textbf {\bibinfo {volume} {5}},\ \bibinfo {pages} {e404} (\bibinfo {year} {2021})},\ \bibinfo {note} {publisher:
  Elsevier}\BibitemShut {NoStop}%
\bibitem [{\citenamefont {Knoblauch}\ \emph {et~al.}(2025)\citenamefont {Knoblauch}, \citenamefont {Heidecke}, \citenamefont {de~A.~Rocha}, \citenamefont {Paolucci~Pimenta}, \citenamefont {Reinmuth}, \citenamefont {Lautenbach}, \citenamefont {Brady}, \citenamefont {Jänisch}, \citenamefont {Resch}, \citenamefont {Biljecki}, \citenamefont {Rocklöv}, \citenamefont {Wilder-Smith}, \citenamefont {Bärnighausen},\ and\ \citenamefont {Zipf}}]{knoblauch_modeling_2025}%
  \BibitemOpen
  \bibfield  {author} {\bibinfo {author} {\bibfnamefont {S.}~\bibnamefont {Knoblauch}}, \bibinfo {author} {\bibfnamefont {J.}~\bibnamefont {Heidecke}}, \bibinfo {author} {\bibfnamefont {A.~A.}\ \bibnamefont {de~A.~Rocha}}, \bibinfo {author} {\bibfnamefont {P.~F.}\ \bibnamefont {Paolucci~Pimenta}}, \bibinfo {author} {\bibfnamefont {M.}~\bibnamefont {Reinmuth}}, \bibinfo {author} {\bibfnamefont {S.}~\bibnamefont {Lautenbach}}, \bibinfo {author} {\bibfnamefont {O.~J.}\ \bibnamefont {Brady}}, \bibinfo {author} {\bibfnamefont {T.}~\bibnamefont {Jänisch}}, \bibinfo {author} {\bibfnamefont {B.}~\bibnamefont {Resch}}, \bibinfo {author} {\bibfnamefont {F.}~\bibnamefont {Biljecki}}, \bibinfo {author} {\bibfnamefont {J.}~\bibnamefont {Rocklöv}}, \bibinfo {author} {\bibfnamefont {A.}~\bibnamefont {Wilder-Smith}}, \bibinfo {author} {\bibfnamefont {T.}~\bibnamefont {Bärnighausen}},\ and\ \bibinfo {author} {\bibfnamefont {A.}~\bibnamefont {Zipf}},\ }\bibfield  {title} {\bibinfo {title} {Modeling {Intraday} {Aedes}-human
  exposure dynamics enhances dengue risk prediction},\ }\href {https://doi.org/10.1038/s41598-025-91950-9} {\bibfield  {journal} {\bibinfo  {journal} {Scientific Reports}\ }\textbf {\bibinfo {volume} {15}},\ \bibinfo {pages} {7994} (\bibinfo {year} {2025})},\ \bibinfo {note} {publisher: Nature Publishing Group}\BibitemShut {NoStop}%
\bibitem [{\citenamefont {Romeo-Aznar}\ \emph {et~al.}(2022)\citenamefont {Romeo-Aznar}, \citenamefont {Picinini~Freitas}, \citenamefont {Gonçalves~Cruz}, \citenamefont {King},\ and\ \citenamefont {Pascual}}]{romeo-aznar_fine-scale_2022}%
  \BibitemOpen
  \bibfield  {author} {\bibinfo {author} {\bibfnamefont {V.}~\bibnamefont {Romeo-Aznar}}, \bibinfo {author} {\bibfnamefont {L.}~\bibnamefont {Picinini~Freitas}}, \bibinfo {author} {\bibfnamefont {O.}~\bibnamefont {Gonçalves~Cruz}}, \bibinfo {author} {\bibfnamefont {A.~A.}\ \bibnamefont {King}},\ and\ \bibinfo {author} {\bibfnamefont {M.}~\bibnamefont {Pascual}},\ }\bibfield  {title} {\bibinfo {title} {Fine-scale heterogeneity in population density predicts wave dynamics in dengue epidemics},\ }\href {https://doi.org/10.1038/s41467-022-28231-w} {\bibfield  {journal} {\bibinfo  {journal} {Nature Communications}\ }\textbf {\bibinfo {volume} {13}},\ \bibinfo {pages} {996} (\bibinfo {year} {2022})},\ \bibinfo {note} {publisher: Nature Publishing Group}\BibitemShut {NoStop}%
\bibitem [{\citenamefont {Pastor-Satorras}\ \emph {et~al.}(2015)\citenamefont {Pastor-Satorras}, \citenamefont {Castellano}, \citenamefont {Van~Mieghem},\ and\ \citenamefont {Vespignani}}]{pastor-satorras_epidemic_2015}%
  \BibitemOpen
  \bibfield  {author} {\bibinfo {author} {\bibfnamefont {R.}~\bibnamefont {Pastor-Satorras}}, \bibinfo {author} {\bibfnamefont {C.}~\bibnamefont {Castellano}}, \bibinfo {author} {\bibfnamefont {P.}~\bibnamefont {Van~Mieghem}},\ and\ \bibinfo {author} {\bibfnamefont {A.}~\bibnamefont {Vespignani}},\ }\bibfield  {title} {\bibinfo {title} {Epidemic processes in complex networks},\ }\href {https://doi.org/10.1103/RevModPhys.87.925} {\bibfield  {journal} {\bibinfo  {journal} {Reviews of Modern Physics}\ }\textbf {\bibinfo {volume} {87}},\ \bibinfo {pages} {925} (\bibinfo {year} {2015})},\ \bibinfo {note} {publisher: American Physical Society}\BibitemShut {NoStop}%
\bibitem [{\citenamefont {Cunze}\ \emph {et~al.}(2016{\natexlab{b}})\citenamefont {Cunze}, \citenamefont {Kochmann}, \citenamefont {Koch},\ and\ \citenamefont {Klimpel}}]{cunze_aedes_2016-1}%
  \BibitemOpen
  \bibfield  {author} {\bibinfo {author} {\bibfnamefont {S.}~\bibnamefont {Cunze}}, \bibinfo {author} {\bibfnamefont {J.}~\bibnamefont {Kochmann}}, \bibinfo {author} {\bibfnamefont {L.~K.}\ \bibnamefont {Koch}},\ and\ \bibinfo {author} {\bibfnamefont {S.}~\bibnamefont {Klimpel}},\ }\bibfield  {title} {\bibinfo {title} {Aedes albopictus and {Its} {Environmental} {Limits} in {Europe}},\ }\href {https://doi.org/10.1371/journal.pone.0162116} {\bibfield  {journal} {\bibinfo  {journal} {PLOS ONE}\ }\textbf {\bibinfo {volume} {11}},\ \bibinfo {pages} {e0162116} (\bibinfo {year} {2016}{\natexlab{b}})},\ \bibinfo {note} {publisher: Public Library of Science}\BibitemShut {NoStop}%
\bibitem [{\citenamefont {Merkenschlager}\ \emph {et~al.}(2025)\citenamefont {Merkenschlager}, \citenamefont {Bangelesa}, \citenamefont {Paeth},\ and\ \citenamefont {Hertig}}]{MERKENSCHLAGER_2025}%
  \BibitemOpen
  \bibfield  {author} {\bibinfo {author} {\bibfnamefont {C.}~\bibnamefont {Merkenschlager}}, \bibinfo {author} {\bibfnamefont {F.}~\bibnamefont {Bangelesa}}, \bibinfo {author} {\bibfnamefont {H.}~\bibnamefont {Paeth}},\ and\ \bibinfo {author} {\bibfnamefont {E.}~\bibnamefont {Hertig}},\ }\bibfield  {title} {\bibinfo {title} {Evolution of the recent habitat suitability area of aedes albopictus in the extended mediterranean area due to land-use and climate change},\ }\href {https://doi.org/https://doi.org/10.1016/j.scitotenv.2025.179202} {\bibfield  {journal} {\bibinfo  {journal} {Science of The Total Environment}\ }\textbf {\bibinfo {volume} {974}},\ \bibinfo {pages} {179202} (\bibinfo {year} {2025})}\BibitemShut {NoStop}%
\bibitem [{\citenamefont {Balcan}\ \emph {et~al.}(2009{\natexlab{b}})\citenamefont {Balcan}, \citenamefont {Colizza}, \citenamefont {Gonçalves}, \citenamefont {Hu}, \citenamefont {Ramasco},\ and\ \citenamefont {Vespignani}}]{balcan_multiscale_2009}%
  \BibitemOpen
  \bibfield  {author} {\bibinfo {author} {\bibfnamefont {D.}~\bibnamefont {Balcan}}, \bibinfo {author} {\bibfnamefont {V.}~\bibnamefont {Colizza}}, \bibinfo {author} {\bibfnamefont {B.}~\bibnamefont {Gonçalves}}, \bibinfo {author} {\bibfnamefont {H.}~\bibnamefont {Hu}}, \bibinfo {author} {\bibfnamefont {J.~J.}\ \bibnamefont {Ramasco}},\ and\ \bibinfo {author} {\bibfnamefont {A.}~\bibnamefont {Vespignani}},\ }\bibfield  {title} {\bibinfo {title} {Multiscale mobility networks and the spatial spreading of infectious diseases},\ }\href {https://doi.org/10.1073/pnas.0906910106} {\bibfield  {journal} {\bibinfo  {journal} {Proceedings of the National Academy of Sciences}\ }\textbf {\bibinfo {volume} {106}},\ \bibinfo {pages} {21484} (\bibinfo {year} {2009}{\natexlab{b}})},\ \bibinfo {note} {publisher: Proceedings of the National Academy of Sciences}\BibitemShut {NoStop}%
\bibitem [{\citenamefont {Kraemer}\ \emph {et~al.}(2025)\citenamefont {Kraemer}, \citenamefont {Tsui}, \citenamefont {Chang}, \citenamefont {Lytras}, \citenamefont {Khurana}, \citenamefont {Vanderslott}, \citenamefont {Bajaj}, \citenamefont {Scheidwasser}, \citenamefont {Curran-Sebastian}, \citenamefont {Semenova} \emph {et~al.}}]{kraemer2025artificial}%
  \BibitemOpen
  \bibfield  {author} {\bibinfo {author} {\bibfnamefont {M.~U.}\ \bibnamefont {Kraemer}}, \bibinfo {author} {\bibfnamefont {J.~L.-H.}\ \bibnamefont {Tsui}}, \bibinfo {author} {\bibfnamefont {S.~Y.}\ \bibnamefont {Chang}}, \bibinfo {author} {\bibfnamefont {S.}~\bibnamefont {Lytras}}, \bibinfo {author} {\bibfnamefont {M.~P.}\ \bibnamefont {Khurana}}, \bibinfo {author} {\bibfnamefont {S.}~\bibnamefont {Vanderslott}}, \bibinfo {author} {\bibfnamefont {S.}~\bibnamefont {Bajaj}}, \bibinfo {author} {\bibfnamefont {N.}~\bibnamefont {Scheidwasser}}, \bibinfo {author} {\bibfnamefont {J.~L.}\ \bibnamefont {Curran-Sebastian}}, \bibinfo {author} {\bibfnamefont {E.}~\bibnamefont {Semenova}}, \emph {et~al.},\ }\bibfield  {title} {\bibinfo {title} {Artificial intelligence for modelling infectious disease epidemics},\ }\href@noop {} {\bibfield  {journal} {\bibinfo  {journal} {Nature}\ }\textbf {\bibinfo {volume} {638}},\ \bibinfo {pages} {623} (\bibinfo {year} {2025})}\BibitemShut {NoStop}%
\bibitem [{\citenamefont {Newton}\ and\ \citenamefont {Reiter}(1992)}]{newton_model_1992}%
  \BibitemOpen
  \bibfield  {author} {\bibinfo {author} {\bibfnamefont {E.~A.~C.}\ \bibnamefont {Newton}}\ and\ \bibinfo {author} {\bibfnamefont {P.}~\bibnamefont {Reiter}},\ }\bibfield  {title} {\bibinfo {title} {A {Model} of the {Transmission} of {Dengue} {Fever} with an {Evaluation} of the {Impact} of {Ultra}-{Low} {Volume} ({ULV}) {Insecticide} {Applications} on {Dengue} {Epidemics}},\ }\href {https://doi.org/10.4269/ajtmh.1992.47.709} {\bibfield  {journal} {\bibinfo  {journal} {The American Journal of Tropical Medicine and Hygiene}\ }\textbf {\bibinfo {volume} {47}},\ \bibinfo {pages} {709} (\bibinfo {year} {1992})},\ \bibinfo {note} {publisher: The American Society of Tropical Medicine and Hygiene Section: The American Journal of Tropical Medicine and Hygiene}\BibitemShut {NoStop}%
\bibitem [{\citenamefont {Keeling}\ and\ \citenamefont {Rohani}(2008)}]{keeling_modeling_2008}%
  \BibitemOpen
  \bibfield  {author} {\bibinfo {author} {\bibfnamefont {M.~J.}\ \bibnamefont {Keeling}}\ and\ \bibinfo {author} {\bibfnamefont {P.}~\bibnamefont {Rohani}},\ }\href {https://doi.org/10.2307/j.ctvcm4gk0} {\emph {\bibinfo {title} {Modeling {Infectious} {Diseases} in {Humans} and {Animals}}}}\ (\bibinfo  {publisher} {Princeton University Press},\ \bibinfo {year} {2008})\BibitemShut {NoStop}%
\bibitem [{\citenamefont {Nguyen}\ \emph {et~al.}(2023)\citenamefont {Nguyen}, \citenamefont {Bartels},\ and\ \citenamefont {Gilligan}}]{nguyen_modelling_2023}%
  \BibitemOpen
  \bibfield  {author} {\bibinfo {author} {\bibfnamefont {V.-A.}\ \bibnamefont {Nguyen}}, \bibinfo {author} {\bibfnamefont {D.~W.}\ \bibnamefont {Bartels}},\ and\ \bibinfo {author} {\bibfnamefont {C.~A.}\ \bibnamefont {Gilligan}},\ }\bibfield  {title} {\bibinfo {title} {Modelling the spread and mitigation of an emerging vector-borne pathogen: {Citrus} greening in the {U}.{S}.},\ }\href {https://doi.org/10.1371/journal.pcbi.1010156} {\bibfield  {journal} {\bibinfo  {journal} {PLOS Computational Biology}\ }\textbf {\bibinfo {volume} {19}},\ \bibinfo {pages} {e1010156} (\bibinfo {year} {2023})},\ \bibinfo {note} {publisher: Public Library of Science}\BibitemShut {NoStop}%
\bibitem [{\citenamefont {Sumner}\ \emph {et~al.}(2017)\citenamefont {Sumner}, \citenamefont {Orton}, \citenamefont {Green}, \citenamefont {Kao},\ and\ \citenamefont {Gubbins}}]{sumner_quantifying_2017}%
  \BibitemOpen
  \bibfield  {author} {\bibinfo {author} {\bibfnamefont {T.}~\bibnamefont {Sumner}}, \bibinfo {author} {\bibfnamefont {R.~J.}\ \bibnamefont {Orton}}, \bibinfo {author} {\bibfnamefont {D.~M.}\ \bibnamefont {Green}}, \bibinfo {author} {\bibfnamefont {R.~R.}\ \bibnamefont {Kao}},\ and\ \bibinfo {author} {\bibfnamefont {S.}~\bibnamefont {Gubbins}},\ }\bibfield  {title} {\bibinfo {title} {Quantifying the roles of host movement and vector dispersal in the transmission of vector-borne diseases of livestock},\ }\href {https://doi.org/10.1371/journal.pcbi.1005470} {\bibfield  {journal} {\bibinfo  {journal} {PLOS Computational Biology}\ }\textbf {\bibinfo {volume} {13}},\ \bibinfo {pages} {e1005470} (\bibinfo {year} {2017})},\ \bibinfo {note} {publisher: Public Library of Science}\BibitemShut {NoStop}%
\bibitem [{\citenamefont {Bosetti}\ \emph {et~al.}(2020)\citenamefont {Bosetti}, \citenamefont {Poletti}, \citenamefont {Stella}, \citenamefont {Lepri}, \citenamefont {Merler},\ and\ \citenamefont {De~Domenico}}]{bosetti_heterogeneity_2020}%
  \BibitemOpen
  \bibfield  {author} {\bibinfo {author} {\bibfnamefont {P.}~\bibnamefont {Bosetti}}, \bibinfo {author} {\bibfnamefont {P.}~\bibnamefont {Poletti}}, \bibinfo {author} {\bibfnamefont {M.}~\bibnamefont {Stella}}, \bibinfo {author} {\bibfnamefont {B.}~\bibnamefont {Lepri}}, \bibinfo {author} {\bibfnamefont {S.}~\bibnamefont {Merler}},\ and\ \bibinfo {author} {\bibfnamefont {M.}~\bibnamefont {De~Domenico}},\ }\bibfield  {title} {\bibinfo {title} {Heterogeneity in social and epidemiological factors determines the risk of measles outbreaks},\ }\href {https://doi.org/10.1073/pnas.1920986117} {\bibfield  {journal} {\bibinfo  {journal} {Proceedings of the National Academy of Sciences}\ }\textbf {\bibinfo {volume} {117}},\ \bibinfo {pages} {30118} (\bibinfo {year} {2020})},\ \bibinfo {note} {publisher: Proceedings of the National Academy of Sciences}\BibitemShut {NoStop}%
\bibitem [{\citenamefont {Marini}\ \emph {et~al.}(2019)\citenamefont {Marini}, \citenamefont {Guzzetta}, \citenamefont {Marques~Toledo}, \citenamefont {Teixeira}, \citenamefont {Rosà},\ and\ \citenamefont {Merler}}]{marini_effectiveness_2019}%
  \BibitemOpen
  \bibfield  {author} {\bibinfo {author} {\bibfnamefont {G.}~\bibnamefont {Marini}}, \bibinfo {author} {\bibfnamefont {G.}~\bibnamefont {Guzzetta}}, \bibinfo {author} {\bibfnamefont {C.~A.}\ \bibnamefont {Marques~Toledo}}, \bibinfo {author} {\bibfnamefont {M.}~\bibnamefont {Teixeira}}, \bibinfo {author} {\bibfnamefont {R.}~\bibnamefont {Rosà}},\ and\ \bibinfo {author} {\bibfnamefont {S.}~\bibnamefont {Merler}},\ }\bibfield  {title} {\bibinfo {title} {Effectiveness of {Ultra}-{Low} {Volume} insecticide spraying to prevent dengue in a non-endemic metropolitan area of {Brazil}},\ }\href {https://doi.org/10.1371/journal.pcbi.1006831} {\bibfield  {journal} {\bibinfo  {journal} {PLOS Computational Biology}\ }\textbf {\bibinfo {volume} {15}},\ \bibinfo {pages} {1} (\bibinfo {year} {2019})},\ \bibinfo {note} {publisher: Public Library of Science}\BibitemShut {NoStop}%
\end{thebibliography}

%

\end{document}

\maketitle
\noindent
\textbf{1} Department of Physics \& Astronomy ‘Galileo Galilei’, University of Padua, Padua, Italy
\\
\textbf{2} Istituto Nazionale di Fisica Nucleare, Sez. Padova, Padua, Italy
\\
\textbf{3} Istituto Nazionale di Fisica Nucleare, Sez. Milano Bicocca, Milan, Italy
\\
\textbf{4} Department of Mathematical Sciences, University of Parma, Parma, Italy
\\
\textbf{5} Padova Neuroscience Center, University of Padua, Padua, Italy
\\
\textbf{6} Department of Informatics, Bioengineering, Robotics, and Systems Engineering, University of Genoa, Genoa, Italy
\\
\textbf{7}  Machine Learning Genoa Center, University of Genoa, Genoa, Italy
\\
\textbf{8} 
Padua Center for Network Medicine, University of Padua, Padua, Italy
\bigskip
\section{Material and Methods}
\linenumbers 

\subsection{Data sources}
All the data used to build the Europe model are listed in Tab.\ref{tab:raw_data}, together with their sources and a succinct description.
\begin{table}[h!]
\begin{adjustwidth}{-0.5in}{0in}
\centering
\begin{tabular}{||c c c c||} 
 \hline
 \thead{Covariate} & \thead{Type} & \thead{Source} & \thead{Description} \\ [0.5ex] 
 \hline\hline
 Daily mean temperature & Climate & \href{https://cds.climate.copernicus.eu/datasets/}{CMIP6 Copernicus} & \makecell{Daily mean temperature \\ projections on a year \\ from 2030 to 2100, \\ at every decade \\ (models: ACCESS-CM2,\\CESM2,MRI-ESM2-0)}  \\ 
 \hline
 Monthly precipitation & Climate & CMIP6 Copernicus & \makecell{Monthly precipitations projections \\ on a year \\ from 2030 to 2100, \\ at every decade \\ (models: ACCESS-CM2,\\CESM2,MRI-ESM2-0)} \\
 \hline
 Population projections & Population & \makecell{\href{https://www.cgd.ucar.edu/sections/iam/modeling/spatial-population}{CGD-NCAR}} & \makecell{World population projection \\ with resolution of $0.125^\circ \times 0.125^\circ$ \\($ \approx15 \times 15$ kms \\ at the equator) \\ scenarios: SSP1,SSP2,SSP5}  \\
 \hline
 Mobility flow   & Mobility & \makecell{\href{https://www.istat.it/notizia/matrice-di-pendolarismo-per-lavoro/}{ISTAT}} & \makecell{Matrix origin-destination \\ for commuting between \\ italian NUTS3 (2021)} \\
 \hline
 Air traffic mobility & Mobility & \href{https://www.oag.com/flight-data-sets}{OAG} & \makecell{Scheduled seats origin-destination \\ in 2014 \\ on commercial lines} \\ 
 \hline
 Load factor & Mobility & \href{https://www.icao.int/world-air-transport-2014}{ICAO} & \makecell{Load factor of \\ european flights in 2014} \\ 
 \hline
 Air traffic projections & Mobility & \href{https://www.eurocontrol.int/publication/eurocontrol-forecast-2024-2050}{EuroControl} & \makecell{Projections of the growth \\ of the European air traffic \\
 from 2024 to 2050} \\ 
 \hline
 Shape and position of NUTS3  & Geography & \href{https://ec.europa.eu/eurostat/web/gisco/geodata/statistical-units/territorial-units-statistics}{Eurostat NUTS}  & \makecell{NUTS3 in 2021,\\ scale 20M, \\CRS 4326}  \\
 \hline
 Current population  & Population & \href{https://ec.europa.eu/eurostat/web/gisco/geodata/population-distribution/population-grids}{Eurostat Population Grid} & \makecell{Grid of the European Union \\ population from the 2021 Census \\ (resolution cell 1 km2)}  \\
 %(make updated population.py)}
 \hline
 Surface NUTS3  & Geography & \href{https://ec.europa.eu/eurostat/databrowser/view/reg_area3/default/map?lang=en}{Eurostat Regional Area}  & \makecell{Surface of the \\ NUTS3 in km2 (2021)} \\  
 \hline
 Daily mean temperature 2024 & Climate & ERA-5 Land Copernicus & \makecell{Historical record  \\daily mean temperature \\ in 2024 \\ with resolution of \\$0.5^\circ \times 0.5^\circ$ \\($ \approx 50 \times 50$ kms \\at the equator)}  \\ 
 \hline
 Monthly precipitation 2024 & Climate &ERA-5 Land Copernicus & \makecell{Historical record  \\monthly precipitations \\  in 2024  \\with resolution  of \\ $0.1^\circ \times 0.1^\circ$ ($ \approx 11 \times 11$ kms \\ at the equator)} \\
 \hline
 Imported cases of dengue & Epidemiology & \href{https://www.ecdc.europa.eu/en/dengue/surveillance/dengue-virus-infections-travellers}{ECDC Dengue}  & \makecell{Number of travel-associated dengue \\ cases reported to ECDC \\ on the period 2019-2023} \\
 \hline
\end{tabular}
\caption{Brief description and references of the primary data sources used in the Europe model. }
\label{tab:raw_data}
\end{adjustwidth}
\end{table}

\subsection{Projected and current population data}

For the European Union, the projected populations and current population in each grid are summed at the NUTS3 level with \textit{geocube} mask in \textit{python} using the shape file provided by Eurostat (see Tab.\ref{tab:raw_data}). Analysis is reduced to the 27 countries in the European Union by 2021 (excluding outermost territories) and the EFTA countries (excluding Iceland) which leads to 1200 NUTS3 nodes. Each NUTS3 is located in space by the centroid of its polygon shape provided by Eurostat. Distance between them is computed with the \emph{distance} method from the \emph{geopandas} library.

\subsection{Surface and correction of the vector-to-host-ratio}
\label{sec:vector_to_host_ratio}
For each patch, the vector abundance model provides the dynamics of the number of vectors per surface unit.  Considering low density areas in terms of population would lead to overestimate the vector and human populations that  are effectively in contact. Following the methodology in \cite{zardini_estimating_2024}, we compute the ratio $r_V^i$ number of cells per NUTS3 patch $i$ where the density of inhabitants is greater than 100 inhabitants over the total number of cells of the patch
from the current gridded population data (resolution $\text{$1\times 1$}$ $km^2$). We then compute the corrected vector population $\kappa^{i,corr} = r_V^i \kappa^{i}$

This gives the correction factor for each patch's surface from which we compute an effective vector abundance.
Similarly we obtain for each patch $i$ the ratio $r_H^i$ of the number of inhabitants living in cells with more than 100 inhabitants over the total number of inhabitants. We then compute the corrected human population as $N_h^{i,corr}=r_H^i N_h^i$. This corrects for the fact that we disregard the inhabitants living in cells with density lower than the bound.
For the Europe model, this leads to keep 92.6 \% of the current population for subsequent analysis.

\subsection{Mobility flow}

\begin{figure}
    \centering
    \includegraphics[width=0.5\linewidth]{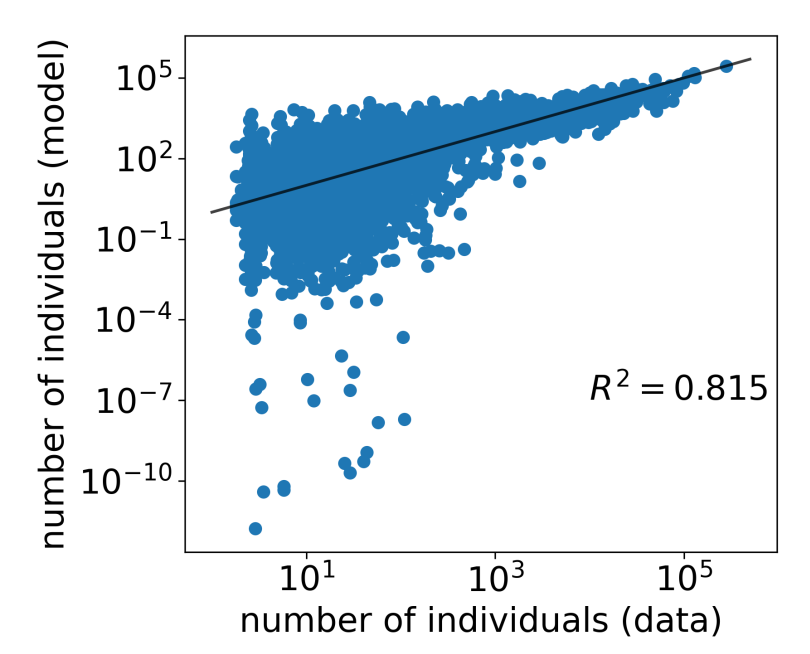}
    \caption{\textbf{Calibration of the gravity model for short-range human mobility.} Comparison between the observed number of individuals commuting daily from italian NUTS3 patches and the predictions of the gravity model calibrated on those data, $R^2=0.815$. Black line is $y=x$.}
    \label{fig:gravity model}
\end{figure}

\begin{figure}
    \centering
    \includegraphics[width=0.5\linewidth]{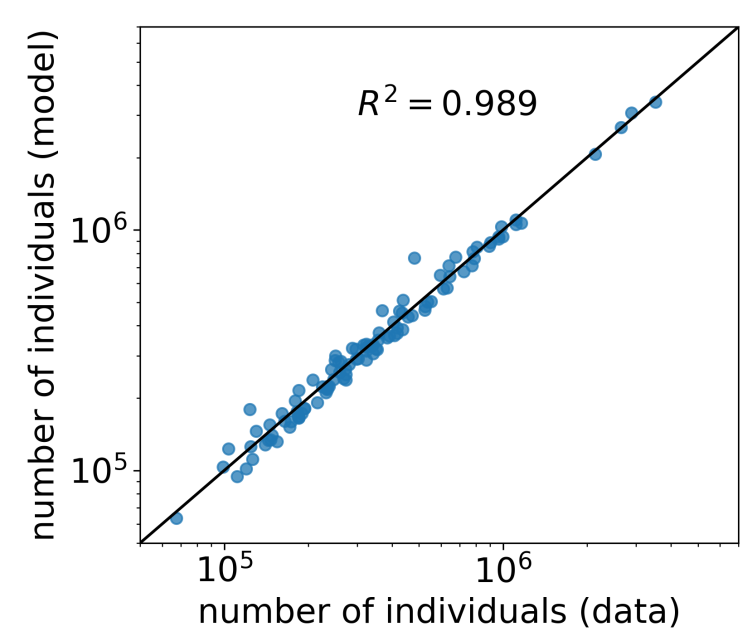}
    \caption{\textbf{Calibration of the power-law model for short-range human mobility.} Comparison between the observed number of individuals remaining daily in the same italian NUTS3 patches and the predictions of the power-law model calibrated on those data, $R^2=0.989$. Black line is $y=x$.}
    \label{fig:powerlaw_stayer}
\end{figure}

First, we estimate the daily short-range mobility flows $P_{ij}^{short}$ between the NUTS3 patches using a gravity model \cite{balcan_multiscale_2009}:
\begin{align}
    P_{ij}^{short} = C \frac{N_i^{\alpha}N_j^{\gamma}}{e^{\frac{d_{ij}}{r}}},
\end{align}
with $N_i$,$N_j$,$r$ the human population in patch $i$, the human population in patch $j$ and the characteristic length of the mobility in the exponential kernel, respectively. The model is calibrated using italian census data of commuting (see Tab.\ref{tab:raw_data}). Parameters read: $C=(2.2\pm0.6)*10^{-5}$, $\alpha=0.60\pm0.01$,$\gamma=1.05\pm0.01$,$r=28155\pm374$ (see Fig. \ref{fig:gravity model})
Using the same italian census data, we estimate the daily population that remains within the same node using a power-law model \cite{bettencourt_2016}:
\begin{align}
    P_{ii} = aN_i^\beta
\end{align}
After calibration, parameters read: $a=0.51 \pm 0.07$,$\beta=1.04\pm0.01$ (see Fig. \ref{fig:powerlaw_stayer})
In order to account for daily long-range mobility flows, we use data from OAG that accounts for the number of scheduled seats between any pair of commercial airports worldwide. Here, we consider only flows between airports located in the Europe model. The airports are associated with the NUTS3 patch whose centroid is the closest from their position. In order to account for the effective number of passengers, we adjust the number of scheduled seats between any two pairs of patches $i$ and $j$, $F_{ij}$ with a load factor $C_f =0.8$ such that $P_{ij}^{long} = C_fF_{ij}$ (see Tab.\ref{tab:raw_data}).Finally $P_{ij}=P_{ij}^{long}+P_{ij}^{short}$ is then normalized such that $\sum_jP_{ij}=1$, to obtain daily transition rates between and within patches. While $P_{ij}^{short}$evolves according to the population projections, $P_{ij}^{long}$ grows 1.1\% per year following the projections of the evolution of the European airline traffic from EuroControl (see Tab. \ref{tab:raw_data}).

\subsection{Air traffic flow and importation rate}
The air traffic data is filtered to only take into account passengers flow from airports located in the 25 first countries or overseas territories that imported cases of dengue in Europe between 2019-2023 to EU airports (see Tab. \ref{tab:raw_data}). The arrival airports are associated with the NUTS3 whose centroid is the closest from their position. The total number of incoming schedule seats by patch $f_i$ is then normalized such that $\sum_i f_i=1$. The daily importation rate of infected cases $N_w$ is determined by dividing the total number of imported cases in the Europe model between 2019 and 2023, as provided by the ECDC by the total number of days between 1st January 2019 and 31st December 2023 (see Tab. \ref{tab:value_model}) accounting that around only $25$\% of imported cases are reported  \cite{hitchings_2025}.
This leads to $N_w=29.37$ daily imported cases of dengue from endemic countries in the Europe model.
Note that $N_w$ grows with the airline traffic at a rate of 1.1 \% per year from the baseline year 2024 following the projections of the evolution of the European airline traffic from EuroControl (see Tab. \ref{tab:raw_data}).

\subsection{Climate projections and climate data}
The subset of climate projections are selected for being independent i.e. coming from different model families \cite{merrifield_climate_2023}, consistent in reproducing historical patterns in summer conditions averaged over mediterranean Europe  and central Europe \cite{palmer_performance-based_2023} and leading to a diversity of climate change patterns. For instance, MRI-ESM2-0 has been consistently associated with a stronger weakening of the Atlantic Meridional Overtuning Circulation (AMOC- a powerful regulator of the european climate) over a variety of scenarios as compared with ACCESS-CM2, for instance \cite{gomez_amoc_2025}. 

All climate projections are down-scaled  to cells of size 0.5°x0.5° (around 50*50 kms and 70 000 points worldwide) before use in epidemiological simulations or in the large-scale migration model.

Climate records for the year 2024 and climate projections were converted in Celsius for the daily mean temperature and in millimeters for the monthly precipitation where needed.

\subsection{Transmission model}

\subsubsection{ODEs}
\begin{align}
    \frac{dS_h^i}{dt} &= \frac{N_h^i-S_h^i}{T_{lh}} - \frac{S_h^ic_{vh}^i}{N_h^i} \sum_j I_v^j \left(\frac{e^{-\frac{d_{j i}}{d_0}}+ p_0P_{j i}}{1+\sum_{i \neq j }e^{-\frac{d_{j i}}{d_0}}+p_0}\right) \\
	\frac{dE_h^i}{dt} &= \frac{S_h^ic_{vh}^i}{N_h^i} \sum_j I_v^j \left(\frac{e^{-\frac{d_{j i}}{d_0}}+ p_0P_{j i}}{1+\sum_{i \neq j }e^{-\frac{d_{j i}}{d_0}}+p_0}\right) - E_h^i(\frac{1}{T_{iit}} + \frac{1}{T_{lh}}) \\
	\frac{dI_h^i}{dt} &= \frac{E_h^i}{T_{iit}} - I_h^i(\frac{1}{T_{id}} + \frac{1}{T_{lh}}) \\ 
	\frac{dR_h^i}{dt} &= \frac{I_h^i}{T_{id}} -  \frac{R_h^i}{T_{lh}} \\
	\frac{dS_v^i}{dt} &= \frac{\kappa^i-S_v^i}{T_{lv}} - \frac{S_v^i c_{hv}^i}{N_h^i} \sum_j \left(I_h^j + \delta_{ij}f_iN_w \right)P_{j i} \\
	\frac{dE_v^i}{dt} &= \frac{S_v^i c_{hv}^i}{N_h^i} \sum_j \left(I_h^j + \delta_{ij}f_iN_w \right)P_{j i}  - E_v^i(\frac{1}{T_{eit}} + \frac{1}{T_{lv}}) \\
	\frac{dI_v^i}{dt} &= \frac{E_v^i}{T_{eit}} - \frac{I_v^i}{T_{lv}}.
\end{align}
\subsubsection{Value of parameters}
The values of the parameters used in the main text are described in Tab.\ref{tab:value_model}. Parameters related to time are in days and distances are in meters.
The scaling factor $p_0$ is determined from the average number of \textit{Aedes albopictus} that are present by car trip departing from the Barcelona metropolitan area as extrapolated from field experiments \cite{eritja_direct_2017}. Other parameters are determined from literature on vector-borne disease described in Tab.\ref{tab:value_model}.

\begin{table}[h!]
\begin{adjustwidth}{-0.5in}{0in}
\centering
\begin{tabular}{|| c c c c c ||} 
 \hline
 \thead{Category} & \thead{Parameter} & \thead{Value} & \thead{Function} & \thead{References}\\ [0.5ex] 
 \hline\hline
Demographics & \makecell{Human life expectancy \\ \\ \\ \\ \\ Vector life expectancy \\  \\ \\ \\ \\ \\ Vector carrying capacity \\ (Vector population)} & \makecell{$T_{lh}=29220$  \\  \\ \\ $T_{lv}$\\ \\ \\  \\ \\ $\kappa$} & \makecell{ / \\ If $T<15$°C: \\ $(\frac{1}{1.1+e^{-4.04+0.576*T}}+0.12)^{-1}$ \\ If $15$°C$\leq T \leq 26.3$°C: \\ $\frac{1}{0.000339T^2  – 0.0189T + 0.336}$ \\ If $T >26.3 $°C: \\ $(\frac{1}{1.065+e^{32.3-0.92T}}+0.0747)^{-1}$ \\ see \cite{zardini_estimating_2024} (supplementary) }  & \makecell{/ \\ \cite{zardini_estimating_2024} \\ \\ \\ \\ \\ \cite{zardini_estimating_2024} } \\ \hline
Mobility & \makecell{Characteristic length \\ (autonomous vector dispersal) \\ Scaling factor from host mobility \\ to the host mediated vector one} & \makecell{$d_0 = 150 m$ \\ \\ $p_0 = 0.005$ \\ } & \makecell{ / \\ \\ / \\ } & \makecell{ \cite{zardini_estimating_2024,marini_estimating_2019} \\ \\ \cite{eritja_direct_2017} \\} \\ \hline
Dengue & \makecell{Extrinsic incubation time \\ \\ Intrinsic incubation time \\ \\ Human infection duration \\ \\ Effective contact rate host to vector \\ \\ Effective contact rate vector to host \\ \\ Daily importation rate } & \makecell{$T_{eit}$  \\ \\ $T_{iit}= 5$ \\ \\ $T_{id}=5$ \\ \\ $c_{hv}$ \\ \\ $c_{vh}$ \\ \\ $N_w=29.37$} & \makecell{  $\frac{1}{0.000109T(T-10.39)\sqrt{43.05-T}}$ \\ \\ \\ \\ \\   $0.6 \times 0.5(0.0043T+0.0943)$ \\ $\times 0.000439T(T-3.62)\sqrt{36.82-T}$  \\ \\ $0.6 \times 0.5(0.0043T+0.0943)$ \\ $\times 0.000735T(T-15.84)\sqrt{36.40-T}$ \\  /} & \makecell{\cite{mordecai_detecting_2017} \\ \\ \cite{manore_comparing_2014} \\ \cite{marini_estimating_2019} \\ \cite{mordecai_detecting_2017}\\} \\ \hline
\end{tabular}
\caption{Epidemiological parameters of \textit{Aedes albopictus} for dengue, integrated in the joint model. Times are in $day$, contact rate in $day^{-1}$,distance in meters, temperatures $T$ in Celsius.}
\label{tab:value_model}
\end{adjustwidth}
\end{table}

\subsection{Large-scale migration model}
In this section we detail the large-scale migration model.
\subsubsection{From temperature and precipitation to human suitability }

In \cite{xu_human_niche_2020}, the Authors show that the vast majority of human population lives under narrow conditions of temperature and precipitation on the earth. Furthermore, they show that this niche stayed relatively stable since Antiquity, suggesting fundamental climatic constraints for human lives. Quantitatively the \textit{human niche} is defined as the current density of human population for a given combination of mean annual temperature (MAT) and mean annual precipitation (MAP). Therefore, the human niche is deterministically determined by the pair MAT-MAP at a given location. Here, we propose to use the evolution of the human niche as a migration potential. To compute the human niche, we adapted the code provided by  \cite{xu_human_niche_2020} at this address (https://datadryad.org/dataset/doi:10.5061/dryad.fj6q573q7).

\begin{figure}[h!]
    \centering
    \includegraphics[width=1\linewidth]{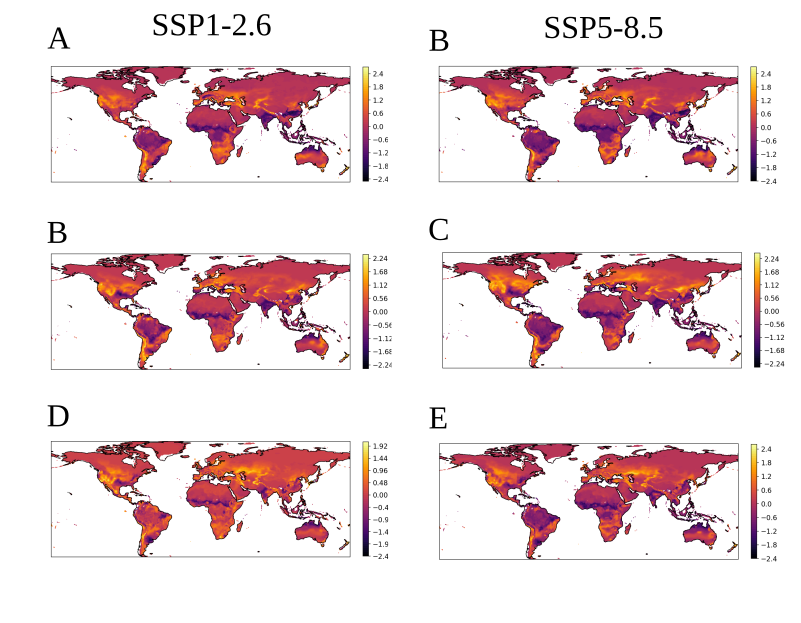}
    \caption{\textbf{Difference $\Delta H$ between the predicted human niche in 2060 and the current human niche  for SSP1-2.6 and SSP5-8.5.}. Human niche is predicted for the ACCESS-CM2 (A-B), the CESM2 (B-C) and the MRI-ESM2-0 model (D-E).}
    \label{fig:2060_suitability}
\end{figure}

\begin{figure}[h!]
    \centering
    \includegraphics[width=1\linewidth]{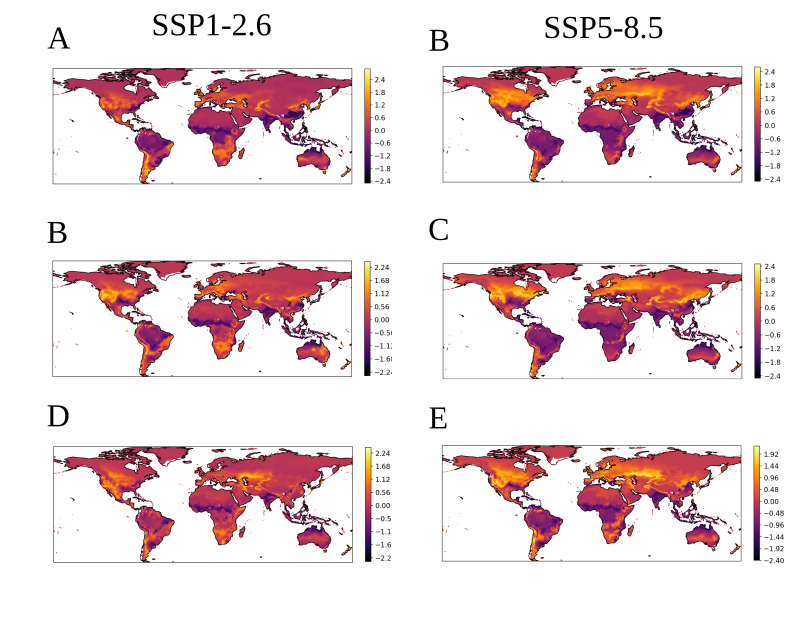}
    \caption{\textbf{Difference between the predicted human niche in 2090 and the current human niche  for SSP1-2.6 and SSP5-8.5.}. Human niche is predicted for the ACCESS-CM2 (A-B), the CESM2 (B-C) and the MRI-ESM2-0 model (D-E).}
    \label{fig:2090_suitability}
\end{figure}

For each considered year, scenario and climate projections, we converted the longitude of daily surface temperature data and daily precipitation into the $-180^\circ,180^\circ$ format. Daily precipitation in kg.m-2.s-1 are converted into mm/day. We compute the MAT and MAP and each combination MAP-MAT is associated with an index of human niche based on the current MAP-MAT human distribution. Since each cell is associated with a MAT-MAP combination every position is also attributed a human niche. Finally the difference in human niche $\Delta H$ is computed between the predicted human niche (for a given year, scenario, model) and the current one for each position. Note that the current human niche is defined for climate representative of 1960-1990. More details on the computation of the current human niche are available at \cite{xu_human_niche_2020}.The population projections  are up-scaled to the same resolution as the climate data i.e. ($0.5^\circ\times0.5^\circ$) before analysis.
An illustration of the evolution of the human niche is shown for the year 2060 (see Fig. \ref{fig:2060_suitability}) and 2090 (see Fig. \ref{fig:2090_suitability}) for the SSP1-2.6 and SSP5-8.5 scenarios.

\subsubsection{From human niche to large-scale migration}
We make the hypothesis that the difference in human niche $\Delta H_n$  at a given location $n$ between predictions and current situation quantifies a climate-related migration potential. In other words, we suppose that individuals tend to move from the places with the most degraded conditions (relative to current situation) to the places with the least degraded or improved conditions. Yet, current migration patterns are still mostly driven by economic opportunities \cite{niva_worlds_2023}. Therefore we suppose that future climate related migration will be a mix between moving to the nearest least degraded environment and reaching the nearest, opportunity-prone urban centers (potentially also least degraded).

To implement these assumptions, we use the $\Delta H_n$ as the fitness measure of the FERM (described in Sec.\ref{subsec:FERM}).
Since we do not have any knowledge on the shape of the benefit distribution, we define it as a Gaussian with mean $\Delta H_n$ and variance $\sigma=1$. When $\sigma \rightarrow 0$ the migration flows are purely driven by $\Delta H_n$ since the population of the nodes does not matter while $\sigma \rightarrow \infty$ reduces to the classical radiation model. Following \cite{raimondo_network_2022}, we set the variance $\sigma = 1$, which is also in the same order of magnitude with observed values of $\Delta H_n$ (see Fig. \ref{fig:2060_suitability} and Fig. \ref{fig:2090_suitability}). This assumption leads to migration flows depending both on change in the human niche and population at a given place (i.e. proxy of economic opportunity).
For each climate model, year, scenario, we compute the migration flows between approximately 70 000 ground
cells using the population projections data described in Tab. \ref{tab:raw_data} upscaled at the resolution $0.5^\circ\times 0.5^\circ$ equivalent to 70 000 ground cells. The migration flow $\mathbf{M}$ is computed on every populated point according to the sampling method described in Sec. \ref{subsec:FERM} (1000 iterations per
point). Then, we normalize the flow so that $\sum_j \mathbf{M}_{ij}=1$. Finally, the projected population after migration in cell $i$, $N_{h,mig}^i$ is updated in a diffusive fashion  $N_{h,mig}^i = \sum_j M_{ji} N_h^j$. Note that this is a major assumption, as it essentially implies that all individuals move.

\begin{figure}
    \centering
    \begin{adjustwidth}{0in}{0.5in}
    \includegraphics[width=1.1\linewidth]{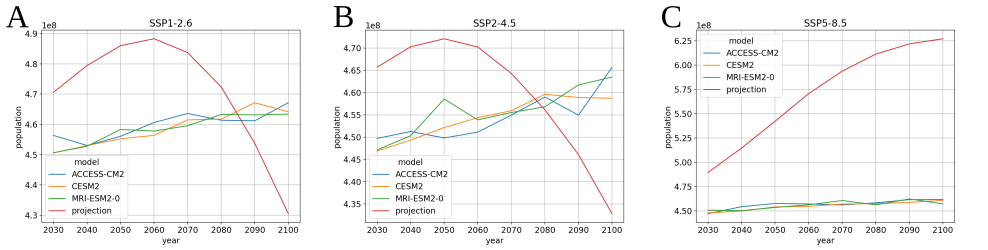}
    \caption{\textbf{Evolution of the total population of the Europe model during the century}.A) SSP1-2.6, B) SSP2-4.5, C) SSP5-8.5. Population projections are the one described in Tab\ref{tab:raw_data}. Climate-based projections are computed for each year based on the population projections.}
    \label{fig:migration_EU}
    \end{adjustwidth}
\end{figure}

To aggregate the predicted populations at the NUT3 level, we down-scaled our prediction to the original resolution of projected population data ($15\times15$ kms), before aggregating. However since the initial down-scaling introduces biases due to border effects, we rescale for each administrative unit the predicted population by a factor equals to the ratio of the projected population at vanilla resolution ($15\times15$ kms) over the projected population at resolution $50\times50$ kms. The projection from the migration models at the system-level can be seen in Fig. \ref{fig:migration_EU}. 

\subsection{FERM: Feature-Enriched Radiation Model}
\label{subsec:FERM}

\begin{figure*}
\centering
    \includegraphics[width=0.75\linewidth]{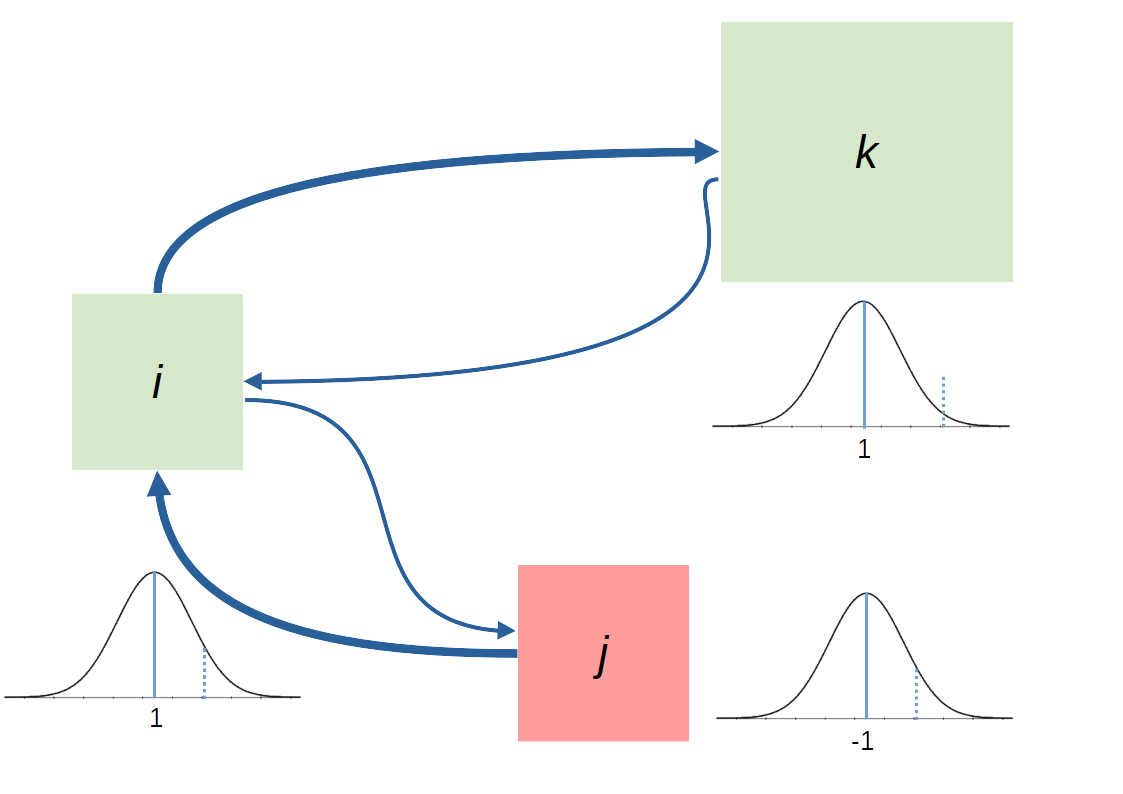}
    \caption{\textbf{Illustration of the mobility flux produced by the FERM model}. Each rectangle represents a point associated with their benefit distributions (here illustrated as gaussian distributions). The plain line represents the mean of the distribution and the dashed lines represents the expected value of the maximum of the distribution with a number of sample proportional to the size of the rectangle. The flux from point $i$ to point $k$ is large because the latter has a bigger size, therefore its distribution is sampled more. On the contrary, from point $j$ to point $i$ the flux is large because the quality of the opportunistic measure is higher although both points have the same size (mean of the gaussian distribution at 1 versus -1). }
    \label{fig:FER_sketch}
    
\end{figure*}

To quantify the migration induced by the change in human suitability we use the FERM, a generalized version of the radiation model \cite{simini_universal_2012}. The radiation model originally provides mobility flux between places based on their respective population which represents the variable of opportunity. Here, in the FERM, the opportunity is extended to any variables, alongside population.
To build the mobility flows, the principle of the model is inspired from emission and absorption processes in physics. In essence, we suppose that any locations in our system emit and absorb identical and independent particles. The more a location absorbs particles from another location, the stronger is the mobility flow between them (see Fig. \ref{fig:FER_sketch}). The emission/absorption processes is as follow:
\begin{itemize}
    \item We fix $m_i$ as the population of location $i$ and we define the probability distribution $F(x_i |\theta_i)$, with $\theta_i$ characterizing the distribution at location $i$ (e.g. the difference in human suitability $\Delta H_i$). We define $z_i$, the absorption threshold, as the maximum number obtained after $m_i$ trials from $F(x_i |\theta_i)$. In our numerical implementation we sample this maximum through adaptive rejection sampling since it is analytically intractable in general.
    \item Then, we consider location $j$,the closest neighbor of location $i$, with population $n_j$. We define $z_j$, the absorbance threshold, as the maximum number obtained after $n_j$ trials from $F(x_j |\theta_j)$. If $z_i<z_j$, we consider that the particle emitted from location $i$ is absorbed in location $j$. We update the mobility flow between location $i$ and location $j$, $M_{ij} \to M_{ij}+1$.
    \item Else, we consider the second closest neighbor in location $k$ and repeat the process of the previous step. The process stops when the particle emitted in location $i$ is finally absorbed in a given location.
    \item Finally, a large number of particle are emitted from location $i$ and from any other location to obtain the mobility flow $M_{ij}$ which is normalized such that $\sum_j M_{ij}=1 $
\end{itemize}

Formally, the quality of the opportunities  is computed by a benefit distribution $F(x)$ (with associated PDF, $\frac{dF(x)}{dx} = p(x)$). Contrary to the classical radiation model, distributions are a priori different for any given point $i$ and $j$ i.e. $F(x_i|\theta_i) \neq F(x_j|\theta_j)$ where $\theta_i$ and $\theta_j$ are the parameter vectors characterizing the two distributions (see Fig. \ref{fig:FER_sketch}). The conditional probability that an individual departing from node $i$ travels to node $j$ is given by $P(1|m_i,n_j,L_{ij},\theta_i,\theta_j)$:
\begin{adjustwidth}{-0.45in}{0in}
    \begin{align}
    P(1|m_i,n_j,L_{ij},\theta_i,\theta_j) = \int_0^\infty \int_0^\infty m_i F(x_i|\theta_i) ^{m_i-1} \frac{dF(x_i|\theta_i) }{dx_i} n_j F(x_j|\theta_j) ^{n_j-1} \frac{dF(x_j|\theta_j) }{dx_j} \prod_{k \in L_{ij}}[1 - F_U(u_{ik})]F_U(u_{ij})dx_idx_j,
\end{align}
\end{adjustwidth}

where $m_i$ and $n_j$ are the populations at point $i$ and $j$ respectively, $L_{ij}$ is the set of the indices of the locations in the circle of radius $r_{ij}$ centered in $i$.  The term in the $x_i$ variable represents the probability 
that the maximum value extracted from $F(x_i|\theta_i)$ after $m_i$ trials is equal to $x_i$. The term in the $x_j$ variable represents the probability 
that the maximum value extracted from $F(x_j|\theta_j)$ after $n_j$ trials is equal to $x_j$. he variable $U$ is the difference between the fitness: $U_{ij} = x_i - x_j$. The product term represents the probability that the values extracted in in location $k\in L_{ij}$ is always less than the values extracted in location $i$. Finally, the last term is the probability that the value extracted in $j$ is greater than the value extracted in $j$ i.e. that the absorbance threshold in $j$ is greater than the absorption threshold in $i$. 

\begin{figure}
\centering
    \includegraphics[width=\linewidth]{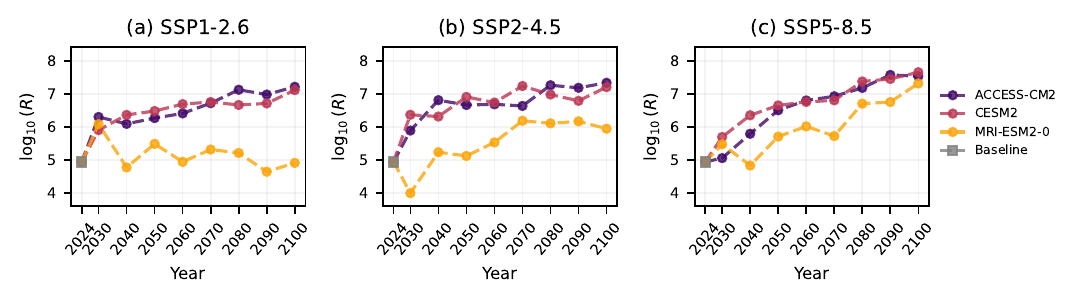}
    \caption{\textbf{Evolution of the population-at-risk index for  dengue infection in Europe from 2024 to 2100  according to the joint model.} Magnitude
of the cumulative recovered individual for the  ACCESS-CM2, CESM2 and MRI-ESM2-0 climate projections on the \textbf{Left)} SSP1-2.6, \textbf{Middle)} SSP2-4.5 and \textbf{Right)} SSP5-8.5 scenarios. Every simulation is performed by decade from 2030 onward to 2100. Historical cumulated recovered individuals are estimated for the year 2024. Population is estimated from every year according to population projections \cite{jones_spatially_2016}. The order of magnitude is computed as the logarithm in base 10 of the total number of recovered individual in a year. Simulations are performed under the value of parameters defined in Tab.\ref{tab:value_model}.
    }
    \label{fig:recovered}
\end{figure}

\begin{figure}
    \centering
    \includegraphics[width=0.5\linewidth]{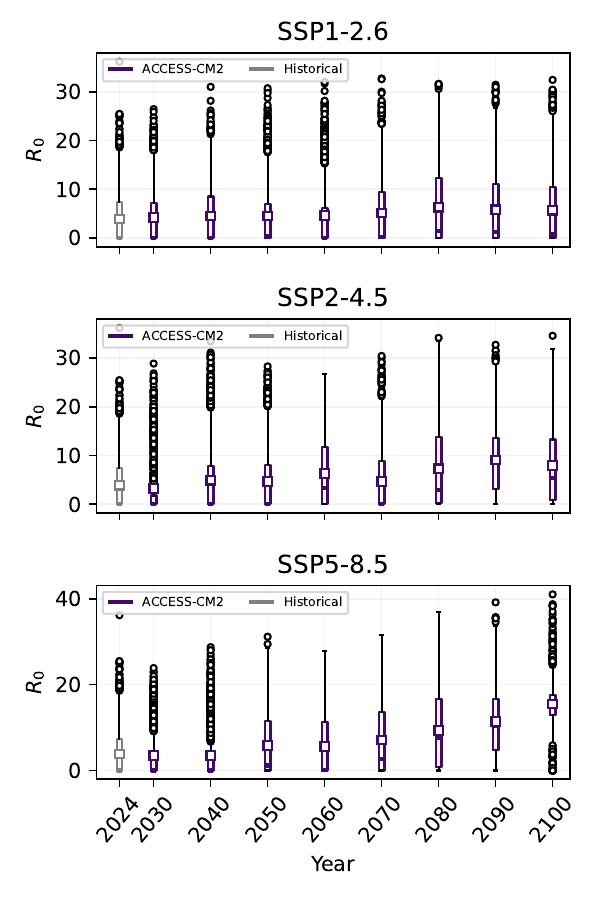}
    \caption{\textbf{Boxplot of the number of weeks the basic reproduction number is in the endemic regime $R_0 >1$ in function of time using the ACCESS-CM2 climate projection .}
    We consider \textbf{Up)} the SSP1-2.6, \textbf{Middle)} the SSP2-4.5 and \textbf{Bottom)} the SSP5-8.5 scenarios.
    The basic reproduction number is computed daily from 1st January to 31st December for each patch.
    All patches are considered in isolation for the computation of the basic reproduction number. Outliers are located at 1.5  times the limits of the interquartile range. Computations are performed under the value of parameters defined in Tab.\ref{tab:value_model}. }
    \label{fig:r0_access}
\end{figure}

\begin{figure}
    \centering
    \includegraphics[width=0.5\linewidth]{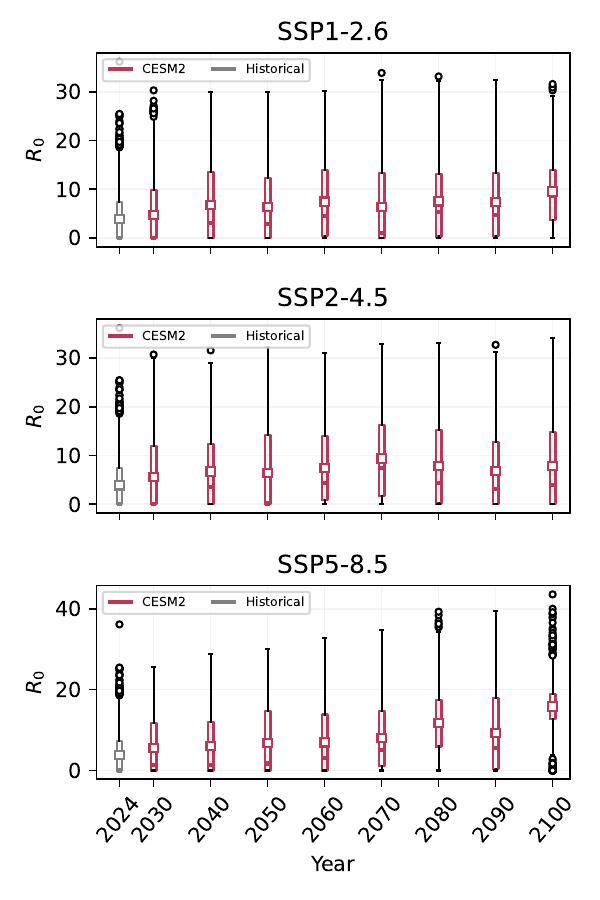}
    \caption{\textbf{Boxplot of the number of weeks the basic reproduction number is in the endemic regime $R_0 >1$ in function of time using the CESM2 climate projection .}
    We consider \textbf{Up)} the SSP1-2.6, \textbf{Middle)} the SSP2-4.5 and \textbf{Bottom)} the SSP5-8.5.
    The basic reproduction number is computed daily from 1st January to 31st December for each patch.
    All patches are considered in isolation for the computation of the basic reproduction number. Outliers are located at 1.5  times the limits of the interquartile range. Computations are performed under the value of parameters defined in Tab.\ref{tab:value_model}. }
    \label{fig:r0_cesm2}
\end{figure}

\begin{figure}
    \centering
    \includegraphics[width=0.5\linewidth]{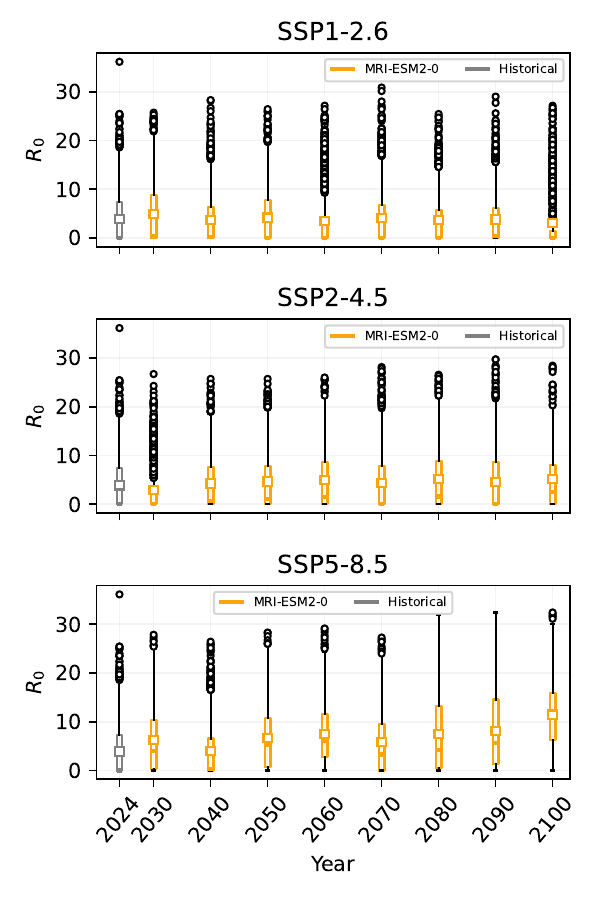}
    \caption{\textbf{Boxplot of the number of weeks the basic reproduction number is in the endemic regime $R_0 >1$ in function of time using the MRI-ESM2-0 climate projection .}
    We consider \textbf{Up)} the SSP1-2.6, \textbf{Middle)} the SSP2-4.5 and \textbf{Bottom)} the SSP5-8.5.
    The basic reproduction number is computed daily from 1st January to 31st December for each patch.
    All patches are considered in isolation for the computation of the basic reproduction number. Outliers are located at 1.5  times the limits of the interquartile range. Computations are performed under the value of parameters defined in Tab.\ref{tab:value_model}. }
    \label{fig:r0_mri}
\end{figure}

\begin{figure}
\centering
    \includegraphics[width=\linewidth]{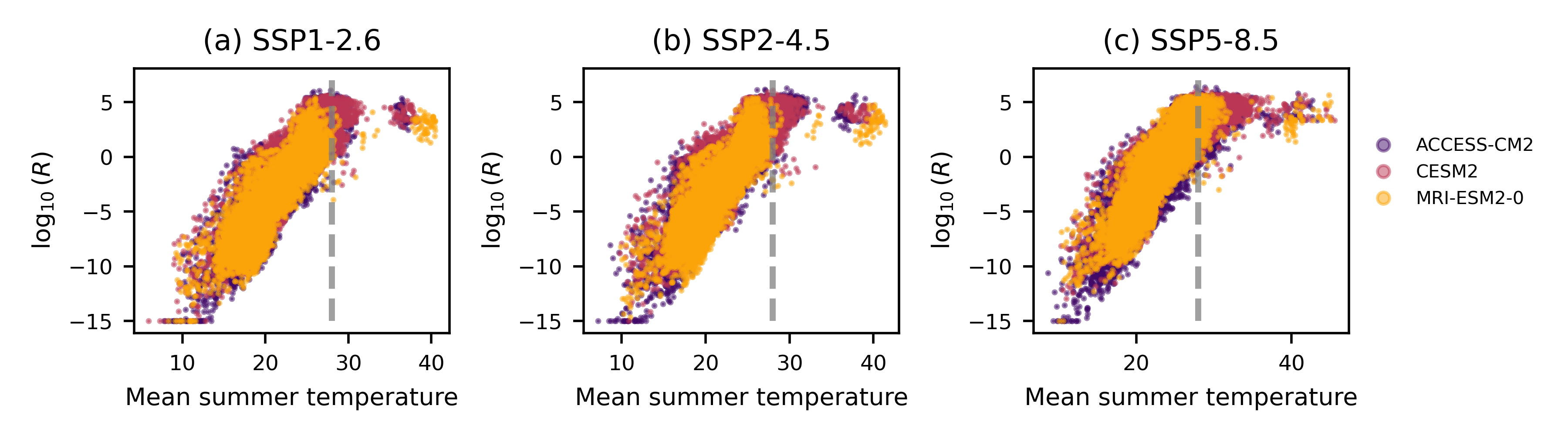}
    \caption{\textbf{Magnitude of the population-at-risk index in function of the daily mean summer temperature for each NUT3 patch of the joint model in each year}. Every decade from 2030 and 2100 is represented. The dashed grey line represents the isotherm \text{$28^\circ$}C which is approximately the upper bound of the optimal transmission temperature for \textit{Aedes albopictus} in the transmission model \cite{mordecai_detecting_2017}. The mean summer temperature is the daily average temperature from 1st June to 1st September (31st May-31st August in leap years). The magnitude of the population-at-risk index is computed as the logarithm in base 10 of the cumulative recovered individuals in a given year.}
    \label{fig:temperature_correlation}.
\end{figure}

\begin{figure}
    \centering
    \includegraphics[width=\linewidth]{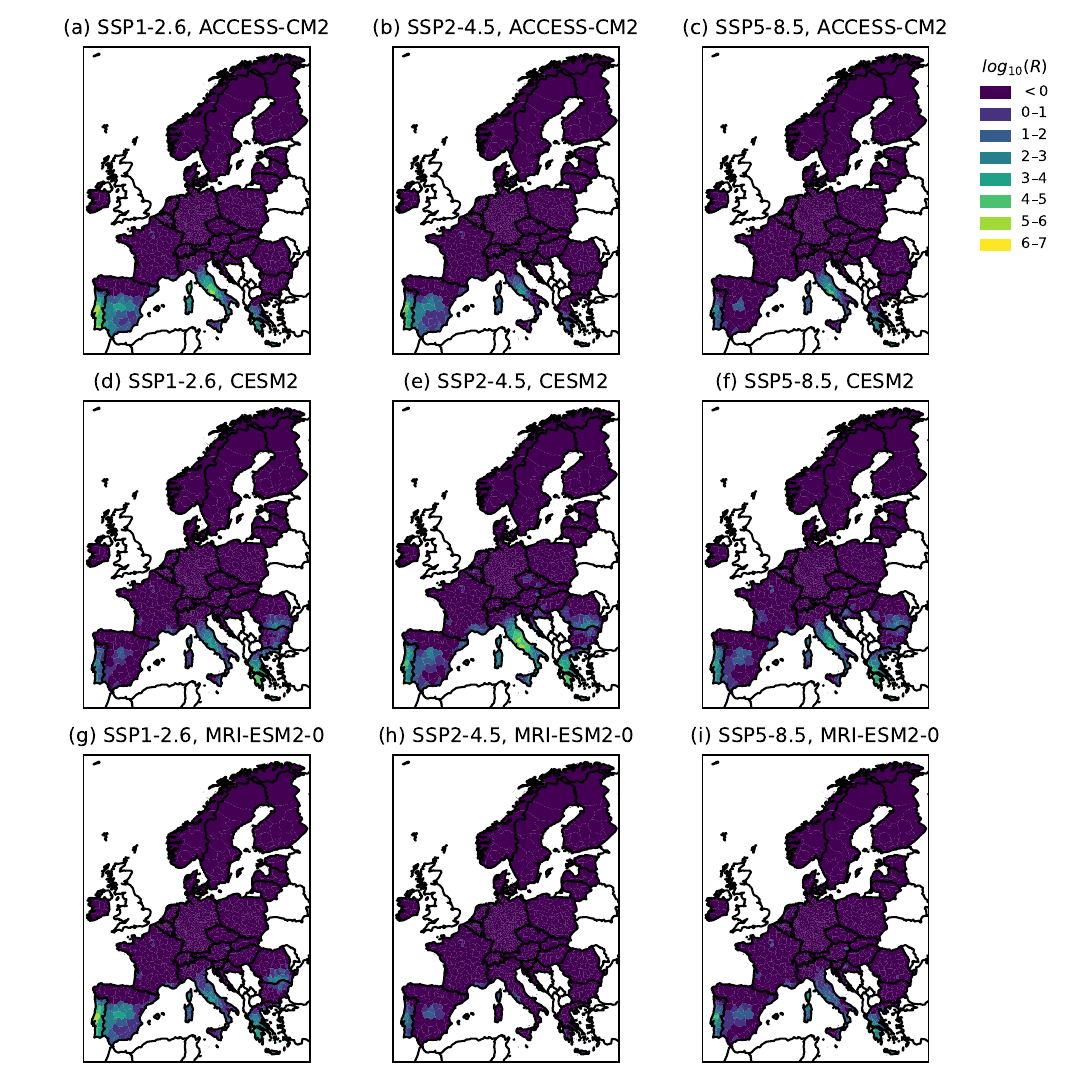}
    \caption{\textbf{Population-at-risk index of dengue infection in 2030 using the joint model}. Magnitude of the cumulative recovered individuals for the ACCESS-CM2 climate projections in the SSP1-2.6 (A), SSP2-4.5 (B) and SSP5-8.5 (C). Magnitude of the cumulative recovered individuals for the CESM2 climate projections in the SSP1-2.6 (D), SSP2-4.5 (E) and SSP5-8.5 (F). Magnitude of the cumulative recovered individuals the MRI-ESM2-0 climate projections in the SSP1-2.6 (G), SSP2-4.5 (H) and SSP5-8.5 (I). The magnitude of the population-at-risk index is computed as the logarithm in base 10 of the cumulative recovered individuals in a given year.
    Simulation are performed under the value of parameters defined in Tab.\ref{tab:value_model}.}
    \label{fig:DENV_sim1_SSP1}
\end{figure}

\begin{figure}
    \centering
    \includegraphics[width=\linewidth]{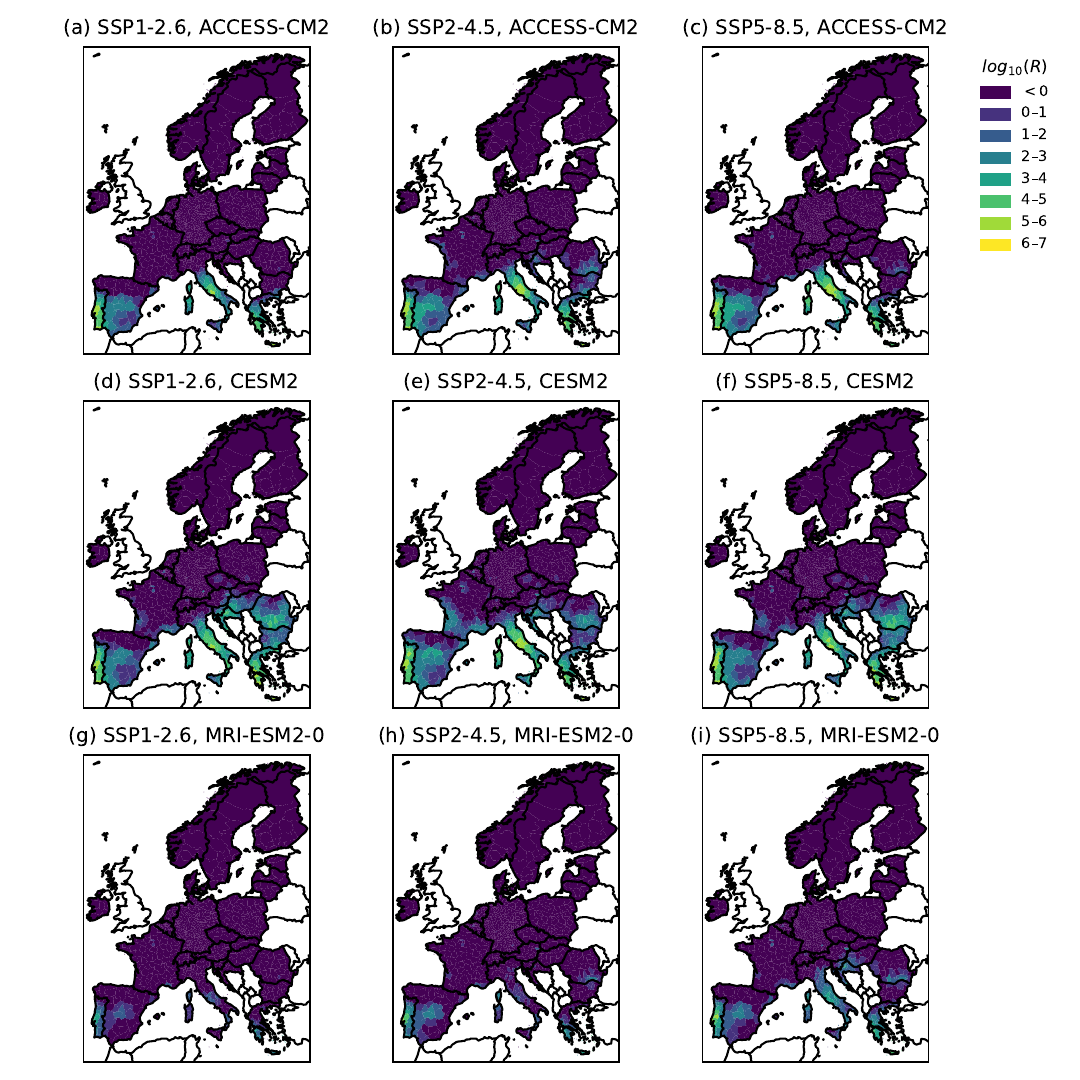}
    \caption{\textbf{Population-at-risk index of dengue infection in 2060 using the joint model}. Magnitude of the cumulative recovered individuals for the ACCESS-CM2 climate projections in the SSP1-2.6 (A), SSP2-4.5 (B) and SSP5-8.5 (C). Magnitude of the cumulative recovered individuals for the CESM2 climate projections in the SSP1-2.6 (D), SSP2-4.5 (E) and SSP5-8.5 (F). Magnitude of the cumulative recovered individuals the MRI-ESM2-0 climate projections in the SSP1-2.6 (G), SSP2-4.5 (H) and SSP5-8.5 (I). The magnitude of the population-at-risk index is computed as the logarithm in base 10 of the cumulative recovered individuals in a given year.
    Simulation are performed under the value of parameters defined in Tab.\ref{tab:value_model}.}
    \label{fig:DENV_sim1_SSP5}
\end{figure}

\begin{figure}
    \centering
    \includegraphics[width=\linewidth]{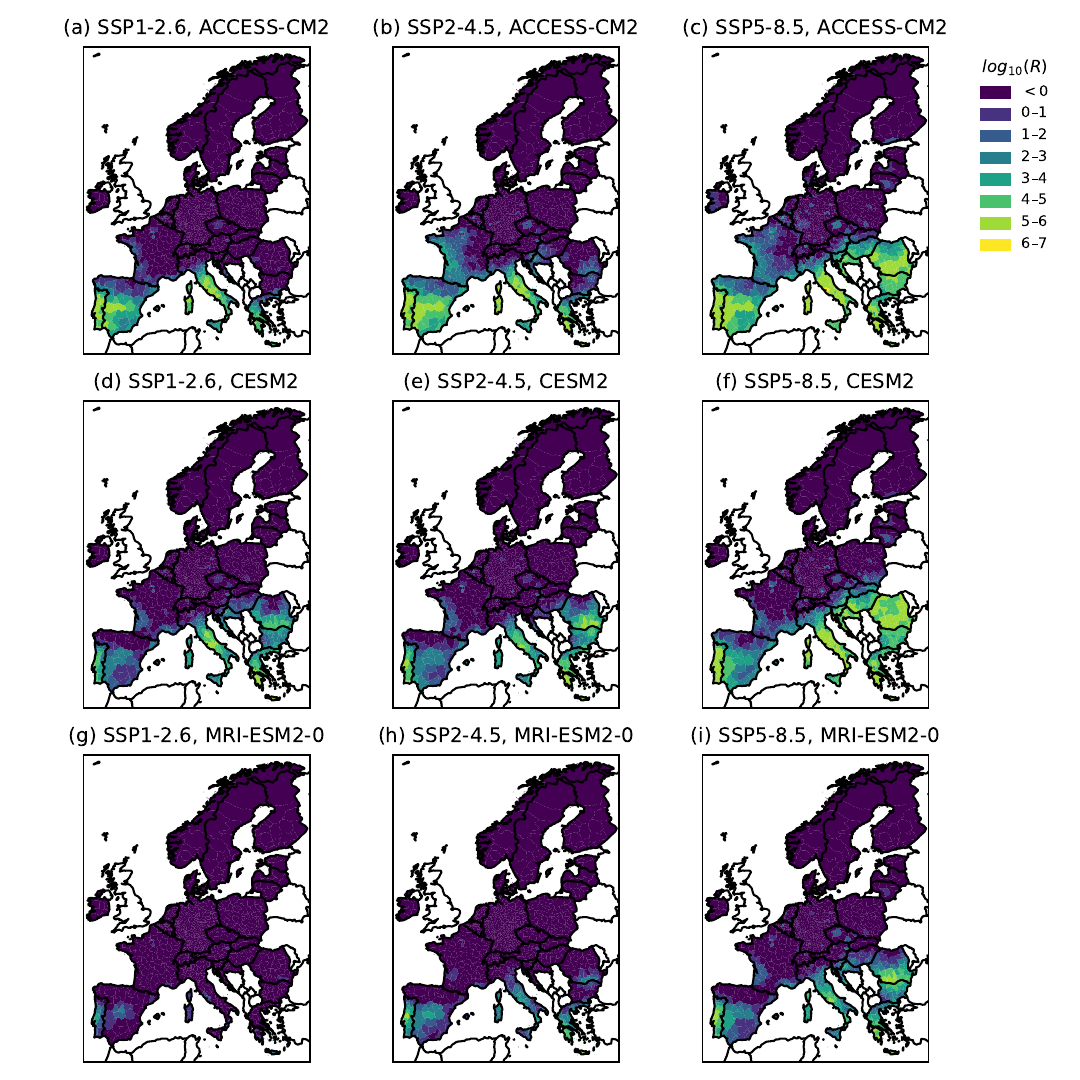}
    \caption{\textbf{Population-at-risk index of dengue infection in 2090 using the joint model}. Magnitude of the cumulative recovered individuals for the ACCESS-CM2 climate projections in the SSP1-2.6 (A), SSP2-4.5 (B) and SSP5-8.5 (C). Magnitude of the cumulative recovered individuals for the CESM2 climate projections in the SSP1-2.6 (D), SSP2-4.5 (E) and SSP5-8.5 (F). Magnitude of the cumulative recovered individuals the MRI-ESM2-0 climate projections in the SSP1-2.6 (G), SSP2-4.5 (H) and SSP5-8.5 (I). The magnitude of the population-at-risk index is computed as the logarithm in base 10 of the cumulative recovered individuals in a given year.
    Simulation are performed under the value of parameters defined in Tab.\ref{tab:value_model}.}
    \label{fig:DENV_sim1A_SSP1}
\end{figure}

\begin{figure}
\centering
    \includegraphics[width=\linewidth]{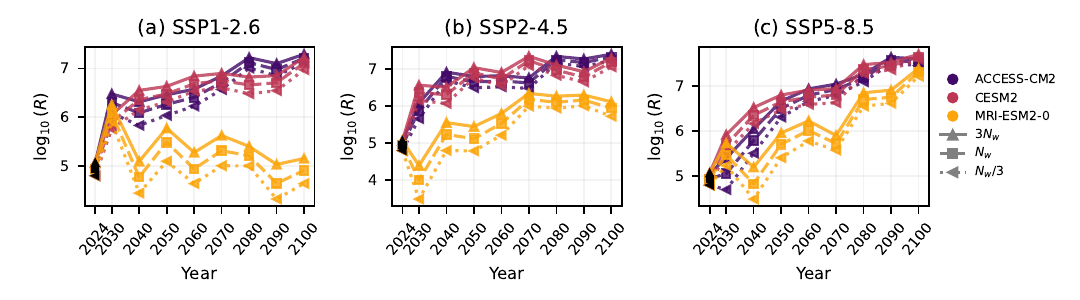}
    \caption{\textbf{Evolution of the estimated population-at-risk index of dengue  in Europe from 2024 to 2100  across different importation scenarios.} Magnitude of the cumulative recovered individual for the  ACCESS-CM2, CESM2 and MRI-ESM2-0 climate projections on the \textbf{Left)} SSP1-2.6, \textbf{Middle)} SSP2-4.5 and \textbf{Right)} SSP5-8.5 scenarios. Together with the base scenario (daily importation rate $N_w$) we consider a low importation scenario (daily importation rate $\frac{N_w}{3}$) and a high importation scenario (daily importation rate $3N_w$). Every simulation is performed by decade from 2030 onward to 2100. Historical cumulated recovered individuals are estimated for the year 2024. Population is estimated from every year according to population projections \cite{jones_spatially_2016}. The order of magnitude is computed as the logarithm in base 10 of the total number of recovered individual in a year. Simulation are performed under the value of parameters defined in Tab.\ref{tab:value_model}.}
    \label{fig:recovered}.
\end{figure}

\begin{figure}
\centering
    \includegraphics[width=\linewidth]{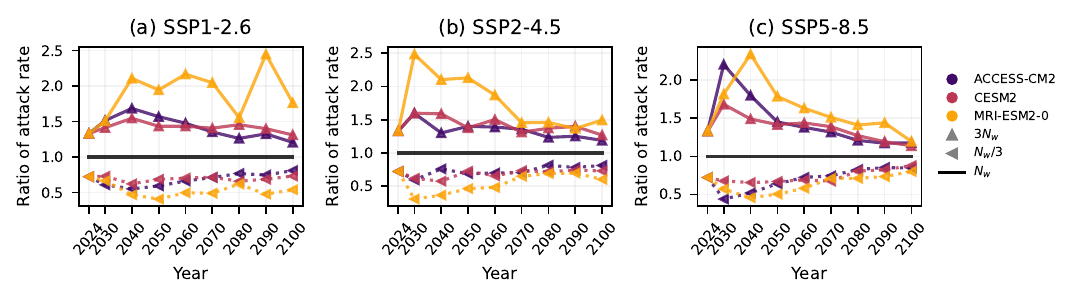}
    \caption{\textbf{Evolution of the proportion of the population-at-risk index in Europe from 2024 to 2100  between different importation scenarios over the base scenario.}
    Ratio of attack rate across importation scenarios for the  ACCESS-CM2, CESM2 and MRI-ESM2-0 climate projections on the \textbf{Left)} SSP1-2.6, \textbf{Middle)} SSP2-4.5 and \textbf{Right)} SSP5-8.5 scenarios. Together with the baseline scenario (daily importation rate $N_w$) we consider a low importation scenario (daily importation rate $\frac{N_w}{3}$) and a high importation scenario (daily importation rate $3N_w$). The base importation scenario is represented with a black line. Together with the baseline scenario (daily importation rate $N_w$) we consider a low importation scenario (daily importation rate $\frac{N_w}{3}$) and a high importation scenario (daily importation rate $3N_w$). Population is estimated from every year according to population projections \cite{jones_spatially_2016}. Simulation are performed under the value of parameters defined in Tab.\ref{tab:value_model}.}
    \label{fig:recovered}.
\end{figure}

\begin{figure}
\centering
    \includegraphics[width=\linewidth]{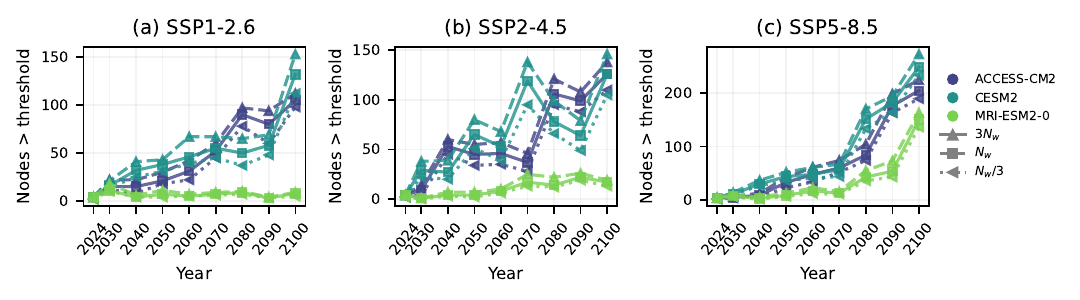}
    \caption{\textbf{Evolution of number of high-risk NUTS3 patches in Europe from 2024 to 2100 across different importation scenarios.} Number of NUTS3 patches whose attack rate exceeds $20\%$ i.e. high-risk patches in the SSP2-4.5 using the ACCESS-CM2, CESM2 and MRI-ESM2-0 climate projections on the \textbf{Left)} SSP1-2.6, \textbf{Middle)} SSP2-4.5 and \textbf{Right)} SSP5-8.5. Together with the base scenario (daily importation rate $N_w$) we consider a low importation scenario (daily importation rate $\frac{N_w}{3}$) and a high importation scenario (daily importation rate $3N_w$). Every simulation is performed by decade from 2030 onward to 2100.  Population is estimated from every year according to population projections \cite{jones_spatially_2016}. Simulation are performed under the value of parameters defined in Tab.\ref{tab:value_model}.}
    \label{fig:number_nodes}
\end{figure}

\begin{figure}
\centering
    \includegraphics[width=\linewidth
]{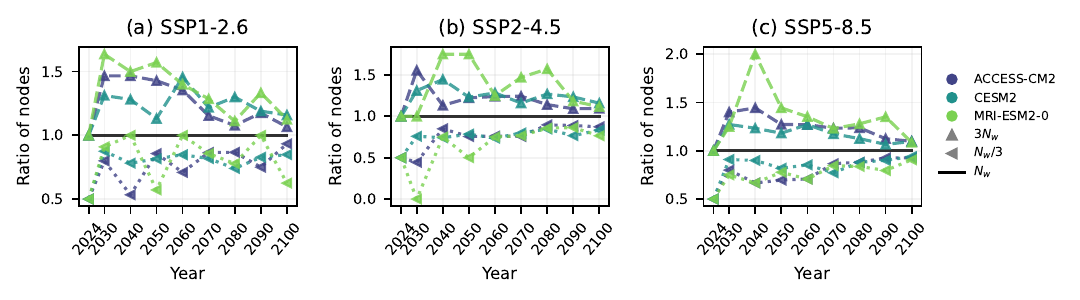}
    \caption{\textbf{Evolution of the ratio of high-risk NUTS3 patches in Europe across different importation scenarios over the base one.} Ratio of the number of high-risk patches (whose attack rate are higher than $20\%$ in the high importation scenario (dashed lines) and the low importation scenario (dotted lines) over the baseline importation scenario using the ACCESS-CM2, CESM2 and MRI-ESM2-0 climate projections on the \textbf{Left)} SSP1-2.6, \textbf{Middle)} SSP2-4.5 and \textbf{Right)} SSP5-8.5. Every simulation is performed by decade from 2030 onward to 2100.  Population is estimated from every year according to population projections \cite{jones_spatially_2016}. Simulation are performed under the value of parameters defined in Tab.\ref{tab:value_model}.}
    \label{fig:ratio_nodes}
\end{figure}

\begin{figure}
    \centering
    \includegraphics[width=\linewidth]{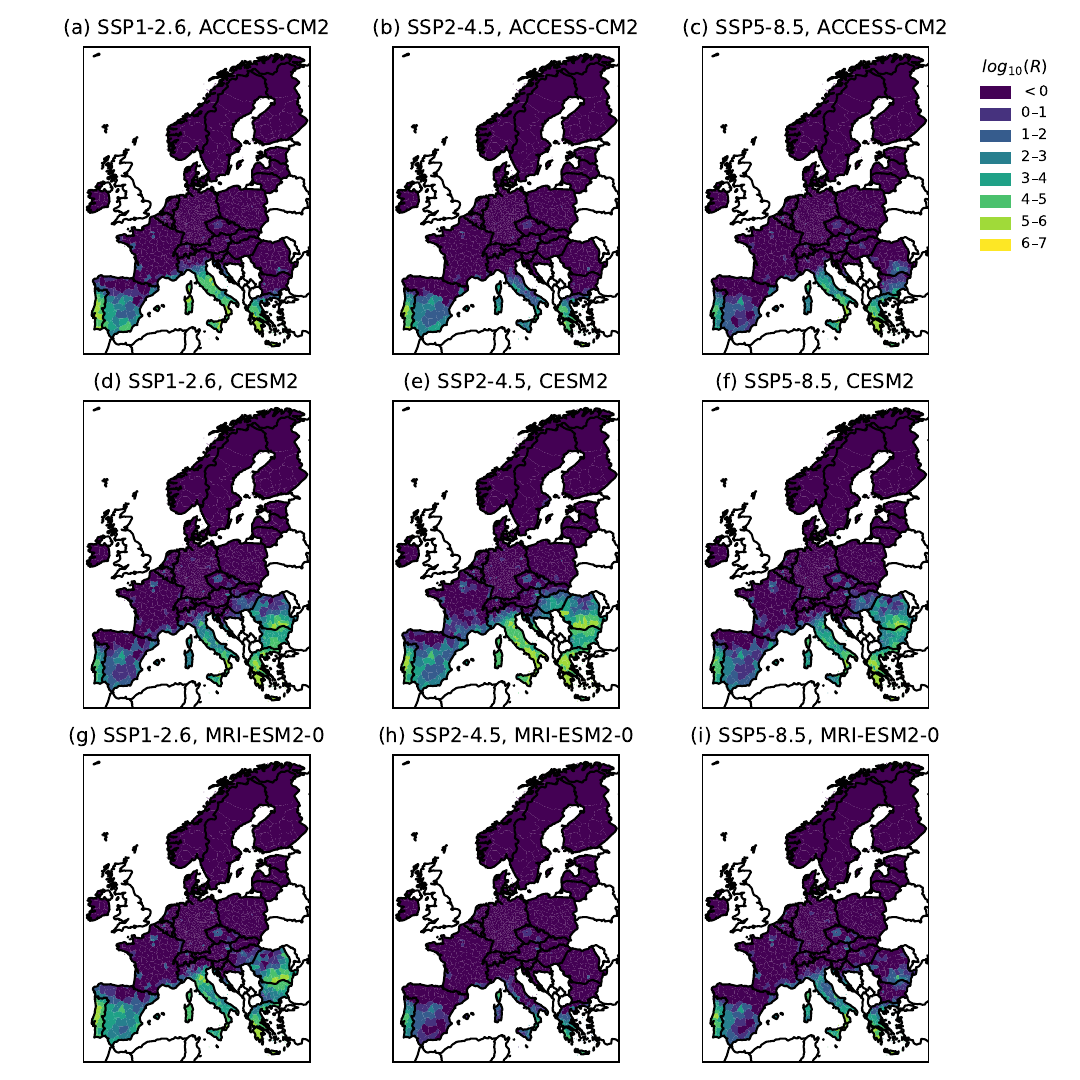}
    \caption{\textbf{Population-at-risk index of dengue infection in 2030 using alternative population projections from the migration model}. Magnitude of the cumulative recovered individuals for the ACCESS-CM2 climate projections in the SSP1-2.6 (A), SSP2-4.5 (B) and SSP5-8.5 (C). Magnitude of the cumulative recovered individuals for the CESM2 climate projections in the SSP1-2.6 (D), SSP2-4.5 (E) and SSP5-8.5 (F). Magnitude of the cumulative recovered individuals the MRI-ESM2-0 climate projections in the SSP1-2.6 (G), SSP2-4.5 (H) and SSP5-8.5 (I). The magnitude of the population-at-risk index is computed as the logarithm in base 10 of the cumulative recovered individuals in a given year.
    Simulation are performed under the value of parameters defined in Tab.\ref{tab:value_model}.}
    \label{fig:DENV_sim1_SSP1}
\end{figure}

\begin{figure}
    \centering
    \includegraphics[width=\linewidth]{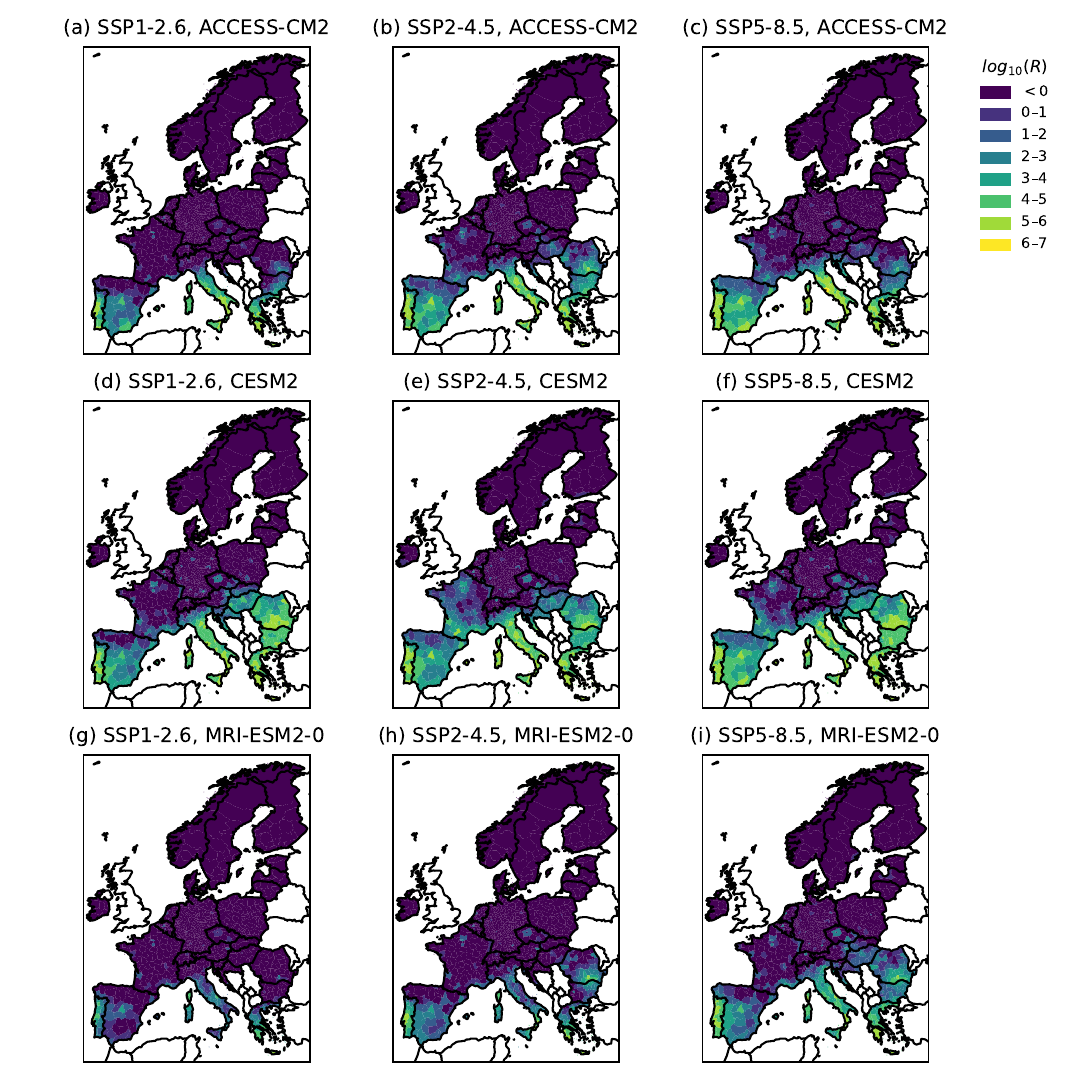}
    \caption{\textbf{Population-at-risk index of dengue infection in 2060 using alternative population projections from the migration model}. Magnitude of the cumulative recovered individuals for the ACCESS-CM2 climate projections in the SSP1-2.6 (A), SSP2-4.5 (B) and SSP5-8.5 (C). Magnitude of the cumulative recovered individuals for the CESM2 climate projections in the SSP1-2.6 (D), SSP2-4.5 (E) and SSP5-8.5 (F). Magnitude of the cumulative recovered individuals the MRI-ESM2-0 climate projections in the SSP1-2.6 (G), SSP2-4.5 (H) and SSP5-8.5 (I). The magnitude of the population-at-risk index is computed as the logarithm in base 10 of the cumulative recovered individuals in a given year.
    Simulation are performed under the value of parameters defined in Tab.\ref{tab:value_model}.}
    \label{fig:DENV_sim1_SSP5}
\end{figure}

\begin{figure}
    \centering
    \includegraphics[width=\linewidth]{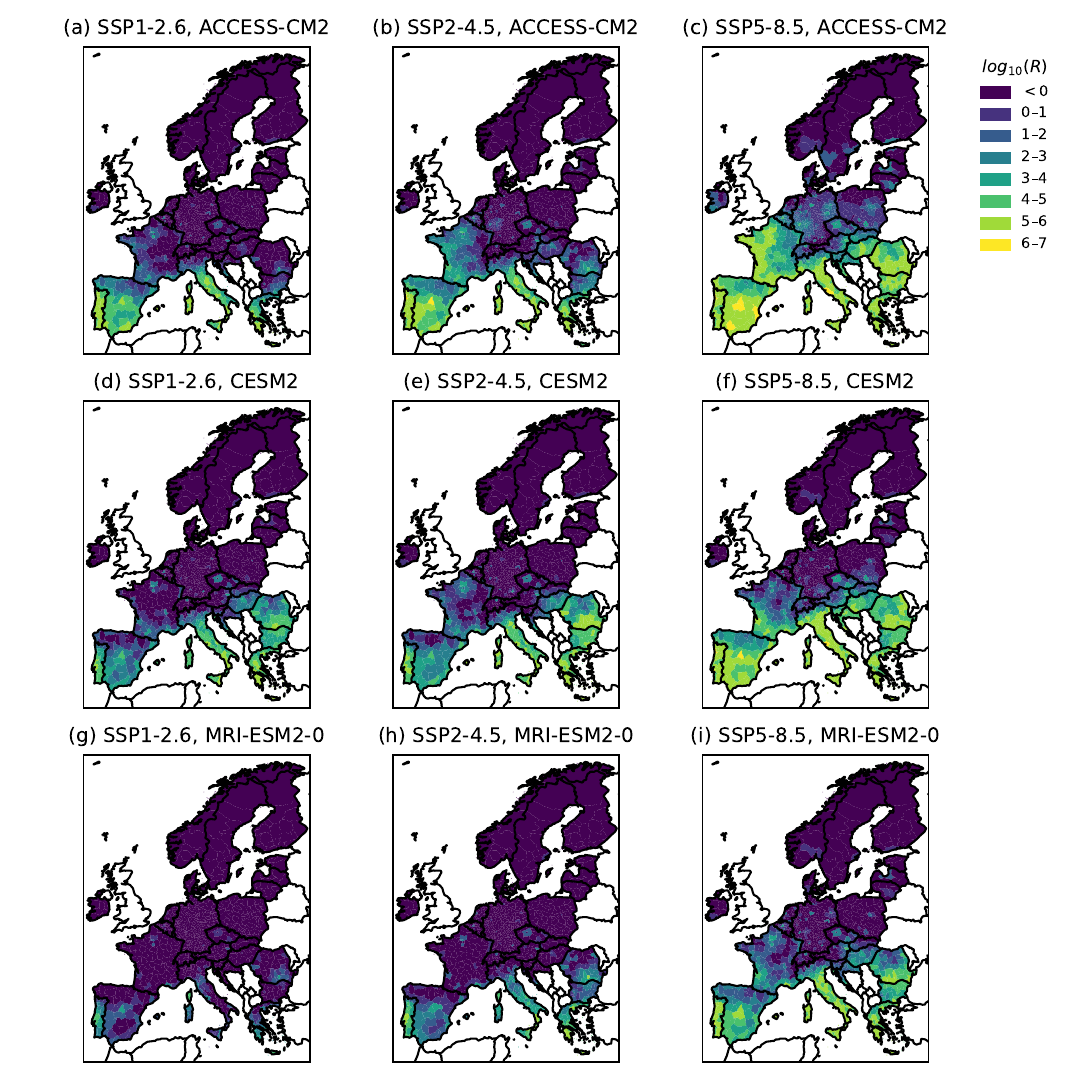}
    \caption{\textbf{Population-at-risk index of dengue infection in 2090 using alternative population projections from the migration model}. Magnitude of the cumulative recovered individuals for the ACCESS-CM2 climate projections in the SSP1-2.6 (A), SSP2-4.5 (B) and SSP5-8.5 (C). Magnitude of the cumulative recovered individuals for the CESM2 climate projections in the SSP1-2.6 (D), SSP2-4.5 (E) and SSP5-8.5 (F). Magnitude of the cumulative recovered individuals the MRI-ESM2-0 climate projections in the SSP1-2.6 (G), SSP2-4.5 (H) and SSP5-8.5 (I). The magnitude of the population-at-risk index is computed as the logarithm in base 10 of the cumulative recovered individuals in a given year.
    Simulation are performed under the value of parameters defined in Tab.\ref{tab:value_model}.}
    \label{fig:DENV_sim1A_SSP1}
\end{figure}
\clearpage
%\bibliographystyle{plain}
%\bibliography{PRIN_bibliography}